\documentclass[a4paper,11pt]{article}

\usepackage[a4paper,left=2.73cm,right=2.7cm,top=3cm,bottom=3.5cm]{geometry}

\usepackage{amsmath,amssymb}

\usepackage[colorlinks=true,linktocpage=true,linkcolor=blue,citecolor=blue,urlcolor=blue]{hyperref}

\usepackage[sort&compress,merge,numbers]{natbib}

\addtocontents{toc}{\protect\setcounter{tocdepth}{2}}
\numberwithin{equation}{section}

\usepackage{epsfig}
\usepackage{graphicx}
\usepackage{epstopdf}
\usepackage{caption}
\usepackage{subcaption}

\def\cA{{\cal A}}
\def\cT{{\cal T}}

\def\cD{{\cal D}}

\def\cQ{{\cal Q}}

\def\wt{\widetilde}

\newcommand*\DAlembert{\mathop{}\!\mathbin\Box}
\newcommand{\Ku}[1]{K^#1\,}
\newcommand{\Gdu}[2]{{g_{#1}}^{#2}}
\newcommand{\Gdd}[2]{g_{#1#2}}
\newcommand{\Guu}[2]{g^{#1#2}}
\newcommand{\hdu}[2]{{\gamma_{#1}}^{#2}}
\newcommand{\hdd}[2]{\gamma_{#1#2}}
\newcommand{\huu}[2]{\gamma^{#1#2}}
\newcommand{\hud}[2]{{\gamma^{#1}}_{#2}}
\newcommand{\pdu}[2]{{\perp_{#1}}^{#2}}
\newcommand{\pud}[2]{{\perp^{#1}}_{#2}}
\newcommand{\pdd}[2]{\perp_{#1#2}}
\newcommand{\puu}[2]{\perp^{#1#2}}
\newcommand{\Bdd}[2]{R_{#1#2}}
\newcommand{\Rdd}[2]{{\cal R}_{#1#2}}

\newcommand{\Pdd}[2]{S_{#1#2}}
\newcommand{\Pdu}[2]{{S_{#1}}^{#2}}

\newcommand{\Kddu}[3]{{K_{#1#2}}{^{#3}}}
\newcommand{\Kddd}[3]{{K_{#1#2#3}}}
\newcommand{\Kuuu}[3]{{K^{#1#2#3}}}
\newcommand{\Kuud}[3]{{K^{#1#2}}{_{#3}}}
\newcommand{\Kudd}[3]{{K^{#1}}{_{#2#3}}}
\newcommand{\Kduu}[3]{{K_{#1}}{^{#2#3}}}
\newcommand{\Kdud}[3]{{{K_{#1}}{^{#2}}}{_{#3}}}
\newcommand{\Cddu}[3]{{C_{#1#2}}{^{#3}}}
\newcommand{\Bddud}[4]{{{R_{#1#2}}{^{#3}}}{}{_{#4}}}
\newcommand{\Buddd}[4]{{{R^{#1}}{_{#2#3#4}}}}
\newcommand{\Bdddd}[4]{R_{#1#2#3#4}}

\newcommand{\Bdduu}[4]{{R_{#1#2}}{^{#3#4}}}

\newcommand{\Ruddd}[4]{{{\cal R}^{#1}}{_{#2#3#4}}}
\newcommand{\Rdddd}[4]{{{\cal R}}_{#1#2#3#4}}

\newcommand{\Wdduu}[4]{{W_{#1#2}}{^{#3#4}}}

\newcommand{\Ouddd}[4]{{\Omega^{#1}}{_{#2#3#4}}}
\newcommand{\Odddd}[4]{\Omega_{#1#2#3#4}}
\newcommand{\oduu}[3]{ {\omega_{#1}}{^{#2#3} }}
\newcommand{\oddu}[3]{ {\omega_{#1#2}}{^{#3} }}
\newcommand{\odud}[3]{ {{\omega_{#1}}^{#2}}{_{#3} }}
\newcommand{\ouuu}[3]{ {\omega}^{#1#2#3} }
\newcommand{\oudd}[3]{ {\omega^{#1}}{_{#2#3} }}

\newcommand{\Tuu}[2]{\boldsymbol{T}^{#1#2}}
\newcommand{\eud}[2]{{e^{#1}}{_{#2}}}
\newcommand{\edu}[2]{{e_{#1}}{^{#2}}}
\newcommand{\edd}[2]{{e_{#1}}{_{#2}}}
\newcommand{\euu}[2]{{e^{#1}}{^{#2}}}
\newcommand{\nud}[2]{{n^{#1}}{_{#2}}}
\newcommand{\ndu}[2]{{n_{#1}}{^{#2}}}
\newcommand{\nuu}[2]{{n^{#1}}{^{#2}}}
\newcommand{\ndd}[2]{{n_{#1}}{_{#2}}}
\newcommand{\Ksddu}[3]{{{\mathcal{K}_{#1#2}}}{^{#3}}}
\newcommand{\Ksduu}[3]{{\mathcal{K}_{#1}}{^{#2#3}}}
\newcommand{\Ksdud}[3]{{{\mathcal{K}_{#1}}^{#2}}{_{#3}}}
\newcommand{\Ksuud}[3]{{{\mathcal{K}}^{#1#2}}{_{#3}}}
\newcommand{\hsdu}[2]{{h_{#1}}{^{#2}}}
\newcommand{\hsdd}[2]{h_{#1#2}}
\newcommand{\hsuu}[2]{h^{#1#2}}
\newcommand{\hsud}[2]{{h^{#1}}{_{#2}}}
\newcommand{\osduu}[3]{ {\varpi_{#1}}{^{#2#3} }}

\newcommand{\psud}[2]{{P^{#1}}{_{#2}}}
\newcommand{\psdd}[2]{P_{#1#2}}
\newcommand{\psuu}[2]{P^{#1#2}}
\newcommand{\tku}[1]{\mathbf{k}^{#1}}
\newcommand{\tkd}[1]{\mathbf{k}_{#1}}
\newcommand{\nnb}{\mathfrak{D}}

\newcommand{\be}{\begin{equation}}
\newcommand{\ee}{\end{equation}}
\newcommand{\bal}{\begin{align}}
\newcommand{\bse}{\begin{subequations}}
\newcommand{\ese}{\end{subequations}}

\begin{document}

\begin{titlepage}

\thispagestyle{empty}

\begin{flushright}
\hfill{}
\end{flushright}

\vspace{40pt}  
	 
\begin{center}

{\LARGE \textbf{On actions for (entangling) surfaces and DCFTs}}
	\vspace{30pt}
		
{\large \bf Jay Armas  and Javier Tarr\'\i o}
		
\vspace{25pt}

{Universit\'e Libre de Bruxelles (ULB) and International Solvay Institutes,\\
Service  de Physique Th\'eorique et Math\'ematique, \\
Campus de la Plaine, CP 231, B-1050, Brussels, Belgium.}

\vspace{20pt}
{\tt 
\href{mailto:jarmas@ulb.ac.be}{jarmas@ulb.ac.be}, 
\href{mailto:jtarrio@ulb.ac.be}{jtarrio@ulb.ac.be}
}

\vspace{40pt}
				
\abstract{
The dynamics of surfaces and interfaces describe many physical systems, including fluid membranes, entanglement entropy and the coupling of defects to quantum field theories. Based on the formulation of submanifold calculus developed by Carter, we introduce a new variational principle for (entangling) surfaces. This principle captures all diffeomorphism constraints on surface/interface actions and their associated spacetime stress tensor. The different couplings to the geometric tensors appearing in the surface action are interpreted in terms of response coefficients within elasticity theory. An example of a surface action with edges at the two-derivative level is studied, including both the parity-even and parity-odd sectors. Its conformally invariant counterpart restricts the type of conformal anomalies that can appear in two-dimensional submanifolds with boundaries. Analogously to hydrodynamics, it is shown that classification methods can be used to constrain the stress tensor of (entangling) surfaces at a given order in derivatives. This analysis reveals a purely geometric parity-odd contribution to the Young modulus of a thin elastic membrane. Extending this novel variational principle to BCFTs and DCFTs in curved spacetimes allows to obtain the Ward identities for diffeomorphism and Weyl transformations. In this context, we provide a formal derivation of the contact terms in the stress tensor and of the displacement operator for a broad class of actions.

}

\end{center}

\end{titlepage}

\tableofcontents

\hrulefill
\vspace{10pt}

\section{Introduction} \label{sec:intro}
The dynamics of surfaces and interfaces describe a wide range of physical systems and physical phenomena. One common example in nature is that of interfaces between different fluids or fluid phases (e.g. soap bubbles). Another example is that of small deformations of thin elastic membranes: at mesoscopic scales, the physical state of lipid membranes is well captured by the geometric degrees of freedom of the membrane \cite{Guven2018}, while at microscopic scales one may describe the coupling between quantum field theories and interfaces/defects by approximating the latter as thin surfaces \cite{McAvity:1993ue, Jensen:2015swa, Billo:2016cpy} or obtain entanglement properties of quantum field theories by extremising surface functionals \cite{Ryu:2006bv, Hubeny:2007xt}. Many of these systems can be modelled by effective theories for the dynamics of surfaces, with numerous applications ranging from cosmic strings \cite{Carter:1997pb}, black hole physics \cite{Emparan:2009cs, Emparan:2009at, Armas:2013hsa, Armas:2015ssd} to the dynamics of D-branes \cite{Johnson:2000ch}, to mention only a few.

The recent interest in some of these topics have prompted us to explore in detail the formal aspects of these functionals within a framework that allows to treat both bulk and surface actions (or vacuum energy functionals) simultaneously, thereby describing this wide range of physical systems. Based on the spacetime approach to submanifold calculus developed by Carter \cite{Carter:1992vb, Carter:1993wy, Carter:1997pb}, this paper formulates a novel variational principle for surface actions. This formulation determines the constraints on surface actions and its associated spacetime stress tensor due to general covariance. These constraints had been largely overlooked and, in a purely geometric setting, restrict the type of contributions that can appear in entanglement entropy functionals. In addition, the variational principle developed here allows for a direct derivation of the spacetime stress tensor and displacement operator associated with surface/defect actions. In turn, this leads to a straightforward extraction of Ward identities and can be used to evaluate correlation functions in conformal field theories (CFTs) and constrain CFT data as in \cite{Billo:2016cpy}. Extensions of this formalism to include background gauge and dilaton fields are only natural.

Although the work that will be presented in this paper is broadly applicable, our original motivation lies in the recent interest in two different research directions. The first one was initiated by the Ryu-Takayanagi proposal \cite{Ryu:2006bv}, and its covariant counterpart \cite{Hubeny:2007xt}, stating that the  extremisation of geometric functionals provides information, via holographic dualities, of the entangling properties of the dual quantum field theory. The nature of these geometric functionals depends on the specifics of the gravitational theory. In particular, in higher derivative gravity theories these geometric functionals may depend on the extrinsic curvature of the surface or the background Riemann tensor \cite{Casini:2011kv, Camps:2013zua, Dong:2013qoa}. The extremisation of these functionals is required in order to extract information about the entanglement entropy of the corresponding dual quantum field theory \cite{Dong:2017xht}. The second research direction is the study of the properties of conformal field theories with boundaries (BCFTs) and of CFTs coupled to defects (DCFTs) \cite{Cardy:1984bb, Cardy:1991tv, McAvity:1993ue, McAvity:1995zd, Liendo:2012hy, Bianchi:2015liz, Jensen:2015swa, Billo:2016cpy, Liendo:2016ymz, Rastelli:2017ecj}.\footnote{In the holographic context, BCFTs and DCFTs have been approached in numerous ways, e.g. \cite{Karch:2000gx, DeWolfe:2001pq, Erdmenger:2002ex, Aharony:2003qf, Erdmenger:2014xya, Erdmenger:2015spo, Bianchi:2015liz, deLeeuw:2015hxa, Buhl-Mortensen:2016pxs, deLeeuw:2017dkd} to mention only a few.} In this context, it is necessary to understand how to correctly couple a quantum field theory living on a boundary/interface or defect, regardless of its shape, to a given bulk CFT. This much is required in order to obtain the Ward identities for such theories \cite{McAvity:1993ue, Jensen:2015swa, Billo:2016cpy}.

These considerations and research directions have lead us to perform a thorough analysis of the constraints on surface functionals with non-trivial edge geometry due to diffeomorphism invariance. The naive expectation, and a common misconception, is that, analogous to spacetime actions and spacetime tensors, covariant surface actions can be built by simply appropriately contracting surface tensors (e.g. contractions of extrinsic curvatures). However, this is not the case, as these diffeomorphism constraints impose stronger restrictions, in particular at the edges of the surface. As we shall see, one of these constraints leads to the shape equation itself, describing the surface dynamics, but many other constraints, which have been largely overlooked in the literature, must be satisfied (see Eqs.\eqref{eq:st}-\eqref{eq:pt}, Eq.~\eqref{eq:genshapetang} and Eq.~\eqref{eq:st1}). In the context of conformal anomalies for submanifolds with boundaries, these constraints restrict the type of anomalies that can be present,
while in the context of DCFTs, they play an important role: once implemented in the contact terms in the stress tensor and in the displacement operator they can be used to correctly evaluate correlation functions and to obtain conserved currents and charges.

When dealing with entanglement entropy functionals, the main interest lies in determining the shape equation and evaluating the on-shell value of the entropy functional at the extrema. In this context, barely no attention is given to the physical interpretation of the different couplings that appear in surface actions. However, from the broader point of view that such actions describe a wide range of physical systems such as lipid membranes \cite{Canham197061,Helfrich1973, doi:10.1080/00018739700101488, 0305-4470-37-47-010}, it is important to understand the physical meaning of these couplings. Following the general approach of \cite{Armas:2013hsa}, we interpret the different structures appearing in surface actions in the context of elasticity theory. In Sec.~\ref{sec:2derivative}, we identify a new purely geometric parity-odd contribution to the surface's Young modulus, which breaks the classical symmetries associated with an elasticity tensor. Couplings to the background Riemann tensor that appear in the context of entanglement entropy \cite{Casini:2011kv, Camps:2013zua, Dong:2013qoa} can be interpreted as quadrupole moments of stress which introduce new force terms in the shape equation and characterise the response of the surface to changes in background curvature.\footnote{The interpretation of these couplings in terms of elasticity theory had been suggested in \cite{Camps:2013zua}.} This leads to an extension of classical elasticity theory \cite{Landau:1959te} for the deformations of thin membranes.\footnote{Elastic membranes are characterised by a thickness scale $\tau$. In order to write effective theories for their deformations, one considers deformations with wavelength much larger than the membrane thickness. In this regime, the membrane geometry can be described by a surface action whose response coefficients are dimensionful. The accompanying membrane stress tensor has a derivative expansion in terms of the membrane thickness.}

\subsubsection*{Different approaches to submanifold calculus} 
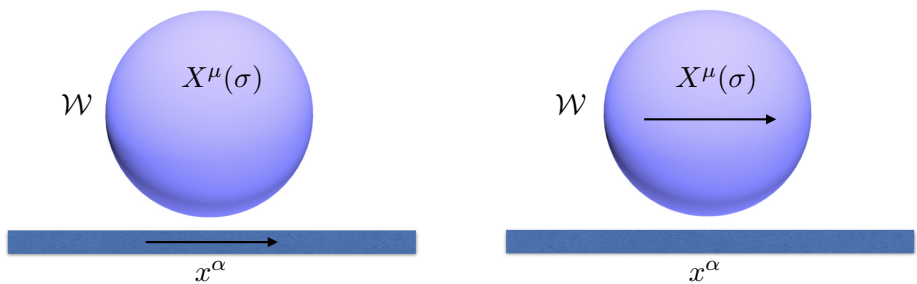
\begin{figure}[!ht]
\centerline{ \includegraphics[scale=0.4]{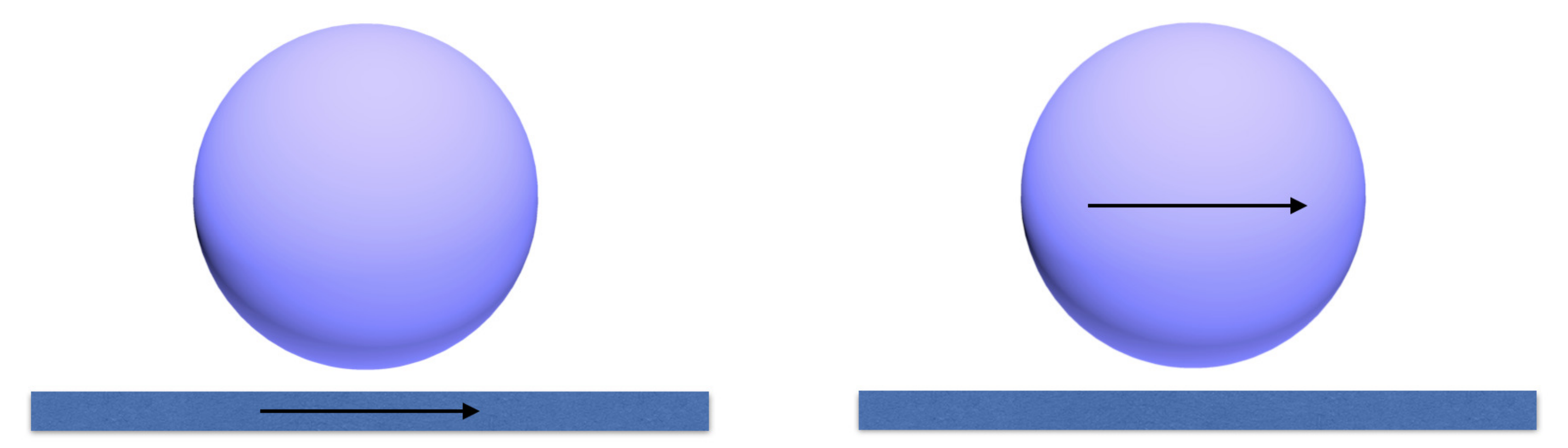}}
 \begin{picture}(0,0)(0,0)
  \put(115,80){ $X^{\mu}(\sigma)$}
   \put(300,80){ $X^{\mu}(\sigma)$}
    \put(120,7){ $x^\alpha$}
      \put(305,7){ $x^\alpha$}
        \put(70,70){ $\mathcal{W}$}
             \put(255,70){ $\mathcal{W}$}
\end{picture} 
\caption{Schematic representation of the two variational methods for the surface $\mathcal{W}$. The figure on the left corresponds to method \textbf{(1)}, where the background coordinates $x^\alpha$ are displaced while the embedding map $X^\mu$ is kept fixed.  The figure on the right corresponds to method \textbf{(2)} where the background coordinates are kept fixed and the embedding map is displaced. The first method can be thought of displacing a "mat" (background) underneath the surface while the second can be thought of displacing the surface above the "mat".} \label{fig:sur}
\end{figure}

This paper deals with equilibrium surfaces, that is, surfaces whose shape is determined by the extrema of an action or vacuum energy functional. We assume that these actions or vacuum energy functionals are functionals of geometric fields only, namely the background metric $\Gdd\mu\nu(x)$, the embedding map of the surface $X^\mu(\sigma)$, and the embedding map of its edges $\wt X^{\mu}(\wt \sigma)$. In order to perform variations of these functionals and obtain the shape equation and remaining diffeomorphism constraints one may either \textbf{(1)} displace the background coordinates by an infinitesimal amount along some vector field $\xi^{\mu}$ such that $x^{\mu}\to x^\mu+\xi^{\mu}(x)$ while keeping the embedding map fixed or \textbf{(2)} displace the embedding map by a small amount $X^{\mu}(\sigma)\to X^{\mu}(\sigma)+\delta X^{\mu}(\sigma)$ while keeping the background coordinates fixed. We show in Sec.~\ref{sec:DCFTs} that these two methods yield the same equations of motion once certain constraints are satisfied. This is illustrated in Fig.~\ref{fig:sur}. 
If the change in background coordinates is compensated by a corresponding change in the embedding map, the surface  is not displaced, though certain constraints must be satisfied. Method \textbf{(2)} formally involves working with a foliation of surfaces, even if just in a local neighbourhood, and it leads to a non-manifestly covariant intermediate calculus.\footnote{Though it is possible to work with covariant deformations as developed by Guven et al. \cite{Guven:1993ex, Capovilla:1994bs}. Other variational methods using auxiliary variables or constrained variations are also available in the literature and deserve further exploration \cite{Guven:2004wd, Guven:2012hn}. } On the other hand, when using method \textbf{(1)}, the variational calculus is covariant and it  is sufficient to work with a single surface. This is shown explicitly in App.~\ref{app:variations}. When dealing with (entangling) surfaces one may use either one of the methods in order to obtain the equations of motion (modulo constraints), but due to the natural covariant properties and simplicity of method \textbf{(1)}, we discard \textbf{(2)}. In the context of DCFTs, however, both methods must be employed since the defect must be coupled regardless of its shape.

The action functional can be cast under two different formulations. One of these formulations, which we refer to as \emph{gauge formulation} due its natural analogue in gauge theory, consists in working with natural quantities defined on the surface and using tangential and transverse indices. For example, the induced metric on the surface $\gamma_{ab}$ has only tangential indices $a,b$ while the extrinsic curvature $\Kddu abi$ has two tangential indices and one transverse index $i$. This is, for instance, the type of approach followed in \cite{Capovilla:1994bs, Arreaga:2000mr, Armas:2013hsa} and in most of the entanglement entropy literature (see \cite{Rangamani:2016dms} and references therein). The other formulation is the one championed by Carter in \cite{Carter:1992vb, Carter:1993wy, Carter:1997pb}, which we refer to as \emph{spacetime formulation}. Within this formulation, it is only necessary to work with a single type of indices, namely spacetime indices $\mu,\nu,\cdots$, such that using the set of tangent and normal vectors to the surface $\{\eud\mu a~,~\nud\mu i\}$ one may define the spacetime analogue of the induced metric and extrinsic curvature as $\huu\mu\nu=\eud\mu a\eud \nu a\gamma^{ab}$ and $\Kddu\mu\nu\rho=\edu\mu a\edu\nu b\nud\rho i\Kddu abi$, respectively. Notice that the latter fields have support on the surface and are \emph{not} fields in spacetime. The necessity of working with two types of indices in a gauge formulation is traded by the necessity of keeping track of the index order in a given geometric structure in a spacetime formulation.

Both formulations introduced above have advantages and disadvantages. In particular, the gauge formulation makes it more accessible to evaluate variations of purely intrinsic quantities, such as the intrinsic Ricci scalar, while the spacetime formulation makes it more accessible to compute variations of background quantities, such as the background Ricci scalar. However, as we will argue and demonstrate, the spacetime formulation of the action functional has several clear advantages. For example, when dealing with composite systems of bulk and surface it is only natural to use spacetime indices since it is senseless to differentiate between tangential and normal indices in the bulk, unless one could introduce a foliation of surfaces or defects in the entire spacetime but this is neither always possible nor necessary.  Furthermore, when working with a single set of indices, general covariance needs to be ensured only in that single set.\footnote{We would like to invite the reader to get acquainted with the non-linear history of index proliferation in submanifold calculus by reading the introductory remarks of \cite{Carter:1992vb}. } In a gauge formulation, general covariance must be required on two different sets of indices separately, namely, on the tangent and normal bundles. Most approaches to variational calculus within a gauge formulation introduce a specific coordinate system \cite{PhysRevA.39.5280, Capovilla:1994bs, Charmousis:2005ey, Fonda:2016ine}, leading to non-manifestly spacetime covariant intermediate steps.\footnote{This issue could be bypassed by simply performing variations using method \textbf{(1)} within this formulation as for instance in \cite{Billo:2016cpy, Armas:2016mes}.} Finally, as we shall demonstrate, specifically in App.~\ref{app:highdev}, the spacetime formulation allows to extract all diffeomorphism constraints on surface actions and their associated spacetime stress tensor while the gauge formulation does not. For these reasons we adopt this spacetime formulation in the core of this paper and develop it further not only by extending its variational calculus but also by formulating a new variational principle for surface actions.

\subsubsection*{Organisation of the material} 
In Sec.~\ref{sec:surfaces} we first introduce the reader to this spacetime formulation of surface geometry, as not only is this paper's intent to be introductory to those who would like to pursue submanifold calculus, but also because the majority of works in both entanglement entropy and DCFTs have adopted the gauge formulation of the action principle. We then introduce the new variational principle for surfaces and obtain the diffeomorphism constraints and shape equation for surfaces/interfaces with non-trivial edges or intersections (see Eqs.\eqref{eq:st}-\eqref{eq:pt}, Eq.~\eqref{eq:genshapetang} and Eq.~\eqref{eq:st1}). This action includes couplings to several geometric structures with at most two-derivatives, though the action itself can contain contributions with an arbitrary number of derivatives.

In Sec.~\ref{sec:stresstensor} we show how to extract the spacetime stress tensor associated with these surface actions. This stress tensor has a multipole expansion in terms of derivatives of the delta function. We analyse its symmetry properties and equivalent formulations. We identify frame-invariant tensor structures which can be used to count independent couplings to the surface action. We then use this spacetime stress tensor to obtain conserved surface/edge currents and charges. 

In Sec.~\ref{sec:2derivative}, we analyse a generic two-derivative action based on a classification of the different couplings that can appear in both the parity-even and parity-odd sectors and interpret them in the context of elasticity theory. We impose the diffeomorphism constraints in order to eliminate some of these contributions, which in the presence of edges  leads to highly non-trivial relations between surface and edge couplings. Using methods analogous to those employed in hydrodynamics, we show in Sec.~\ref{sec:2derivative} that these methods can be used to constrain the stress tensor of (entangling) surfaces.  We furthermore study the constraints imposed by Weyl invariance and comment on their implications for conformal anomalies of two-dimensional submanifolds. 

In Sec.~\ref{sec:DCFTs} we extend the variational principle within this spacetime formulation to BCFTs and DCFTs. In this context, we obtain Ward identities for defects with edges coupled to bulk CFTs. We furthermore derive explicitly the contact terms in the spacetime stress tensor. We encounter a mismatch with other ad-hoc forms of the stress tensor in previous literature. We also derive a full-fledged displacement operator in curved spacetime for a broad class of actions.

Finally, in Sec.~\ref{sec:discussion} we conclude with a brief summary and future research directions. We also provide App.~\ref{app:variations}, which contains variational formulae for geometric tensors while App.~\ref{app:highdev} contains further details on the different types of variational principles.

\section{Geometric actions for (entangling) surfaces and interfaces}\label{sec:surfaces}

This section introduces a new variational principle for surfaces/interfaces based on the notion of Lagrangian variations (method \textbf{(1)} introduced above) within the spacetime formulation championed by Carter \cite{Carter:1992vb, Carter:1993wy, Battye:1995hv, Carter:1997pb}. The advantage of this strategy resides in its potential to capture all diffeomorphism constraints on surface actions and to yield directly the components of the spacetime stress tensor associated to these surfaces. This formalism is applicable to many physical systems and is useful for the study of extremal surfaces that describe, via holographic dualities, the entangling properties of the dual quantum field theory.

The first part of this section introduces the reader to the geometric quantities associated to surfaces in the spacetime formulation as well as the notation that is used throughout the paper. We then proceed and show how the surface/interface dynamics and shape equations emerge from the requirement of diffeomorphism invariance, including the possibility of non-trivial edges and intersections. The latter requires considerable attention due to the several constraints imposed by the well-definiteness of the variational principle. 


\subsection{Geometry of submanifolds and geometric tensors} \label{sec:geometry}

We consider a $D$-dimensional spacetime $\mathcal{M}$ endowed with a non-degenerate metric $g_{\mu\nu}(x^{\alpha})$, where $x^{\alpha}$ denotes the background spacetime coordinates and the Greek indices $\mu,\nu,\alpha,\cdots$ denote spacetime indices. In this spacetime we place a $p$-dimensional surface $\mathcal{W}$ with edges $\partial\mathcal{W}$. The location of this surface of codimension $n=D-p$ is described by the embedding map $X^{\mu}(\sigma^{a})$, where $\sigma^{a}$ ($a=1,\cdots,p$) denote the coordinates on the surface, collectively denoted by $\sigma$. 

Given the mapping functions, the tangent vectors take the form $\eud\mu a=\partial_a X^{\mu}$. In turn, one may introduce the induced metric on the surface
\be
\gamma_{ab}= \Gdd\mu\nu\, \eud\mu a  \eud\nu b \ ,
\ee
with inverse matrix components $\gamma^{ab}$. We assume that neither $\gamma_{ab}$ nor its inverse are null at any point on the surface.
The set of normal vectors $\nud \mu i$, where $i$ denotes the transverse $n$ directions is implicitly defined via the relations
\be \label{eq:relations}
\nuu \mu i \ndu \mu j=\delta^{ij} \ , \quad \eud\mu a \ndu \mu i=0 \ .
\ee
These conditions, though sufficient to describe the surface, do not fix entirely the normal vectors, as they allow for the freedom of shifting the normal vectors by a sign or by a rotation $\ndu \mu i\to {\omega^{i}}_j \ndu \mu j$ where $\omega^{ij}$ is an anti-symmetric matrix in O($n$).\footnote{\label{foot:O}We are choosing an orthonormal frame in the transverse space. In the context of entangling surfaces, which are codimension $n=2$ surfaces, the time-like direction can be Wick rotated so that the transverse space is still purely spatial. One may also work with a time-like transverse space for which $\nuu \mu i \ndu \mu j=\eta^{ij}$ and the matrix $\omega^{ij}$ is now an anti-symmetric matrix in O(1,$n-1$). Our results can be straightforwardly applied to the latter case by replacing $\delta^{ij}\to\eta^{ij}$. }

A spacetime covariant approach can be formulated by appropriately contracting geometric tensors with the tangential and normal vectors. This avoids the use of tangential and orthogonal indices, $a$ and $i$, in favour of spacetime indices $\mu$. In particular, the induced metric and transverse metric can be expressed as
\be \label{eq:metrics}
\huu\mu\nu=\eud\mu a\eud\nu b\gamma^{ab} \ , \quad \puu\mu\nu=\nud \mu i\nuu \nu i \ .
\ee
Given the conditions \eqref{eq:relations}, these two structures are obviously orthogonal to each other, i.e. $\huu\mu\nu\pud\lambda\mu=0$. This decomposition implies that we have chosen to describe the surface in an adapted frame for which the background metric, restricted to the surface, can be decomposed as 
\be
\Gdd\mu\nu=\hdd\mu\nu+\pdd\mu\nu \ .
\ee
This naturally breaks the diffeomorphism symmetries of the background spacetime into general coordinate transformations on the $p$-dimensional surface and (generalised) rotations in the $n$-dimensional transverse space to the surface. This can only be expected since the presence of the surface in the spacetime will naturally break some of the background symmetries. 


\subsubsection{Covariant differentiation}

The tensors $\hdd\mu\nu$ and $\pdd\mu\nu$  in \eqref{eq:metrics}, as well as the several geometric tensors that we introduce below, have support only on the surface, i.e., they are not well defined anywhere else in spacetime, with the only exceptions of background fields, such as the background metric $\Gdd\mu\nu$ and its derivatives, which are well defined everywhere in spacetime. It is possible to extend all geometric tensors to the entirety of spacetime by working with a foliation of surfaces \cite{Capovilla:1994yk, Carter:1996wr, Charmousis:2005ey},  however this is beyond the scope of this paper and unnecessary for the calculus of variations of surface actions.

As a result of the necessity to restrict to tensors with support on the surface, covariant differentiation of tensors with support on the surface is not a well defined operation: only its tangential projection is. Therefore we introduce the surface covariant derivative 
\be
\overline\nabla_\lambda=\hud\mu\lambda\nabla_\mu \ ,
\ee
where $\nabla_\mu$ is the spacetime covariant derivative associated with $\Gdd\mu\nu$ and its corresponding Christoffel connection $\Gamma^{\rho}_{\mu\nu}$. When applied to tensors with support only on the surface the operator $\overline\nabla_\lambda$ is meaningful, whereas the orthogonal projection of the background covariant derivative, $\pud\mu\lambda\nabla_\mu$, is not. On the other hand, if the covariant derivative acts on a tensor with support on the whole background spacetime both projections are well defined.


\subsubsection{The extrinsic curvature tensor}

Given a well defined notion of covariant differentiation, one may introduce several geometric objects of interest, characterising how the submanifold is embedded in the background spacetime. These can be obtained by acting with the operator $\overline\nabla_\mu$ on tangent and normal vectors (and contractions thereof). 

The first example is the extrinsic curvature of the surface
\be \label{eq:extc}
\Kddu\mu\nu\rho=\hud\sigma\nu\overline\nabla_\mu\hud\rho\sigma=-\hud\sigma\nu\overline\nabla_\mu\pud\rho\sigma \ ,
\ee 
which describes the rate of change of the normal vectors along surface directions. The extrinsic curvature \eqref{eq:extc} transforms as a tensor in its spacetime indices and is invariant under changes of the tangential and normal vectors that satisfy \eqref{eq:relations}. Opting for a spacetime covariant formalism, in which there is no reference to surface and transverse indices, requires keeping fixed the order of the indices in any given geometric structure. In particular, the extrinsic curvature is tangential and symmetric in its first two indices and orthogonal in its last index\footnote{Throughout this paper we use symmetrisation with weight $2\Kddu{(\mu}{\nu)}\rho=\Kddu{\mu}{\nu}\rho+\Kddu\nu\mu\rho$ and equivalently for the anti-symmetrisation.}
\be\label{eq:extcsyms}
\Kddu\mu\nu\rho=\Kddu{(\mu}{\nu)}\rho \ , \quad \hud\lambda\rho\,\Kddu\mu\nu\rho=\pud\mu\lambda\,\Kddu\mu\nu\rho=0 \ .
\ee
For many practical purposes it is useful to keep track of the number of derivatives associated with a given tensor. Any tensor built from the operator $\overline\nabla_\mu$ acting on zero-derivative tensors (tangent vectors, normal vectors and contractions thereof) is a one-derivative tensor. This is the case for \eqref{eq:extc} and also for the contraction
\be\label{eq.Ktrace}
K^\rho=\huu\mu\nu\Kddu\mu\nu\rho \ ,
\ee
which denotes the mean extrinsic curvature of the submanifold and inherits the orthogonality property in its index from \eqref{eq:extcsyms}. If the surface has codimension $n=1$, then there is only one normal direction, $i=1$, and hence only one normal vector, i.e. $\nud\mu i=\nud\mu 1=n^\mu$. In this case, both the extrinsic curvature tensor and the mean extrinsic curvature have only one transverse direction and instead one may work with the symmetric tensor $\Kddu\mu\nu\rho n_\rho$ and the scalar $K^\rho n_\rho$.


\subsubsection{The external rotation tensor}
Another one-derivative object of interest is obtained  from tangential covariant differentiation of the normal vectors. This object is referred to as external rotation tensor and takes the form
\be \label{eq:extt}
{{\omega_\mu}^{\nu}}{_\rho}=\pud\nu\sigma\, \ndu\rho i\, \overline\nabla_\mu \nud \sigma i \ .
\ee
One may clearly observe, from  \eqref{eq:extt}, that the definition of the external rotation tensor necessarily includes the appearance of transverse indices $i$. In fact, the external rotation tensor is not invariant under rotations of the normal vectors that satisfy \eqref{eq:relations} and, indeed, it transforms as a connection in the indices $i$, though it is fully tensorial in its spacetime indices. For this reason, it is usually referred to as a \emph{pseudo-tensor}, and it can be understood as a normal spin connection in the transverse space, characterising how a given pair of normal vectors is being twisted around as one moves along surface directions.

The external rotation tensor is tangential in its first index, and transverse and anti-symmetric in its last two indices
\be
\oduu\mu\nu\rho=\oduu\mu{[\nu}{\rho]} \ , \quad \pud\mu\lambda\oduu\mu\nu\rho=\hud\lambda\nu\oduu\mu\nu\rho=0 \ .
\ee
When coupling the external rotation tensor to a surface action, the resulting scalars, with a few exceptions, do not in general satisfy the requirements of diffeomorphism invariance. However, one may construct the associated curvature to the external rotation tensor, namely the outer curvature tensor\footnote{We note that there is a minus sign typo in \cite{Carter:1997pb} in the second term of \eqref{eq:oint} and \eqref{eq:rint}. Furthermore, we have changed the notation slightly for the outer curvature tensor. The first two indices here are transverse to the surface, while the first two indices in \cite{Carter:1997pb} are tangential.}
\be\label{eq:oint}
\Ouddd \mu\nu\kappa\lambda=2\, \pud\mu\sigma \pud\tau\nu \hud\pi{[\lambda} \overline \nabla_{\kappa]} {\omega_\pi}{^{\sigma}}{_\tau}+2\,{\omega{_{[\lambda}}}{^{\mu\pi}}\omega{_{\kappa]\pi\nu}} \ ,
\ee
which is invariant under rotations of the normal vectors and hence fully tensorial in both the spacetime and internal transverse indices (see Eqn.~\eqref{eq:RV} bellow). The outer curvature tensor is a two-derivative tensor structure which satisfies the following properties 
\be
\Odddd \mu\nu\kappa\lambda=\Odddd {\mu}{\nu}{[\kappa}{\lambda]}=\Odddd {[\mu}{\nu]}{\kappa}{\lambda} \ , \quad \pud\kappa\alpha \, \Odddd\mu\nu \kappa\lambda=\hud\mu\alpha \, \Odddd\mu\nu \kappa\lambda=0 \ .
\ee
Note that the outer curvature tensor only has a subset of the symmetries of a Riemann tensor. 

When the codimension of the surface is $n=2$, the transverse rotation group is Abelian and one may use the Levi-Civita tensor in the transverse space $\epsilon^{\mu\nu}_{\perp}$ in order to construct the normal fundamental 1-form as
\be
\omega_\mu=\frac{1}{2}\epsilon_{\nu\rho}^\perp\, \oduu\mu\nu\rho \ , \qquad (\text{for }n=2) \ ,
\ee
which inherits the tangentiality property in its index from the external rotation tensor. In this case the outer curvature tensor \eqref{eq:oint} becomes a field strength for the 1-form $\omega_\mu$, that is
\be
\Omega_{\kappa\lambda}=4\gamma^{\pi}_{[\lambda}\overline \nabla_{\kappa]}\omega_\pi \ , \qquad (\text{for }n=2) \ ,
\ee
and is tangential and anti-symmetric in its two indices.
When $p=n=2$ then one may contract the field strength with the surface Levi-Civita tensor $\epsilon^{\kappa\lambda}_{||}$ in order to obtain the outer curvature scalar
\be
\Omega=\epsilon^{\kappa\lambda}_{||}\Omega_{\kappa\lambda}={\epsilon}^{\mu\nu\kappa\lambda}\Odddd \mu\nu\kappa\lambda=4\overline\nabla_\mu\left(\epsilon^{\mu\nu}_{||}\omega_\nu\right) \ ,\qquad (\text{for }p=n=2) \ ,
\ee
where ${\epsilon}^{\mu\nu\kappa\lambda}$ is the $D=4$ background Levi-Civita tensor. It is, therefore, clear from the above that $\Omega$ is purely topological. 
If the surface dimensionality is $p=1$ (with $n=2$), then there is only one tangent vector $\eud\mu a=\eud\mu 1=u^\mu |\hdd11|$, with $\hdd11$ being the single metric component in the tangential direction, which can be interpreted as the unnormalised point-particle's velocity in Lorentzian signature. One may then use it to construct a scalar $u^{\mu}\omega_\mu$, which has applications for spinning point-particle actions and the Post-Newtonian approximation in General Relativity (see e.g. \cite{Porto:2016pyg}).


\subsubsection{Internal rotation tensor}

The external rotation tensor describes how normal vectors twist when one moves along the surface. It is convenient to introduce its internal counterpart, the internal rotation tensor, as
\be \label{eq:intt}
{{\rho_{\mu}}^{\nu}}{_\rho}=\hud\nu\sigma \edu\rho a\overline\nabla_\mu \eud\sigma a  \ .
\ee
The definition of the internal rotation tensor includes the appearance of tangential indices. As in the case of the external rotation tensor, this tensor, though fully tensorial in its spacetime indices, is also a \emph{pseudo-tensor}, as it is not invariant under changes of the tangent vectors. All its indices are tangential and its last two indices are anti-symmetric, that is
\be
{{\rho_{\mu}}^{\nu\rho}}={{\rho_{\mu}}^{[\nu\rho]}} \ , \quad \pud\mu\lambda \, {{\rho_{\mu}}^{\nu\rho}}=\pud\lambda\nu\, {{\rho_{\mu}}^{\nu\rho}}=0 \ .
\ee
Couplings to the internal rotation tensor to surface actions do not generally satisfy the diffeomorphism constraints due to the fact that it transforms as a connection in its tangential indices. However, its associated curvature, namely the intrinsic Riemann tensor
\be \label{eq:rint}
\Ruddd\mu\nu\kappa\lambda=2 \, \hud\mu\sigma \hud\tau\nu \hud\pi{[\lambda} \overline\nabla_{\kappa]}{{\rho_\pi}{^{\sigma}}}{_\tau} + 2 \, {\rho_{[\lambda}}{^{\mu\pi}} \rho_{\kappa]\pi\nu} \ ,
\ee
is fully tensorial and invariant under coordinate transformations in the internal indices (see Eqn.~\eqref{eq:GC} bellow). All the indices of the intrinsic Riemann tensor are tangential and, in addition, it has the same symmetry properties as the Riemann tensor associated with the background metric, $\Buddd\mu\nu\kappa\lambda$. The definition \eqref{eq:rint} makes it apparent that the internal rotation tensor may be seen as a connection  associated to the intrinsic geometry. In particular the contraction $\eud\mu a \, \edu\nu c \, \eud\rho b \,{{\rho_{\mu}}^{\nu}}{_\rho}$ is the Christoffel connection associated with the induced metric $\gamma_{ab}$.

When the surface dimensionality is $p=2$, it is possible to define the tangential one-form ${2\rho_\mu=\epsilon_{||}{_{\nu\rho}}\,{{\rho_{\mu}}^{\nu\rho}}}$. The contraction of the intrinsic Riemann tensor with $\epsilon^{\mu\nu}_{||}$ becomes a field strength for $\rho_\mu$, while a further contraction yields the intrinsic Ricci scalar $\mathcal{R}=\hdu\mu\kappa\huu\nu\lambda\Ruddd\mu\nu\kappa\lambda$, that is
\be \label{eq:bricci}
\epsilon^{\mu\nu}_{||}\Rdddd\mu\nu\kappa\lambda=4\hud\pi{[\lambda}\overline\nabla_{\kappa]}\rho_\pi \ , \quad \epsilon^{\mu\nu}_{||}\epsilon^{\kappa\lambda}_{||}\Rdddd\mu\nu\kappa\lambda=4\overline\nabla_\mu\left(\epsilon^{\mu\nu}_{||}\rho_\nu\right)=2\mathcal{R} \ , \qquad (\text{for }p=2) \ .
\ee
As it is well known, this makes it clear that the intrinsic Ricci scalar for $p=2$ is topological. 


\subsubsection{Integrability conditions}

Given an extrinsic curvature, an intrinsic Riemann tensor and an outer curvature tensor of the embedding, the fundamental theorem of surfaces states that for theses tensors to be supported in an embedded surface there are three equations to be satisfied (see e.g. \cite{aminov2001geometry}). These are the Gauss-Codazzi equation
\be\label{eq:GC}
\Ruddd \mu\nu\kappa\lambda =  \Kduu \kappa\mu\tau \Kddd \lambda\nu\tau  - \Kduu \lambda\mu\tau \Kddu \kappa\nu\tau + \hdu \lambda\rho \hdu \kappa\sigma \hdu \tau\mu \hdu \nu\alpha \Bddud \rho\sigma\tau\alpha \  \ ,
\ee
the Ricci-Voss equation
\be \label{eq:RV}
\Ouddd\mu\nu \kappa\lambda =  \Kddu \kappa\rho\mu \Kdud \lambda\rho\nu - \Kddu \lambda\rho\mu \Kdud \kappa\rho\nu + \hdu \kappa\rho \hdu \lambda\sigma \pdu \tau\mu \pdu \nu\alpha \Bddud \rho\sigma\tau\alpha \  \ ,
\ee
and the Codazzi-Mainardi equation
\be \label{eq:CM}
2\hud\sigma\mu\pud\nu\tau\hud\rho{[\lambda}\overline\nabla_{\kappa]}\Kddu\rho\sigma\tau=\hud\rho\kappa\hud\sigma\lambda\hud\tau\mu\pud\nu\alpha \Buddd\alpha\tau\rho\sigma \ .
\ee
In particular, \eqref{eq:GC} and \eqref{eq:RV} make explicit the fully tensorial character of the intrinsic Riemann tensor and the outer curvature tensor, since they can be expressed in terms of other fully tensorial quantities. These integrability conditions will play a crucial role in proving conservation laws, as well as in the counting of  independent terms that can appear in surface actions.

\subsubsection{Conformal tensors} 
In order to address the Weyl-invariant properties of surface actions, it is useful to define the background Weyl tensor for $D\ge3$ as
\be\label{eq.bgWeyl}
\Wdduu \mu\nu\lambda\rho \equiv  \Bdduu \mu\nu\lambda\rho - \Gdu \mu\lambda \Pdu \nu\rho - \Gdu \nu\rho \Pdu \mu\lambda + \Gdu \mu\rho \Pdu \nu\lambda + \Gdu \nu\lambda \Pdu \mu\rho \ ,
\ee
which is trace-free and has the same symmetry properties as the Riemann tensor. Here we have introduced the background Schouten tensor, defined as
\be\label{eq.bgSchouten}
\Pdd \mu\nu = \frac{1}{D-2} \left( \Bdd \mu\nu  - \frac{R}{2(D-1)}  \Gdd \mu\nu \right) \ ,
\ee
which encodes all the information about the curvature scales of the manifold. It is also useful to define the pull-back of the Weyl tensor onto the surface as
\be
\widehat W_{\mu\nu\rho\sigma} \equiv \hdu\mu\alpha \hdu\nu\beta \hdu\rho\gamma \hdu\sigma\delta W_{\alpha\beta\gamma\delta}~~.
\ee
Finally, we introduce the conformation tensor
\be\label{eq:conformation}
{C_{\mu\nu}}^\rho = \Kddu \mu\nu\rho - \frac{1}{p} \hdd\mu\nu K^\rho \ ,
\ee
which is a well known Weyl-invariant tensor \cite{Carter:1992vb} (see Eqn.~\eqref{eq.weylconformation}). As we shall see, explicitly in App.~\ref{app:variations}, other Weyl-invariant tensors include the outer curvature ${\Omega^\mu}_{\nu\lambda\rho}$ and the external rotation tensor $\odud\mu\nu\rho$.

\subsubsection{Edge geometry}\label{sec:edgegeometry}
The edges $\partial \mathcal{W}$  are the $(p-1)$-dimensional boundaries of the surface $\mathcal{W}$. The location of these codimension $(n+1)$ edges in the ambient background metric $\Gdd\mu\nu$ is given by the set of mapping functions $\wt X^{\mu}(\wt\sigma)$. From the point of view of the surface $\mathcal{W}$, the edges are characterised by a single normal vector $\wt n^\rho$ to the surface. This normal vector is such that the induced metric on the surface can be written as
\be
\hdd\mu\nu=\hsdd\mu\nu+\wt n_\mu \wt n_\nu \ ,
\ee
where $\hsdd\mu\nu$ is the projected metric on $\partial \mathcal{W}$. The spacetime metric, when restricted to the edge, decomposes  as
\be
\Gdd\mu\nu=\hsdd\mu\nu+\psdd\mu\nu \ , \quad \psdd\mu\nu= \,\pdd\mu\nu +\, \wt n_\mu \wt n_\nu \ ,
\ee
where $\psdd\mu\nu$ is the transverse metric to $\partial \mathcal{W}$. Analogously to the surface, covariant differentiation of edge tensors is only well defined via the operator $\wt \nabla_\mu=\hsud\lambda\mu\nabla_\lambda$. In light of this consideration, one may introduce the edge one-derivative tensors
\be \label{eq:1dedge}
\Ksddu\mu\nu\rho=\hsud\sigma\nu\wt\nabla_\mu\hsud\rho\sigma \ , \quad {{\varpi_\mu}^{\nu}}{_\rho}=\psud\nu\sigma \, {\wt{n}_\rho}^i \, \wt\nabla_\mu {\wt{n}^\sigma}_i \ , \quad {{\varrho_{\mu}}^{\nu}}{_\rho}=\hsud\nu\sigma\, {\wt{e}_\rho}^a \, \wt\nabla_\mu {\wt{e}^\sigma}_a \ ,
\ee
which denote the edge extrinsic curvature tensor, external rotation and internal rotation tensor, respectively. Here, ${\wt{n}_\rho}^i= \left(n_\rho^i ,\wt n_\rho \right)$ denotes the set of normal vectors to the edge\footnote{Notice that we have abused slightly our notation, since $i$ in $\wt{n}_\rho^i$ runs from 1 to $n+1$ whereas in $n_\rho^i$ runs just to $n$. The reader can immediately identify the range by the symbol it is accompanying. Similar remarks can be made about ${e}^\rho_a$ and ${\wt{e}^\rho}_a$. Since in this work we always use spacetime indices, this distinction should not cause any confusion.} while ${\wt{e}^\rho}_a$ denotes the 
set of tangent vectors. We also define the edge mean extrinsic curvature as $\mathcal{K}^{\rho}=\hsuu\mu\nu\Ksddu\mu\nu\rho$. The components of the edge extrinsic curvature along the normal direction to the surface can be obtained via the contraction with the normal vector, that is
\be
\Ksddu\mu\nu\rho \, \wt n_\rho \ , \quad \mathcal{K}^{\rho} \, \wt n_\rho \ .
\ee
From \eqref{eq:1dedge} one may define their respective curvatures, which satisfy analogous integrability conditions as \eqref{eq:GC}-\eqref{eq:CM}. However, we will not need to consider these in this work. 

When describing the edge dynamics by means of an action, it is possible to consider couplings beyond the ones to the edge geometry. In the case of the surface action, one can also consider couplings to the background metric and its derivatives besides couplings to surface geometric tensors. In the case of the edge, however, all the tensors characterising the surface geometry are background fields from the edge point of view. In particular, for these surface fields, the covariant derivative $\wt n^\mu \nabla_\mu$ is well defined. One example of a possible coupling that can appear in this way is $\wt n^\alpha\nabla_\alpha \Kddu\mu\nu\rho$. These considerations hint at the complexity of describing the edge dynamics. In this work, we have only considered couplings to the edge fields and the background metric. While many of these implicitly include couplings to the surface fields, we do not wish to claim that our work is exhaustive in this respect. On the other hand, it is sufficient to exhibit the richness of the edge dynamics.

\subsection{Surface dynamics and shape equation} \label{sec:shapeequations}

Now we proceed to find the dynamics of surfaces, assuming that such dynamics follow from an action, which we consider to be a functional of the set of geometric fields that we collectively denote by $\Phi(\sigma)$, that is\footnote{We have explicitly considered the geometric fields $\Rdddd\mu\nu\lambda\rho$ and $\Odddd\mu\nu\lambda\rho$ for practical purposes, as one may find them more convenient to use instead of others. However, these are not independent from the remaining fields due to the integrability conditions \eqref{eq:GC}-\eqref{eq:RV}. \label{footnote:gc}}
\be \label{eq:fields}
\Phi(\sigma)=\{X^{\mu}(\sigma),\hdd\mu\nu(\sigma),\pud\mu\nu(\sigma),\pdd\mu\nu\!(\sigma),\Kddu\mu\nu\rho(\sigma), \oduu\mu\lambda\rho(\sigma), \Rdddd\mu\nu\lambda\rho(\sigma), \Odddd\mu\nu\lambda\rho(\sigma), \Bdddd\mu\nu\lambda\rho(\sigma)\} \ .
\ee
For $n=1$ we can consider the normal vector $n^\mu$ and trade it for $\pud\mu\nu=n^\mu n_\nu$, which in this case is obviously not independent of $\pdd\mu\nu$. For $n>1$ the consideration of both $\pud\mu\nu$ and $\pdd\mu\nu$ in \eqref{eq:fields} as independent fields follows from the projective nature of $\pdd\mu\nu$. Indeed, since this tensor has zero eigenvalues there is no strict inverse that can be constructed, and $\pud\mu\nu\neq\delta^\mu{_\nu}$.\footnote{A similar argument holds for $\hud\mu\nu=\delta^\mu{_\nu}-\pud\mu\nu$ but not for $\hud a b$, which is restricted to the surface's worldvolume and thus has no vanishing eigenvalues in general.} Ultimately, this implies that variations with respect to $\pud\mu\nu$ and $\pdd\mu\nu$, when $n>1$ give independent results, as shown explicitly in Eqn.~\eqref{eq:mvarcore} below.

Before proceeding, notice that it is possible to consider more general actions with arbitrary couplings to the one- and two-derivative geometric tensors ${{\rho_\mu}^{\nu}}{_\rho}$, $\wt\nabla_\lambda{{\rho_\mu}^{\nu}}{_\rho}$ and $\wt\nabla_\lambda{\omega_\mu}^{\nu\rho}$, but up to second order in derivatives such couplings do not lead to covariant actions, except for the specific couplings that are considered here. 

Consider now a surface geometric action that takes the generic form
\be \label{eq:genact}
S[\Phi(\sigma)]=\int_{\mathcal{W}}d^{p}\sigma\mathcal{L}[\Phi(\sigma)] \ ~,
\ee
where $\mathcal{L}[\Phi(\sigma)]$ is a Lagrangian density.  As mentioned in Sec.~\ref{sec:intro}, there are two ways of obtaining the resulting dynamics. One way consists of slightly deforming the surface such that $X^{\mu}(\sigma)\to X^{\mu}(\sigma)+\delta X^{\mu}(\sigma)$ for some small deformation $\delta X^{\mu}(\sigma)$, without displacing the background coordinates. However, this approach, as we shall see in Sec.~\ref{sec:DCFTs}, deals with non-manifestly covariant expressions and non-manifestly covariant intermediate steps. Here, instead, we follow the approach by Carter \cite{Carter:1992vb, Carter:1993wy, Battye:1995hv, Carter:1997pb} and introduce a new type of variational principle that leads to non-trivial constraints on the spacetime stress tensor, does not require extending the surface to a foliation and always provides manifestly covariant expressions. This method employs Lagrangian variations, in which the background coordinates are displaced by a small amount $x^{\mu}\to x^{\mu}+\xi^{\mu}(x)$ while the mapping functions $X^{\mu}(\sigma)$ are held fixed. Under such infinitesimal displacements, parametrised by the flows of the vector field $\xi^{\mu}$, the background metric changes by a Lie derivative,
\be
\delta_\xi\, g_{\mu\nu}=2\nabla_{(\mu}\xi_{\nu)} \ .
\ee
Lagrangian variations are equivalent to infinitesimal diffeomorphism transformations with fixed mapping functions. As explained in App.~\ref{app:variations}, a complete set of Lagrangian variations of $\Gdd\mu\nu$ in terms of $\hdd\mu\nu$ and $\pdd\mu\nu$ is given by the three independent projections
\be\label{eq:mvarcore}
\delta_\xi \huu\mu\nu=-\huu\mu\lambda\huu\nu\rho\delta_\xi\Gdd\lambda\rho \ , \quad \delta_\xi \pdd\mu\nu=\pud\lambda\mu\pud\rho\nu\delta_\xi\Gdd\lambda\rho \ , \quad \delta_\xi\pud\mu\nu=-\huu\mu\lambda\pud\rho\nu\delta_\xi\Gdd\lambda\rho \ .
\ee

From the three variations in \eqref{eq:mvarcore} we can define the surface's worldvolume stress tensor $\mathcal{T}_{\mu\nu}$, the mixed tangential-transverse stress tensor ${\mathcal{P}^{\mu}}_{\nu}$, the transverse stress tensor $\mathcal{B}^{\mu\nu}$ as
\be\label{eq:def1a}
\mathcal{T}_{\mu\nu}=-\frac{2}{\sqrt{|\gamma|}}\frac{\delta_\xi \mathcal{L}}{\delta_\xi \huu\mu\nu} \ ,
\quad {\mathcal{P}_{\mu}}^{\nu}=-\frac{1}{\sqrt{|\gamma|}}\frac{\delta_\xi \mathcal{L}}{\delta_\xi \pud\mu\nu}  \ ,
\quad  \mathcal{B}^{\mu\nu}=\frac{2}{\sqrt{|\gamma|}}\frac{\delta_\xi \mathcal{L}}{\delta_\xi \pdd\mu\nu} \ ,
\ee
where $|\gamma|$ denotes the absolute value of the determinant of the metric $\hdd ab$. Naively, this may appear to be in contradiction with the strategy of the current paper, which consists of considering tensorial objects with spacetime (Greek) indices. However it must be noted that Lagrangian variations are taken with $\delta_\xi e^\mu{_a}=0$ (see App.~\ref{app:variations}), such that the variation of the determinant is given by
\be
\huu ab \delta_\xi \hdd ab = \Guu\mu\nu e_\mu{^a}e_\nu{^b}\delta_\xi \hdd ab = \Guu\mu\nu  e_\mu{^a}e_\nu{^b} e^\rho{_a}e^\sigma_{b} \delta_\xi \Gdd\rho\sigma = -\hdd\alpha\beta \delta_\xi \huu\alpha\beta \ ,
\ee 
and therefore can be traded by the variation of the projector $\hdd\mu\nu$ with spacetime indices.
All the tensor structures introduced in \eqref{eq:def1a} inherit their symmetry properties and index structure from their variational counterparts. In particular, $\mathcal{T}_{\mu\nu}$ and $\mathcal{B}^{\mu\nu}$ are symmetric, with $\mathcal{T}_{\mu\nu}$ being tangential and $\mathcal{B}^{\mu\nu}$ being transverse in their indices. Furthermore, ${\mathcal{P}^{\mu}}_{\nu}$ is tangential in its first index and transverse in its second index.

The mixed tangential-transverse and transverse stress tensors describe the normal components of the full spacetime stress tensor of the surface, as shown in Sec.~\ref{sec:stresstensor}. These objects are not independent quantities and, as it will be shown, are given in terms of the bending moment, spin current and curvature moments to be defined next. The precise relation between them turns out to be a requirement of diffeomorphism invariance of \eqref{eq:genact}.

For codimension $n=1$ the tangential-transverse stress tensor ${\mathcal{P}^{\mu}}_{\nu}$ is identically zero, since we have eliminated from the action any dependence on $\pud\mu\nu$ in favour of the orthogonal vector $n^\mu$ and the projector $\pdd\mu\nu$. Although the projector is not independent of the orthogonal vector, since $\pdd\mu\nu=n_\mu n_\nu$, it is convenient to work with variations of the orthogonal vector with risen indices only and using the projector for the lowered ones. We therefore introduce the tensor $\mathcal{V}_\mu$ to account for couplings to $n^\mu$ such that\footnote{For $p=1$ one could consider introducing couplings to the single normalised tangent vector $u^\mu$. However, for variations that keep the embedding map fixed $\delta \eud\mu a=0$ this is not necessary.}
\be
\mathcal{V}_\mu=\frac{1}{\sqrt{|\gamma|}}\frac{\delta_\xi \mathcal{L}}{\delta_\xi n^\mu} \ .
\ee

The variations with respect to objects with one derivative allow us to define the bending moment, ${\mathcal{D}^{\mu\nu}}_{\rho}$, and the spin current, ${\mathcal{S}^{\mu}}_{\lambda\rho}$, as\footnote{We have introduced here the variation $\wt\delta_\xi$ associated with certain geometric tensors. As explained in App.~\ref{app:variations}, this is a 
Lagrangian variation for which we have stripped off the components that can be incorporated into ${\mathcal{P}^{\mu}}_{\nu}$ and $\mathcal{B}^{\mu\nu}$.}
\be\label{eq:def1b} 
{\mathcal{D}^{\mu\nu}}_{\rho}=\frac{1}{\sqrt{|\gamma|}}\frac{\delta_\xi \mathcal{L}}{\delta_\xi \Kddu\mu\nu\rho}~,~{\mathcal{S}^{\mu}}_{\lambda\rho}=\frac{1}{\sqrt{|\gamma|}}\frac{\wt \delta_\xi \mathcal{L}}{\wt \delta_\xi \oduu\mu\lambda\rho}  \ .
\ee
It is clear from these definitions that the bending moment encodes responses of the surface due to bending, whereas the spin current encodes surface motion in the transverse space. The bending moment ${\mathcal{D}^{\mu\nu}}_{\rho}$ is symmetric and tangential in its two first indices and transverse in its last index, and the spin current ${\mathcal{S}^{\mu}}_{\lambda\rho}$ is tangential in its first index and transverse and anti-symmetric in the last two. 

Furthermore, variations with respect to the curvature tensors define the surface curvature moment, ${\mathcal{I}_{\mu}}^{\nu\lambda\rho}$, the outer curvature moment, ${\mathcal{H}_{\mu}}^{\nu\lambda\rho}$, and the background curvature quadrupole moment, ${\mathcal{Q}_{\mu}}^{\nu\lambda\rho}$, from the expressions
\be\label{eq:def1c} 
{\mathcal{I}_{\mu}}^{\nu\lambda\rho}=\frac{1}{\sqrt{|\gamma|}}\frac{\wt \delta_\xi \mathcal{L}}{\wt \delta_\xi \Ruddd\mu\lambda\nu\rho}~,~{\mathcal{H}_{\mu}}^{\nu\lambda\rho}=\frac{1}{\sqrt{|\gamma|}}\frac{\wt \delta_\xi \mathcal{L}}{\wt \delta_\xi \Ouddd\mu\lambda\nu\rho}~,~{\mathcal{Q}_{\mu}}^{\nu\lambda\rho}=\frac{1}{\sqrt{|\gamma|}}\frac{\delta_\xi \mathcal{L}}{\delta_\xi \Buddd\mu\lambda\nu\rho} \ .
\ee
The curvature moments encode responses due to the intrinsic, outer and background curvature. The curvature moments inherit the symmetries of the Riemann tensor, in particular, ${\mathcal{Q}}^{\mu\nu\lambda\rho}=-{\mathcal{Q}}^{\mu\rho\lambda\nu}$. The surface curvature moment ${\mathcal{I}}^{\mu\nu\lambda\rho}$ is purely tangential while the outer curvature moment ${\mathcal{H}}^{\mu\nu\lambda\rho}$ is transverse in its first and third indices and tangential in its second and fourth indices.

Given the definitions in Eqn.~\eqref{eq:def1a}-\eqref{eq:def1c} the action \eqref{eq:genact} transforms under a Lagrangian variation as
\be \label{eq:varL}
\begin{split}
\delta_\xi S[\Phi(\sigma)]=\int_{\mathcal{W}}d^{p}\sigma\sqrt{|\gamma|}\Big(&-\frac{1}{2}\mathcal{T}_{\mu\nu} \, \delta_\xi\huu\mu\nu-{\mathcal{P}_{\mu}}^{\nu} \, \delta_\xi \pud\mu\nu+\frac{1}{2}\mathcal{B}^{\mu\nu} \, \delta_\xi\! \pdd\mu\nu +\mathcal{V}_\mu\delta_\xi n^{\mu}\, \delta_{n,1} \\
&+{\mathcal{D}^{\mu\nu}}_{\rho} \, \delta_\xi\Kddu\mu\nu\rho+{\mathcal{S}^{\mu}}_{\lambda\rho} \, \wt \delta_\xi\oduu\mu\lambda\rho +{\mathcal{Q}_\mu}^{\nu\lambda\rho} \, \delta_\xi\Buddd\mu\lambda\nu\rho \\
&+{\mathcal{I}_\mu}^{\nu\lambda\rho} \, \wt \delta_\xi\Ruddd\mu\lambda\nu\rho+{\mathcal{H}_\mu}^{\nu\lambda\rho} \, \wt \delta_\xi\Ouddd\mu\lambda\nu\rho \Big) \ .
\end{split}
\ee
 We do not give considerable attention, due to the reasons explained in footnote \ref{footnote:gc}, to couplings to $\Ruddd\mu\nu\lambda\rho$ and $\Ouddd\mu\nu\lambda\rho$. For that reason we will not consider the last line in Eqn.~\eqref{eq:varL} in the core of this paper, with exception of a few comments in passing, and refer the reader to App.~\ref{app:highdev} for this generalisation.

The result of the variation in \eqref{eq:varL}, after integration by parts and using formulae in App.~\ref{app:variations}, can be expressed in terms of the vector field $\xi^{\mu}$ and its normal derivatives. Terms that involve normal derivatives cannot be further integrated by parts and so need to vanish independently for the variational principle to be well defined. This sets constraints on the type of actions that can be constructed. Schematically, in terms of $\xi^{\mu}$ and its derivatives, we find that\footnote{In principle, we should consider terms up to three normal derivatives of $\xi_\mu$. However, for the couplings we consider, such terms automatically vanish.}
\be \label{eq:varL1}
\begin{split}
\delta_\xi S[\Phi(\sigma)]=&\int_{\mathcal{W}}d^{p}\sigma\sqrt{|\gamma|}\left(B^{\mu\nu}\pdu\mu\rho\nabla_\rho\xi_\nu+B^{\mu}\xi_\mu\right)\\
&+\int_{\partial\mathcal{W}}d^{p-1}\wt\sigma\sqrt{|h|} \,\wt n_\mu\left(\wt B^{\mu\nu\rho}\psud\lambda\nu\nabla_\lambda\xi_\rho+\wt B^{\mu\nu}\xi_\nu\right) \ ,
\end{split}
\ee
where we have assumed that the edges of the surface do not have edges themselves. Each of the above terms must vanish independently. The last two terms are boundary terms and we will give them special attention at the end of this section, where we describe  edge dynamics. 

\subsubsection{Surface equations of motion and diffeomorphism constraints}

Equating the first term in \eqref{eq:varL1} to zero implies the following relation
\be \label{eq:c1}
\begin{split}
&{\mathcal{P}^{\mu}}_\nu\puu\nu\alpha \hdu\mu\sigma+\mathcal{B}^{\alpha\sigma}-\mathcal{D}^{\mu\lambda\sigma}\Kddu\mu\lambda\alpha+\mathcal{S}^{\mu\lambda\alpha}\Kdud\mu\sigma\lambda-\pdu\lambda\sigma\pdu\rho\alpha\overline\nabla_\mu \mathcal{S}^{\mu\lambda\rho}\\
&\qquad =\pdu\lambda\alpha\Pi^{\sigma\lambda}+\left(\mathcal{V}_\mu\puu\mu\sigma n^\alpha+\mathcal{V}_\mu\huu\mu\sigma n^\alpha\right)\delta_{n,1} \ ,
\end{split}
\ee
where we have defined the tensor
\be\label{eq:pidef}
\Pi^{\sigma\lambda}=\mathcal{Q}^{\mu\nu\lambda\rho}\Buddd\sigma\mu\nu\rho+\mathcal{Q}^{\sigma\nu\mu\rho}\Buddd\lambda\mu\nu\rho+2\mathcal{Q}^{\mu\lambda\nu\rho}\Buddd\sigma{\rho}{\nu}\mu \ .
\ee
By projecting \eqref{eq:c1} orthogonally in the index $\sigma$ and anti-symmetrising the two free indices one finds
\be \label{eq:st}
\mathcal{D}^{\mu\lambda[\sigma}\Kddu\mu\lambda{\alpha]}+\pdu\lambda\sigma\pdu\rho\alpha\overline\nabla_\mu \mathcal{S}^{\mu\lambda\rho}+\pdu\nu{[\sigma}\pdu\lambda{\alpha]}\Pi^{\nu\lambda}=0 \ .
\ee
This equation expresses the violation of spin conservation and is a generalisation of that found in \cite{Armas:2013hsa} in order to account for possible couplings to background curvature. By projecting \eqref{eq:c1} orthogonally in the index $\sigma$ and symmetrising both free indices leads to
\be \label{eq:bt}
\mathcal{B}^{\alpha\sigma}=\mathcal{D}^{\mu\lambda(\sigma}\Kddu\mu\lambda{\alpha)}+\pdu\nu{(\sigma}\pdu\lambda{\alpha)}\Pi^{\nu\lambda}+\mathcal{V}_\mu\puu\mu\sigma n^\alpha \, \delta_{n,1} \ .
\ee
As advertised earlier, this equation expresses the fact that the transverse stress tensor $\mathcal{B}^{\alpha\sigma}$ is not independent but given in terms of the bending moment and the quadrupole moment when $n>1$. Finally, projecting \eqref{eq:c1} tangentially along the index $\sigma$ leads to
\be\label{eq:pt}
{\mathcal{P}^{\sigma}}_\alpha=\mathcal{S}^{\mu}{_\alpha}{^\lambda}\Kdud\mu\sigma\lambda+\hdu\nu\sigma\pdd\lambda\alpha\Pi^{\nu\lambda}+\mathcal{V}_\mu\huu\mu\sigma n_\alpha \, \delta_{n,1}  \ ,
\ee
which expresses the fact that the mixed worldvolume-transverse stress tensor is not independent but given in terms of the spin current and the quadrupole moment when $n>1$. Eqs.~\eqref{eq:bt}-\eqref{eq:pt} state that one cannot trade off the bending moment, spin current and quadrupole moment by  the stresses $\mathcal{B}^{\alpha\sigma}$ and ${\mathcal{P}^{\mu}}_\nu$ as the latter do not contain enough information to determine the former. As a reminder, for codimension $n=1$ one should write ${\mathcal{P}^{\sigma}}_\alpha=0$.

Consider now the vanishing of the second term in \eqref{eq:varL1}. This leads to the set of equations
\be \label{eq:genshape}
\begin{split}
&\overline\nabla_\lambda\left(\mathcal{T}^{\lambda\sigma}+\hdu\mu\lambda\left(\pdu\nu\sigma\Pi^{\mu\nu}-\Pi^{\sigma\mu}\right)-\hud\lambda\nu\overline\nabla_\mu\mathcal{D}^{\mu\nu\sigma}-2{{\mathcal{S}^{\mu}}_\alpha}^{\sigma}\Kduu\mu\lambda\alpha\right) \\
&=\left(\mathcal{S}^{\mu\lambda\rho}-\mathcal{D}^{\mu\lambda\rho}\right)\Buddd\sigma\mu\lambda\rho+2\mathcal{Q}^{\mu\nu\lambda\rho}\nabla_\nu\Buddd\sigma\rho\mu\lambda \ ,
\end{split}
\ee
where we have made used of the constraint \eqref{eq:pt} in order to eliminate ${\mathcal{P}^{\mu}}_\nu$.
The equations obtained in \cite{Armas:2013hsa} correspond to the case in which the curvature quadrupole moment vanishes and therefore $\mathcal{Q}^{\mu\nu\lambda\rho}=0$. 
  
The tangential projection of \eqref{eq:genshape} along the $\sigma$ index leads to a conservation equation that must be automatically satisfied for any action that is reparameterisation invariant
\be \label{eq:genshapetang}
\begin{split}
&\hud\rho\sigma\overline\nabla_\lambda\left(\mathcal{T}^{\lambda\sigma}+\hdu\mu\lambda\left(\pdu\nu\sigma\Pi^{\mu\nu}-\Pi^{\sigma\mu}\right)-\hud\lambda\nu\overline\nabla_\mu\mathcal{D}^{\mu\nu\sigma}\right) \\
&=\mathcal{S}^{\mu\lambda\sigma}\Ouddd\rho\mu\lambda\sigma-\mathcal{D}^{\mu\lambda\alpha}\hud\rho\sigma\Buddd\sigma\mu\lambda\alpha+2\mathcal{Q}^{\mu\nu\lambda\alpha}\hud\rho\sigma\nabla_\nu\Buddd\sigma\alpha\mu\lambda \ ,
\end{split}
\ee
where we have used the Ricci-Voss equation \eqref{eq:RV}.

The orthogonal projection in turn yields the non-trivial dynamics and the resulting equation is often called the shape equation, which takes the form
\be \label{eq:genshapeperp}
\begin{split}
\mathcal{T}^{\lambda\sigma}\Kddu\lambda\sigma\rho=&-\pud\rho\sigma\overline\nabla_\lambda\left(\hdu\mu\lambda\left(\pdu\nu\sigma\Pi^{\mu\nu}-\Pi^{\sigma\mu}\right)-\hud\lambda\nu\overline\nabla_\mu\mathcal{D}^{\mu\nu\sigma}-2{{\mathcal{S}^{\mu}}_\alpha}^{\sigma}\Kduu\mu\lambda\alpha\right) \\
&+\pud\rho\sigma\left(\mathcal{S}^{\mu\lambda\alpha}-\mathcal{D}^{\mu\lambda\alpha}\right)\Buddd\sigma\mu\lambda\alpha+2\mathcal{Q}^{\mu\nu\lambda\alpha}\pud\rho\sigma\nabla_\nu\Buddd\sigma\alpha\mu\lambda \ .
\end{split}
\ee
This equation can be seen as a force balance equation on the surface, where the bending moment, spin current and curvature quadrupole moment introduce sources of stress on the surface. 

When both the bending moment and the curvature quadrupole moment vanish, these equations reduce to those obtained by Papapetrou for spinning point-particles \cite{Papapetrou:1951pa}, while if we restrict to codimension-1, this equation is a generalisation of membrane dynamics in classical elasticity theory \cite{Armas:2013hsa}. The simplicity of \eqref{eq:genshape}-\eqref{eq:genshapeperp} is stunning, since the tensor structures involved such as $\mathcal{Q}^{\mu\nu\lambda\rho}$ can contain any contraction of an arbitrary number of copies of all the geometric tensors involved, e.g. $\Bdddd\sigma\rho\mu\lambda$. In particular, the equations of motion arising from any type of background Lovelock theory, surface Lovelock theory or transverse Lovelock theory are included.


\subsubsection{Edge equations of motion and diffeomorphism constraints}
The edge terms in \eqref{eq:varL} deserve special attention. In the variational principle described in \eqref{eq:varL} we have not assumed the existence of extra sources of stress at the edges, for  example the effect due to considering an edge tension in the action \eqref{eq:genact}. The boundary conditions we describe here apply therefore only to the case of free edges, such as open strings without extra sources of matter attached to its ends. We postpone more general boundaries to Sec.~\ref{sec:sedges}. The third term in \eqref{eq:varL1} leads to the boundary conditions, upon projecting with $\pud\alpha\lambda$ and $\wt n_\lambda$, respectively
\be \label{eq:b1}
\wt n_\mu \mathcal{S}^{\mu\lambda\rho}|_{\partial\mathcal W}=0 \ , \quad \wt n_\lambda \wt n_\nu \mathcal{D}^{\lambda\nu\sigma}|_{\partial\mathcal W}=0 \ ,
\ee
while the last term in \eqref{eq:varL1}  leads to the equation of motion for the boundary dynamics
\be \label{eq:b2}
\wt \nabla_\lambda\left(\wt n_\mu\hsud\lambda\nu\mathcal{D}^{\mu\nu\sigma}\right)-\wt n_\lambda\left(\mathcal{T}^{\lambda\sigma}+\hdu\mu\lambda\left(\pdu\nu\sigma\Pi^{\mu\nu}-\Pi^{\sigma\mu}\right)-\hud\lambda\nu\overline\nabla_\mu\mathcal{D}^{\mu\nu\sigma}-2{{\mathcal{S}^{\mu}}_\alpha}^{\sigma}\Kduu\mu\lambda\alpha\right)\bigg|_{\partial\mathcal W}=0~, \,
\ee
where we have again used \eqref{eq:pt} in order to eliminate ${\mathcal{P}^{\mu}}_\nu$. The first two boundary conditions state that there should not be any flow of spin and bending moment along the normal components to the boundary. Eq.~\eqref{eq:b2} can be interpreted as the conservation of the non-symmetric boundary stress tensor $\wt n_\nu\mathcal{D}^{\nu\lambda\sigma}$ and can be projected in different ways. The projection onto the transverse $n$-dimensional space yields
\be
\pud\rho\sigma\wt\nabla_\lambda\left(\wt n_\nu\hsud\lambda\nu\mathcal{D}^{\nu\nu\sigma}\right)+\wt n_\lambda\pud\rho\sigma\left(\hud\lambda\mu \left(\Pi^{\sigma\mu}-\pdu\nu\sigma\Pi^{\mu\nu}\right)+\pud\rho\sigma\hud\lambda\nu\overline\nabla_\mu\mathcal{D}^{\mu\nu\sigma}-2{{\mathcal{S}^{\mu}}_{\alpha}}^{\rho}\Kduu\mu\lambda\alpha\right)=0 \ ,
\ee
while the projection along the normal to the boundary yields
\be
\wt n_\nu \mathcal{D}^{\nu\lambda\sigma}\wt\nabla_\lambda \wt n_\sigma=\wt n_\lambda \wt n_\sigma\left(-\mathcal{T}^{\lambda\sigma}+\Pi^{\lambda\sigma}+\overline\nabla_\mu\mathcal{D}^{\mu\lambda\sigma}\right) \ ,
\ee
which, by subtracting $\wt n_\nu \mathcal{D}^{\nu[\lambda\sigma]}\wt\nabla_\lambda \wt n_\sigma$ from the left hand side, can be interpreted as a Young-Laplace law since $\wt\nabla_{(\lambda} \wt n_{\sigma)}$ is (minus) the component of the extrinsic curvature normal to the boundary but along the surface.
On the other hand, the tangential projection along boundary directions, using $\hsud\rho\sigma$, results in
\be
\wt n_\nu\mathcal{D}^{\nu\lambda\sigma}\Ksdud\lambda\rho\sigma=\wt n_\lambda\hsud\rho\sigma\left(-\mathcal{T}^{\lambda\sigma}+\Pi^{\sigma\lambda}+\hud\lambda\nu\overline\nabla_\mu \mathcal{D}^{\mu\nu\sigma}\right) \ .
\ee 
This equation appears  to be a Young-Laplace equation for the stress $\wt n_\nu\mathcal{D}^{\nu\lambda\sigma}$ and with a pressure term equal to the r.h.s.  but the free index is tangential.  As a final remark, we will show in Sec.~\ref{sec:sedges} that introducing new degrees of freedom on the boundary modifies the r.h.s. of \eqref{eq:b1}-\eqref{eq:b2}, and therefore their projections.


\subsubsection{An interpretation for the constraints}\label{sec:constraints}
The constraints \eqref{eq:st}-\eqref{eq:pt} can be understood as consequences of the invariance of the action under local coordinate transformations. Choosing Riemann-normal coordinates in the neighbourhood of a point $q$, such that $\Gamma^{\lambda}_{\mu\nu}|_{q}=0$, a linear coordinate transformation can be decomposed as
\be
\xi_{\mu}(x)=\left({\omega_{\mu\nu}}+{\Lambda_{\mu\nu}}\right) x^{\nu} \ ,
\ee
where ${\Lambda_{\mu\nu}}$ is a matrix in the Lorentz group and $\omega_{\mu\nu}$ a symmetric matrix with constant coefficients in the completion of the group of general coordinate transformations. Under this restricted variation, one finds 
\be \label{eq:localtrans}
\begin{split}
\delta_{\omega} S[\Phi(\sigma)]|_{q}=&\int_{\mathcal{W}}d^{p}\sigma\sqrt{|\gamma|}B^{\mu\nu}\left(\omega_{\mu\nu}+{\Lambda_{\mu\nu}}\right)+\int_{\partial\mathcal{W}}d^{p-1}\wt\sigma\sqrt{|h|}~\wt n_\mu\wt B^{\mu\nu\rho}\left(\omega_{\mu\nu}+{\Lambda_{\mu\nu}}\right) \quad \\
&\int_{\mathcal{W}}d^{p}\sigma\sqrt{|\gamma|}B^{\mu}\xi_\mu+\int_{\partial\mathcal{W}}d^{p-1}\wt\sigma\sqrt{|h|}~\wt n_\mu \wt B^{\mu\nu}\xi_\nu \ .
\end{split}
\ee
On-shell, when the equations of motion \eqref{eq:genshape} and \eqref{eq:b2} are satisfied, one has that $B^{\mu}=\wt n_\mu \wt B^{\mu\nu}=0$ and the remaining terms above yield the constraints. In particular, the transverse part in the two indices of the Lorentz matrix leads to 
\be
\pud\mu\lambda\pud\nu\rho B^{[\lambda\rho]}=0 \ , \quad \pud\mu\lambda\pud\nu\rho  \wt n_\alpha\wt B^{\alpha[\lambda\rho]}|_{\partial\mathcal{W}}=0 \ ,
\ee
which leads to \eqref{eq:bt} and to the first equation in \eqref{eq:b1}. Therefore, this constraint can be interpreted as a consequence of invariance under local rotations of the normal coordinates.  In turn, the different projections of the symmetric part in \eqref{eq:localtrans} yield the constraints \eqref{eq:bt}, \eqref{eq:pt} and the second constraint in \eqref{eq:b1}. This expresses the fact that this formulation incorporates all requirements of general covariance.

When no couplings to the background curvature are present, i.e. $\Pi^{\sigma\lambda}=0$, \eqref{eq:b1} and \eqref{eq:bt} were also obtained by requiring the action \eqref{eq:genact} to be invariant under infinitesimal rotations of the normal vectors when using a gauge formulation of the variational principle \cite{Armas:2014rva}.

We have just shown that the constraints \eqref{eq:bt} can be understood as a consequence of invariance of the action under local rotations of the normal coordinates. However, in general, arbitrary tangential diffeomorphisms lead to the constraints \eqref{eq:bt}-\eqref{eq:pt} and \eqref{eq:b1}. For example, in order to get the constraint \eqref{eq:bt}, one may consider a tangential diffeomorphism $\xi_{\mu}=\xi^{||}_\mu$ for which $\pud\mu\alpha\pud\nu\sigma\partial_{[\mu}\xi^{||}_{\nu]}$ is non-vanishing. Alternatively, it suffices to consider a diffeomorphism that vanishes at the surface but whose derivatives do not. This is sufficient for obtaining only the first and third terms in \eqref{eq:varL1}. This implies that for the action \eqref{eq:genact} to be well-defined, the constraints \eqref{eq:bt}-\eqref{eq:pt}, \eqref{eq:b1} and the tangential projection of the equations of motion \eqref{eq:genshapetang} must be off-shell satisfied, i.e. they must be satisfied for all shape configurations whether or not they are solutions to \eqref{eq:genshapeperp}. In other words, the constraints \eqref{eq:bt}-\eqref{eq:pt} and \eqref{eq:b1} are non-dynamical and only the shape equation \eqref{eq:genshapeperp} is non-trivial. As we shall see, the constraint $\eqref{eq:bt}$ can restrict the type of terms that can compose the action \eqref{eq:genact}.


\subsection{Surfaces with non-trivial edges and intersections} \label{sec:sedges}
In the previous sections we considered surfaces with trivial edges, i.e. edges with no extra sources of stress. However, the possibility of adding such extra sources of stress deserves to be explored as it can have different physical applications. These include the dynamics of strings with spinning point-particles attached, edge tension or entanglement entropy, among others. The edges of the surface can be seen as codimension $n=1$ surfaces embedded in the $p$-dimensional surface. However, the most natural and general viewpoint, given the covariant approach taken here, is to view the boundary of the surface of codimension $n$ as a surface of codimension $n+1$ localised at the edges of the $p$-dimensional surface.

We therefore consider extending the action \eqref{eq:genact} to an action that also depends on the set of edge fields
\be
\Phi_{e}(\wt \sigma)=\{\wt X^{\mu}(\wt\sigma),\hsuu\mu\nu(\wt\sigma), \psud\nu\mu(\wt\sigma),\psdd\mu\nu(\wt\sigma),\Ksddu\mu\nu\rho(\wt\sigma),\osduu\mu\lambda\rho(\wt\sigma),\Bdddd\mu\nu\lambda\rho(\wt\sigma), \wt n^\mu(\wt \sigma)\} \ ,
\ee
and that takes the general form
\be \label{eq:genactedge}
S[\Phi(\sigma),\Phi_{e}(\wt\sigma)]=\int_{\mathcal{W}}d^{p}\sigma\mathcal{L}[\Phi(\sigma)]+\int_{\partial \mathcal{W}}d^{p-1}\wt \sigma\mathcal{L}_e[\Phi_e(\wt\sigma)] \ ~,
\ee
where the first contribution is the surface action $S_{s}$ as in \eqref{eq:genact} while the second contribution is the edge action $S_{e}$. The variation of $S_{s}$ is given in \eqref{eq:varL} while the Lagrangian variation of the edge action, keeping $\wt X^{\mu}(\wt\sigma)$ fixed, is organised analogously to \eqref{eq:varL} as 
\be \label{sec:edgegeomtr}
\begin{split}
\delta_\xi S_{e}[\Phi_{e}(\wt \sigma)]=\int_{\partial\mathcal{W}}d^{p-1}\wt\sigma\sqrt{|h|}\Big(&-\frac{1}{2}\wt{\mathcal{T}}_{\mu\nu}\delta_\xi \hsuu\mu\nu-{\wt{\mathcal{P}}_{\mu}}^{~\nu}\delta_\xi \psud\mu\nu+\frac{1}{2}\wt{\mathcal{B}}^{\mu\nu}\delta_\xi \psdd\mu\nu+\wt{\mathcal{V}}_\mu\delta_\xi \wt n^\mu \\
&+{\wt{\mathcal{D}}^{\mu\nu}}_{ \quad \rho}\delta_\xi\Ksddu\mu\nu\rho+{\wt{\mathcal{S}}^\mu}_{~\lambda\rho}\delta_\xi \osduu\mu\lambda\rho+{\wt{\mathcal{Q}}_\mu}^{~\nu\lambda\rho}\delta_\xi\Buddd\mu\lambda\nu\rho\Big)\ ,
\end{split}
\ee
where the different tensor structures characterising the edges are defined as in \eqref{eq:def1a}-\eqref{eq:def1c} and have the same interpretation but now applied to the edge. In particular, $\wt{\mathcal{T}}_{\mu\nu}$ is the edge tangential stress tensor. The total variation of \eqref{eq:genactedge} takes the form
\be \label{eq:varLedge}
\begin{split}
\delta_\xi S[\Phi(\sigma), \Phi_{e}(\wt \sigma)]=&\int_{\mathcal{W}}d^{p}\sigma\sqrt{|\gamma|}\left(B^{\mu\nu}\pdu\mu\rho\nabla_\rho\xi_\nu+B^{\mu}\xi_\mu\right)\\
&+\int_{\partial\mathcal{W}}d^{p-1}\wt\sigma\sqrt{|h|}\left(\left(\wt B^{\nu\rho}_e+\wt n_\mu\wt B^{\mu\nu\rho}\right)\psud\lambda\nu\nabla_\lambda\xi_\rho+\left(\wt B^{\nu}_e+\wt n_\mu\wt B^{\mu\nu}\right)\xi_\nu\right) \ ,
\end{split}
\ee
where $\wt B^{\nu\rho}_e$ and $\wt B^{\nu}_e$ denote the contributions due to the edge action and we have again assumed that the surface's edges do not have edges themselves. The first line in Eqn.~\eqref{eq:varLedge}, evaluated on $\mathcal{W}$, leads to the constraints \eqref{eq:st}-\eqref{eq:pt} and the equation of motion \eqref{eq:genshape}. The second line in Eqn.~\eqref{eq:varLedge}, evaluated on $\partial\mathcal{W}$, leads to constraints and equations of motion analogous to those of \eqref{eq:st}-\eqref{eq:genshape} but they take into account the sources \eqref{eq:b1}-\eqref{eq:b2}. In particular, from the first boundary term, one finds the constraints
\be \label{eq:st1}
\begin{split}
&\wt{\mathcal{D}}^{\mu\lambda[\sigma}\Ksddu\mu\lambda{\alpha]}+\psud\sigma\lambda\psud\alpha\rho\wt \nabla_\mu \wt{\mathcal{S}}^{\mu\lambda\rho}+\psud{[\sigma}\nu\psud{\alpha]}\lambda\wt{\Pi}^{\nu\lambda}=\wt n_\lambda\wt n_\mu \mathcal{D}^{\lambda\mu[\sigma}\wt{n}^{\alpha]}+\wt n_\lambda \psud\alpha\mu\psud\sigma\nu\mathcal{S}^{\lambda\nu\mu} \ ,\\
&\wt{\mathcal{B}}^{\alpha\sigma}=\wt{\mathcal{D}}^{\mu\lambda(\sigma}\Ksddu\mu\lambda{\alpha)}+\psud{(\sigma}\nu\psud{\alpha)}\lambda\wt \Pi^{\nu\lambda}-\wt n_\lambda\wt n_\mu \mathcal{D}^{\lambda\mu(\sigma}\wt{n}^{\alpha)}+\wt{\mathcal{V}}_\mu\psuu\mu{(\alpha}\wt n^{\sigma)}  \ ,\\
&{\wt{\mathcal{P}}^{\sigma}}{_{\alpha}} =\wt{\mathcal{S}}^{\mu}{_\alpha}{^\lambda}\Ksdud\mu\sigma\lambda+\hsud\sigma\nu\psdd\alpha\lambda\wt\Pi^{\nu\lambda}+\wt{\mathcal{V}}_\mu\huu\mu\sigma \wt n_\alpha  \ ,
\end{split}
\ee
where we have defined $\wt \Pi^{\mu\nu}$ as in \eqref{eq:pidef} but with $\mathcal{Q}^{\mu\nu\lambda\rho}$ replaced by $\wt{\mathcal{Q}}^{\mu\nu\lambda\rho}$. The first equation expresses the violation of the conservation of the spin current on the edges and, besides the usual terms also present in \eqref{eq:st}, extra sources due to the surface appear. The second equation exhibits a new contribution to the transverse stress tensor due to the presence of the surface bending moment while the third equation shows that the mixed tangential-transverse stress tensor remains unchanged. 

The equation of motion at the edges takes the form
\be \label{eq:genshapeedge}
\begin{split}
&\wt\nabla_\lambda\left(\wt{\mathcal{T}}^{\lambda\sigma}+\hsud\lambda\mu\left(\psud\sigma\nu\wt{\Pi}^{\mu\nu}-\wt{\Pi}^{\sigma\mu}\right)-\hsud\lambda\nu\wt\nabla_\mu\wt{\mathcal{D}}^{\mu\nu\sigma}-2{\wt{\mathcal{S}^{\mu}}{_\alpha}}{^{\sigma}}\Ksduu\mu\lambda\alpha+\wt n_\mu\hsud\lambda\nu\mathcal{D}^{\mu\nu\sigma}\right) \\
&=\left(\wt{\mathcal{S}}^{\mu\lambda\rho}-\wt{\mathcal{D}}^{\mu\lambda\rho}\right)\Buddd\sigma\mu\lambda\rho+2\wt{\mathcal{Q}}^{\mu\nu\lambda\rho}\nabla_\nu\Buddd\sigma\rho\mu\lambda \quad \\
&~~~~+\wt n_\lambda\left(\mathcal{T}^{\lambda\sigma}+\hdu\mu\lambda\left(\pdu\nu\sigma\Pi^{\mu\nu}-\Pi^{\sigma\mu}\right)-\hud\lambda\nu\overline\nabla_\mu\mathcal{D}^{\mu\nu\sigma}-2{{\mathcal{S}^{\mu}}{_\alpha}}{^{\sigma}}\Kduu\mu\lambda\alpha\right) \ ,
\end{split}
\ee
and one may observe that the effect of the sources \eqref{eq:b1}-\eqref{eq:b2} is to add a contribution $\wt n_\nu\mathcal{D}^{\nu\lambda\sigma}$ to an effective stress tensor on the surface and to add a pressure term on the right hand side of the equation, composed of a linear combination of surface contributions.

\subsubsection{Intersections}
The results presented above for surfaces with edges can be easily generalised to any surface/brane complex and its intersections. An intersection is a $(p-1)$-dimensional surface that acts as the edge/boundary of an arbitrary number of $p$-dimensional surfaces. Slightly abusing the notation, we now consider the set of edge fields $\Phi_e(\sigma)$ to be the set of geometric fields living on the intersection, also denoted by $\partial\mathcal{W}$. We consider an $l$ number of surfaces $\mathcal{W}_{(i)}$ and their intersection $\partial\mathcal{W}$. The fields living on each $p$-dimensional surface with coordinates $\sigma_{(i)}$ are denoted by $\Phi_{(i)}(\sigma_{(i)})$ and they consist of the set \eqref{eq:fields} with the subscript $(i)$. The action takes the following form
\be \label{eq:genactinter}
S[\Phi_{(i)}(\sigma_{(i)}),\Phi_{e}(\wt\sigma)]=\sum_{i=1}^{l}\int_{\mathcal{W}_{(i)}}d^{p}\sigma_{(i)}\mathcal{L}[\Phi_{(i)}(\sigma_{(i)})]+\int_{\partial \mathcal{W}}d^{p-1}\wt \sigma\mathcal{L}_e[\Phi_e(\wt\sigma)] \ ~.
\ee
For each surface $\mathcal{W}_{(i)}$ there will be an analogous shape equation to the one derived above for a single surface, while for the intersection, the equations of motion read
\be \label{eq:genshapeint}
\begin{split}
&\wt\nabla_\lambda\left(\wt{\mathcal{T}}^{\lambda\sigma}+\hsud\lambda\mu\left(\psud\sigma\nu\wt{\Pi}^{\mu\nu}-\wt{\Pi}^{\sigma\mu}\right)-\hsud\lambda\nu\wt\nabla_\mu\wt{\mathcal{D}}^{\mu\nu\sigma}-2{\wt{\mathcal{S}^{\mu}}{_\alpha}}{^{\sigma}}\Ksduu\mu\lambda\alpha+\hsud\lambda\nu\sum_{i=1}^{l}\wt n_\mu^{(i)}\mathcal{D}^{\mu\nu\sigma}_{(i)}\right) \\
&=\left(\wt{\mathcal{S}}^{\mu\lambda\rho}-\wt{\mathcal{D}}^{\mu\lambda\rho}\right)\Buddd\sigma\mu\lambda\rho+2\wt{\mathcal{Q}}^{\mu\nu\lambda\rho}\nabla_\nu\Buddd\sigma\rho\mu\lambda \quad \\
&~~~~+\sum_{i=1}^{l}\left[\wt n_\lambda^{(i)}\left(\mathcal{T}^{\lambda\sigma}_{(i)}+{\hdu\mu\lambda}_{(i)}\left({\pdu\nu\sigma}_{(i)}\Pi^{\mu\nu}_{(i)}-\Pi^{\sigma\mu}_{(i)}\right)-{\hud\lambda\nu}_{(i)}\overline\nabla_\mu^{(i)}\mathcal{D}^{\mu\nu\sigma}_{(i)}-2{{{\mathcal{S}^{\mu}}{_\alpha}}{^{\sigma}}}_{(i)}{\Kduu\mu\lambda\alpha}_{(i)}\right)\right] \ ,
\end{split}
\ee
where we have introduced the normal vectors to each $p$-dimensional surface's edge $\wt n_\lambda^{(i)}$ and the surface covariant derivate $\overline\nabla_\mu^{(i)}={\hud\alpha\mu}^{(i)}\nabla_\alpha$ on each $p$-dimensional surface. The effect of each $p$-dimensional surface on the intersection is to contribute with an effective pressure on the right hand side and with an effective stress tensor on the left hand side of the equation of motion. It is straightforward to generalise this to more complicated surface complexes where each of the $p$-dimensional surfaces considered above may also be the edges of other $(p+1)$-dimensional surfaces.

\subsection{Interfaces} \label{sec:interfaces}

In this section we  analyse in detail the diffeomorphism constraints and shape equations for an interior region $\mathcal{B}_{\text{int}}$ in the spacetime $\mathcal{M}$ enclosed by an interface/surface $\mathcal{W}$. This type of actions are of interest, for example, in the context of soap bubbles, fluid membranes and fluid droplets.

Consider actions for an interior spacetime region $\mathcal{B}_{\text{int}}$ enclosed by a dynamical interface $\mathcal{W}$ separating another exterior region $\mathcal{B}_{\text{ext}}$.\footnote{Note that this analysis is different than that considered in the context of boundary terms for variational principles in General Relativity. In that context, the induced metric $\gamma_{ab}$ on the boundary/interface is held fixed under Lagrangian variations (see e.g.~\cite{Lehner:2016vdi}).} We write this general action as the sum of three contributions
\be \label{eq:intact}
S[\Phi_{\text{int}}(x),\Phi(\sigma),\Phi_\text{ext}(x)]=\int_{\mathcal{B}_{\text{int}}}d^{D}x\mathcal{L}[\Phi_{\text{int}}(x)] +\int_{\mathcal{W}}d^{p}\sigma\mathcal{L}[\Phi(\sigma)] +\int_{\mathcal{B}_{\text{text}}}d^{D}x \mathcal{L}[\Phi_\text{ext}(x)]\ ~,
\ee
where the first contribution represents the enclosed internal region and the last contribution the exterior region. The collective set of surface fields $\Phi(\sigma)$ is the same as in \eqref{eq:fields}, while for the interior and exterior regions we consider to be functions of the metric and the background Riemann tensor
\be \label{eq:fieldsint}
\Phi_{\text{int}}(x)=\{\Gdd\mu\nu^{\text{int}}(x)~,~\Bdddd\mu\nu\lambda\rho^{\text{int}}(x)\} \ , \quad \Phi_{\text{ext}}(x)=\{\Gdd\mu\nu^{\text{ext}}(x)~,~\Bdddd\mu\nu\lambda\rho^{\text{ext}}(x)\} \ .
\ee
Here $\Gdd\mu\nu^{\text{int}}$ and $\Bdddd\mu\nu\lambda\rho^{\text{int}}$ denote the background metric and background Riemann tensor, respectively, in the enclosed region while $\Gdd\mu\nu^{\text{ext}}$ and $\Bdddd\mu\nu\lambda\rho^{\text{ext}}$ denote the same quantities in the exterior one. 
From \eqref{eq:intact} it is clear that the  stress tensor will have bulk and exterior components besides the surface components that we have dealt with in the previous section. However, there can be an inflow of energy-momentum from the interior/exterior to the surface. In what follows, we will drop the labels \emph{int}/\emph{ext} from the background fields for simplicity and work as if only the enclosed region in \eqref{eq:intact} was present. However, one can at any time easily restore the contribution from the exterior region by simply subtracting equivalent terms to all contributions arising from the interior of the enclosed region.

Focusing on the action for the enclosed region, we organise its variation according to
\be \label{eq:varintact}
\delta_\xi S[\Phi_{\text{int}}(x),\Phi(\sigma)]=\frac{1}{2}\int_{\mathcal{B}_{\text{int}}}d^{D}x\sqrt{|g|}\left(T^{\mu\nu}_{\text{int}}\delta_\xi \Gdd\mu\nu+{L_\mu}^{\nu\lambda\rho}\delta_\xi\Buddd\mu\lambda\nu\rho\right)+\delta_\xi S[\Phi(\sigma)] \ ,
\ee
while the variation of the interface $\delta_\xi S[\Phi(\sigma)]$ is given by \eqref{eq:varL}. Here, $T^{\mu\nu}_{\text{int}}$ encodes responses to variations of $\Gdd\mu\nu$, as is the case of a simple volume term, while ${L_\mu}^{\nu\lambda\rho}$ encodes the response of changes in the background curvature. Formally, the variation \eqref{eq:varintact} can be expressed as a single variation with respect to $\delta_\xi \Gdd\mu\nu$ but we have chosen to denote by $T^{\mu\nu}_{\text{int}}$ the couplings to $\Gdd\mu\nu$ that appear explicitly in the action \eqref{eq:intact}. These structures are defined according to
\be
T^{\mu\nu}_{\text{int}}=\frac{2}{\sqrt{|g|}}\frac{\delta_\xi \mathcal{L}}{\delta_\xi \Gdd\mu\nu} \ ,~~{L_\mu}^{\nu\lambda\rho}=\frac{2}{\sqrt{|g|}}\frac{\delta_\xi \mathcal{L}}{\delta_\xi \Buddd\mu\lambda\nu\rho} \ .
\ee
The bulk curvature moment ${L_\mu}^{\nu\lambda\rho}$ inherits the symmetries of the Riemann tensor, similarly to the curvature moments introduced in the case of surfaces. The equations of motion that will follow from \eqref{eq:varintact} include any type of theory built out from arbitrary contractions of the background Riemann tensor such as Lovelock gravity.

Using that for Lagrangian variations $\delta_\xi\Gdd\mu\nu=2\nabla_{(\mu}\xi_{\nu)}$, together with App.~\ref{app:variations}, the total variation of the action, including the surface part, can be organised as
\be \label{eq:varintorg}
\begin{split}
\delta_\xi S[\Phi_{\text{int}}(x),\Phi(\sigma)]=&\int_{\mathcal{B}_{\text{int}}}d^{D}x\sqrt{|g|}\mathring B^{\mu}\xi_\mu\\
&+\int_{\mathcal{W}}d^{p}\sigma\sqrt{|\gamma|}\left(B^{\mu\nu\rho}\pdu\nu{(\lambda}\pdu\rho{\alpha)}\nabla_\alpha \nabla_\lambda\xi_\mu+ B^{\mu\nu}\pdu\mu\rho\nabla_\rho\xi_\nu+ B^{\mu}\xi_\mu\right)~,
\end{split}
\ee
for which, besides the appearance of a new term in the first line, there is a new term in the surface variation proportional to $B^{\mu\nu\rho}$ when compared to \eqref{eq:varL1}.\footnote{This term did not appear in \eqref{eq:varL1} simply because all the variations taken there automatically satisfied the constraint that arises from $B^{\mu\nu\rho}$.} All the terms above must vanish individually. From the bulk part we obtain the bulk equation of motion
\be \label{eq:eqbulk}
\nabla_\mu\left(T^{\mu\nu}_{\text{int}}-2\nabla_\lambda\nabla_\rho L^{\lambda\rho(\mu\nu)}\right)\bigg|_{\mathcal{B}_{\text{int}}}=0 \ ,
\ee
which not surprisingly is simply the conservation of the bulk components of the spacetime stress tensor, as we will see in the next section. In turn, on the interface, we find the diffeomorphism constraint
\be
2n_\sigma L^{\mu\sigma\rho\nu}\pdu\nu{(\lambda}\pdu\rho{\alpha)}|_{\mathcal{W}}=0 \ ,
\ee
which is always trivially satisfied since in this case $\pdd\mu\nu=n_\mu n_\nu$ and $L^{\mu\sigma\rho\nu}$ is anti-symmetric in the indices $\sigma,\nu$. The non-trivial diffeomorphism constraint that arises from the second term in the second line of \eqref{eq:varintorg} leads to the same two constraints \eqref{eq:bt} and \eqref{eq:pt}. In particular we have that
\be
\hdu\nu\sigma\pdd\lambda\alpha\Pi^{\nu\lambda}+\mathcal{V}_\mu\huu\mu\sigma n_\alpha\delta_{n,1}=0 \ .
\ee
We remind the reader that since the interface is codimension $n=1$ then $\mathcal{P}^{\mu\nu}$ is absent and the constraint \eqref{eq:st} is automatically satisfied. Finally, the equation of motion on the surface takes the form
\be \label{eq:interfull}
\begin{split}
&\overline\nabla_\lambda\left(\mathcal{T}^{\lambda\sigma}+\hdu\mu\lambda\left(\pdu\nu\sigma\Pi^{\mu\nu}-\Pi^{\sigma\mu}\right)-\hud\lambda\nu\overline\nabla_\mu\mathcal{D}^{\mu\nu\sigma}+\hud\lambda\mu\left(\mathring T^{\mu\sigma}-\overline\nabla_\rho\left(\hud\rho\alpha\mathring T^{\alpha\sigma\mu}\right)\right)\right) \\
&=\mathcal{D}^{\mu\lambda\rho}\Buddd\sigma\mu\rho\lambda+2\mathcal{Q}^{\mu\nu\lambda\rho}\nabla_\nu\Buddd\sigma\rho\mu\lambda+\left(\mathring T^{\mu\alpha\rho}+\frac{1}{2}\pud\rho\tau\mathring T^{\tau\alpha\mu}\right)\Buddd\sigma\alpha\mu\rho+n_\lambda \left(T^{\lambda\sigma}_{\text{int}}-2\nabla_\mu\nabla_\rho L^{\mu\rho(\lambda\sigma)}\right)  \ ,
\end{split}
\ee
where we have set the spin current ${\mathcal{S}_\mu}^{\nu\rho}=0$ since it vanishes for codimension $n=1$. We have also defined 
\be
\mathring T^{\mu\nu}=2 n_\lambda\nabla_\rho L^{\lambda\rho(\mu\nu)}-2\overline\nabla_\lambda\left(\hud\lambda\rho n_\sigma L^{(\mu|\sigma\rho|\nu)}\right) \ , \quad \mathring T^{\mu\nu\rho}=2n_\sigma L^{(\mu|\sigma\lambda|\nu)}\pud\rho\lambda \ .
\ee
These definitions were introduced due to their frame-invariant properties, as explained in App.~\ref{app:highdev}. The last term in \eqref{eq:interfull} is the usual contribution that takes into account the effect of the bulk pressure in the Young-Laplace law. If the bulk action is only composed of the volume term $\sqrt{|g|}P$ for constant P, then the normal projection of the last term in \eqref{eq:interfull} yields the bulk pressure $P$. The last contribution in the first line and the third term in the second line in \eqref{eq:interfull} take into account the inflow of energy-momentum from the bulk to the interface.


\section{Spacetime stress tensor} \label{sec:stresstensor}

The variational principle cast in \eqref{eq:varL} in terms of Lagrangian variations is in essence a variational principle in terms of variations of the underlying background metric. In particular,  the tensorial objects defined in \eqref{eq:def1a}-\eqref{eq:def1c} are nothing but a convenient packaging with a clear physical meaning from the point of view of the embedded surface, thus aiding in the presentation of the resultant dynamics.
In this section we relate explicitly these tensorial objects to the stress tensor obtained from the variation of the surface action with respect to the background metric $\Gdd\mu\nu$. The construction of the spacetime stress tensor allows to define conserved currents and charges associated with a given surface.

\subsection{Spacetime stress tensor for surfaces} \label{sec:st}

One can recast the action \eqref{eq:genact} as an integral over the whole manifold $\mathcal{M}$ of a Lagrangian dependent only  on $\Phi(\sigma)=\{\Gdd\mu\nu(X), X^{\mu}(\sigma)\}$ with the help of the reparametrisation invariant delta function
\be
\widehat\delta(x) = \frac{\sqrt{|\gamma|}}{\sqrt{|g|}} \, \delta^{(n)}(x^{\alpha}-X^{\alpha}) \ ,
\ee
that is\footnote{\label{foot:stress}The form of \eqref{eq:genact2} implies that we have chosen to work in the static gauge $x^{1}=\sigma^{1},\cdots,x^{p}=\sigma^{p}$. However, this is only for convenience and does not affect the analysis carried out here. In full generality one could rewrite \eqref{eq:genact2} as $S[\Phi(\sigma)]=\int_{\mathcal{M}}\sqrt{|g|}d^{D}x \,\bar{\mathcal{L}}\left[\Gdd\mu\nu(X), X^{\mu}(\sigma)\right]$ where $\bar{\mathcal{L}}\left[\Gdd\mu\nu(X), X^{\mu}(\sigma)\right]=\int_{\mathcal{W}}d^p\sigma \widehat\delta(x)^{(D)}\mathcal{L}\left[\Gdd\mu\nu(X), X^{\mu}(\sigma)\right]$ with $\widehat\delta^{(D)}(x)=\sqrt{|\gamma|}\delta^{(D)}(x^\alpha-X^\alpha)/\sqrt{|g|}$. We will explicitly write the stress tensor in this general form in \eqref{eq:stgen2} as derived from here.}
\be \label{eq:genact2}
S[\Phi(\sigma)]=\int_{\mathcal{M}}d^{D}x \,\widehat\delta(x) \, \mathcal{L}\left[\Gdd\mu\nu(X), X^{\mu}(\sigma)\right] \ .
\ee
With this reformulation, a Lagrangian variation where $\delta_\xi X^{\mu}(\sigma)=0$ takes the usual form
\be \label{eq:varL2}
\delta_\xi S[\Phi(\sigma)]=\frac{1}{2}\int_{\mathcal{M}}d^{D}x\sqrt{|g|}\left(\Tuu\mu\nu\delta_\xi \Gdd\mu\nu\right)~~,~~ \Tuu\mu\nu=\frac{2}{\sqrt{|g}|}\frac{\delta S}{\delta_\xi \Gdd\mu\nu}\ ,
\ee
which upon using that $\delta_\xi g_{\mu\nu}=2\nabla_{(\mu}\xi_{\nu)}$ leads to the equations of motion
\be \label{eq:stress}
\nabla_\mu \Tuu\mu\nu=0 \ , \quad \Tuu\mu\nu\eta_\mu|_{\partial \mathcal{M}}=0 \ ,
\ee
where we have introduced the unit-normalised normal co-vector $\eta_{\mu}$ to the spacetime boundary. We assume that this boundary condition is satisfied or that the spacetime has no boundaries. The simplicity of the reformulation of the dynamics \eqref{eq:stress} in terms of the conservation of a spacetime stress tensor is traded by the necessity of dealing with the singular character of the stress tensor, which is now formulated as a multipole expansion in derivatives of $\widehat\delta(x)$. For the case of the actions that we are dealing with in Eqn.~\eqref{eq:genact}, where couplings involve geometric tensors with at most two derivatives, the corresponding stress tensor will at most involve two derivatives of $\widehat\delta(x)$. As such, it can be expressed as
\be \label{eq:stgen}
\Tuu\mu\nu=T^{\mu\nu}\widehat\delta(x)-\nabla_\rho\left(T^{\mu\nu\rho} \,\widehat\delta(x)\right)+\nabla_\lambda\nabla_\rho\left(T^{\mu\nu\rho\lambda} \,\widehat\delta(x)\right) \ ,
\ee
where the coefficients $T^{\mu\nu}$, $T^{\mu\nu\rho}$ and $T^{\mu\nu\rho\lambda}$ are only functions of the surface coordinates $\sigma^{a}$. As we shall see in Sec.~\ref{sec:DCFTs}, the terms $T^{\mu\nu}$, $T^{\mu\nu\rho}$ and $T^{\mu\nu\rho\lambda}$ are also known as contact terms in the context of DCFTs. In essence, $T^{\mu\nu\rho}$ and $T^{\mu\nu\rho\lambda}$ represent the dipole and quadrupole moments of stress, respectively, besides the monopole source $T^{\mu\nu}$.

The stress tensor \eqref{eq:stgen} transforms as a tensor and, in its present form, also do the components $T^{\mu\nu}$, $T^{\mu\nu\rho}$ and $T^{\mu\nu\rho\lambda}$. This can easily be show by evaluating the scalar functional 
\be
T[f]=\int_{\mathcal{M}}d^{D}x \, \sqrt{|g|} \, \Tuu\mu\nu f_{\mu\nu} 
\ee
for an arbitrary tensor field $f_{\mu\nu}(x^{\alpha})$ of compact support \cite{Vasilic:2007wp}. The invariance of the scalar functional dictates the transformation properties of each of the components of the stress tensor. An alternative basis for \eqref{eq:stgen} is also common in the literature
\be \label{eq:stgen1}
\Tuu\mu\nu=\widehat T^{\mu\nu}\widehat\delta(x)+\widehat T^{\mu\nu\rho} \,\nabla_\rho\widehat\delta(x)+\widehat T^{\mu\nu\rho\lambda}\nabla_\lambda\nabla_\rho\widehat\delta(x) \ ,
\ee
where the coefficients in \eqref{eq:stgen1} are related to those in \eqref{eq:stgen} according to
\be \label{eq:stbasis}
\widehat T^{\mu\nu}=T^{\mu\nu}-\nabla_\rho T^{\mu\nu\rho}+\nabla_\lambda\nabla_\rho T^{\mu\nu\rho\lambda} \ , \quad \widehat T^{\mu\nu\rho}=-T^{\mu\nu\rho}+2\nabla_\lambda T^{\mu\nu(\rho\lambda)} \ , \quad \widehat T^{\mu\nu\rho\lambda}=T^{\mu\nu\rho\lambda} \ .
\ee
It is not only clear from the properties of $T[f]$ but also from the relation \eqref{eq:stbasis} that the components of  \eqref{eq:stgen1} do not transform like tensors. Since one has assumed that each component is only a function of $\sigma^{a}$ and hence that $\partial_\rho T^{\mu\nu\rho}=0$, the derivatives in \eqref{eq:stbasis} lack their covariant properties. For some practical purposes, as describing fluids living on surfaces or fluid droplets, one may wish to allow the components of stress to be also functions of $X^{\mu}$ as in \cite{Armas:2013goa} or to extend them to a foliation of such surfaces as in \cite{Armas:2015ssd}. In the latter case, the components in both basis \eqref{eq:stgen} and \eqref{eq:stgen1} transform covariantly. But since most applications do not require such extensions, we opt for the basis \eqref{eq:stgen}.

By using the variation formulae in App.~\ref{app:variations}, the stress tensor following from Eqn.~\eqref{eq:genact} takes the form given in \eqref{eq:stgen} with components \footnote{In writing $T^{\mu\nu\rho\lambda}$ we have assumed that $\mathcal{Q}^{\lambda\rho\mu\nu}$ inherits all symmetries of the Riemann tensor while in the writing of the equations of motion we have only assumed the symmetry $\mathcal{Q}^{\lambda\rho\mu\nu}=-\mathcal{Q}^{\lambda\nu\mu\rho}$. If only the latter symmetry is assumed $T^{\mu\nu\rho\lambda}$ can be written as $T^{\mu\nu\rho\lambda}=2\left(\mathcal{Q}^{(\mu|\rho\lambda|\nu)}+\mathcal{Q}^{(\mu|\rho|\nu)\lambda}-\mathcal{Q}^{\lambda\rho(\mu\nu)}\right)$.}
\be \label{eq:stact2}
\begin{split}
&T^{\mu\nu}=\mathcal{T}^{\mu\nu}+2 \, {\mathcal{P}^{\lambda}}{_\rho}\puu\rho{(\mu}\hdu\lambda{\nu)}+\mathcal{B}^{\mu\nu}-\left(\mathcal{V}_\lambda n^{\lambda}n^\mu n^\nu +2\mathcal{V}_\lambda\huu\lambda{(\mu}n^{\nu)}\right)\delta_{n,1} \ ,\\
&T^{\mu\nu\rho}=2\, \mathcal{D}^{\rho(\mu\nu)}-\mathcal{D}^{\mu\nu\rho}+2 \, \mathcal{S}^{(\mu\nu)\rho} \ , \\
&T^{\mu\nu\rho\lambda}=-4\, \mathcal{Q}^{\lambda\rho(\mu\nu)} \ .
\end{split}
\ee
It is clear from \eqref{eq:stact2} that the structures $\mathcal{T}^{\mu\nu}$, ${\mathcal{P}^{\lambda}}{_\rho}$ and $\mathcal{B}^{\mu\nu}$ for $n>1$ characterise the different components of the monopole part of the stress tensor,  $\mathcal{D}^{\mu\nu\rho}$ and $\mathcal{S}^{\mu\nu\rho}$ characterise the dipole part and  $\mathcal{Q}^{\mu\nu\lambda\rho}$ characterises the quadrupole part. A posteriori, the equations in \eqref{eq:stact2} justify the definitions of these quantities in the previous section. In the case $n=1$ for which ${\mathcal{P}^{\lambda}}{_\rho}=0$, the terms involving $\mathcal{V}_\lambda$ contribute to both the mixed tangential-transverse components and to the fully transverse components. 

Using the constraints \eqref{eq:bt}-\eqref{eq:pt} we can rewrite the stress tensor in terms of its independent components according to
\be \label{eq:stact3}
\begin{split}
\Tuu\mu\nu=&\left[\mathcal{T}^{\mu\nu}-2\mathcal{S}^{\alpha\lambda(\mu}\Kdud\alpha{\nu)}\lambda+2\hdu\rho{(\mu}\pdu\lambda{\nu)}\Pi^{\rho\lambda}+\mathcal{D}^{\alpha\lambda(\mu}\Kddu\alpha\lambda{\nu)}+\pdu\rho{(\mu}\pdu\lambda{\nu)}\Pi^{\rho\lambda}\right]\widehat\delta(x) \\
&-\nabla_\rho\left[\left(2\mathcal{D}^{\rho(\mu\nu)}-\mathcal{D}^{\mu\nu\rho}+2\mathcal{S}^{(\mu\nu)\rho}\right)\widehat\delta(x)\right]-4\nabla_\lambda\nabla_\rho\left[\mathcal{Q}^{\lambda\rho(\mu\nu)}\widehat\delta(x)\right] \ .
\end{split}
\ee
It is worth noting that the constraints \eqref{eq:bt}-\eqref{eq:pt} were derived in \cite{Vasilic:2007wp} with $T^{\mu\nu\rho\lambda}=0$ by integrating \eqref{eq:stress} over spacetime using an analog of a Gaussian pillbox, that is, considering the integral $\int_{\mathcal{M}}d^{D}x\sqrt{|g|}\nabla_\mu \Tuu\mu\nu f_\nu$ for some arbitrary vector field $f_\mu(x^{\alpha})$ with compact support. Here we have generalised these constraints to the case where $T^{\mu\nu\rho\lambda}$ is non-trivial and furthermore shown that a Lagrangian variational principle captures all of them. By the same token, the equations of motion that follow from \eqref{eq:stress} must be equivalent to those obtained in \eqref{eq:genshape}. This has been shown to be the case in \cite{Armas:2013hsa} when $T^{\mu\nu\rho\lambda}=0$. It remains to be shown, for the purpose of completeness, that by considering $\int_{\mathcal{M}}d^{D}x\sqrt{|g|}\nabla_\mu \Tuu\mu\nu f_\nu$ one obtains \eqref{eq:stress} with a non-trivial $T^{\mu\nu\rho\lambda}$. But this is not in doubt as consistency of \eqref{eq:varL1} with \eqref{eq:varL2} requires it.

\subsubsection{Edge contributions}

The presence of non-trivial edges induces new contributions to the spacetime stress tensor
\be \label{eq:stgenform}
\boldsymbol{T}^{\mu\nu}=\boldsymbol{T}^{\mu\nu}_s+\boldsymbol{T}^{\mu\nu}_e \ ,
\ee
where $\boldsymbol{T}^{\mu\nu}_s$ is the surface contribution evaluated in \eqref{eq:stact3} and $\boldsymbol{T}^{\mu\nu}_e$ is the contribution due to the non-zero edge action. This latter contribution has also an expansion in derivatives of a delta function as in \eqref{eq:stgen} but with $\widehat \delta (x)$ replaced by 
\be
\widehat \delta_e (x) = \frac{\sqrt{|h|}}{\sqrt{|g|}} \, \delta^{(n)}(x^{\alpha}-\wt X^{\alpha})
\ee
such that the total stress tensor has the form
\be \label{eq:sts+e}
\begin{split}
\Tuu\mu\nu=&~T^{\mu\nu}\widehat\delta(x)-\nabla_\rho\left(T^{\mu\nu\rho}\widehat\delta(x)\right)+\nabla_\lambda\nabla_\rho\left(T^{\mu\nu\rho\lambda}\widehat\delta(x)\right) \quad \\
&+\wt T^{\mu\nu}\widehat\delta_e(x)-\nabla_\rho\left(\wt T^{\mu\nu\rho}\widehat\delta_e(x)\right)+\nabla_\lambda\nabla_\rho\left(\wt T^{\mu\nu\rho\lambda}\widehat\delta_e(x)\right) \ ,
\end{split}
\ee
where $\wt T^{\mu\nu}$, $\wt T^{\mu\nu\rho}$ and $\wt T^{\mu\nu\rho\lambda}$ are the edge monopole, dipole and quadrupole sources of stress, respectively. For the specific action \eqref{eq:genactedge}, the surface components were given in \eqref{eq:stact2} while the edge components read
\be\label{eq:stactedge}
\begin{split}
&\wt T^{\mu\nu}= \wt{\mathcal{T}}^{\mu\nu}+2 \, {\wt{\mathcal{P}}^{\lambda}}{_\rho}\puu\rho{(\mu}\hdu\lambda{\nu)}+\wt{\mathcal{B}}^{\mu\nu} - \wt n_\lambda\wt n_\rho \mathcal{D}^{\lambda\rho(\mu}\wt{n}^{\nu)}-\wt{\mathcal{V}}_\lambda\psuu\lambda{(\mu}\wt n^{\nu)}-2\wt{\mathcal{V}}_\lambda\hsuu\lambda{(\mu}\wt n^{\nu)} \ ,\\
&\wt T^{\mu\nu\rho}=2 \, \wt{\mathcal{D}}^{\rho(\mu\nu)}-\wt{\mathcal{D}}^{\mu\nu\rho}+2 \, \wt{\mathcal{S}}^{(\mu\nu)\rho} \ ,\\
&\wt T^{\mu\nu\rho\lambda}=-4 \, \wt{\mathcal{Q}}^{\lambda\rho(\mu\nu)} \ .
\end{split}
\ee
The difference between the edge contribution and its surface counterpart is the appearance of the last three terms in $\wt T^{\mu\nu}$.


\subsubsection{Frame choices}\label{sec:framechoices}

The expression for the stress tensor given in Eqn.~\eqref{eq:stgen} contains redundant components, which in turn leads to many equivalent descriptions of the stress tensor for a given surface, and therefore to a symmetry.\footnote{Beyond the symmetry discussed here, in a perturbative setting there is a perturbative symmetry that supervenes on the invariance of the stress tensor under surface displacements $X^{\mu}(\sigma)\to X^{\mu}(\sigma)+\delta X^{\mu}(\sigma)$. However, since we are not concerned with a perturbative analysis here we let the avid reader see \cite{Vasilic:2007wp, Armas:2013hsa, Armas:2013goa} for a discussion of this symmetry.} This redundancy is rooted in the fact that one may make the expression \eqref{eq:stgen} manifestly covariant in the embedding functions $X^{\mu}$ by integrating over the $p$-surface directions\footnote{This form of the stress tensor can also be obtained directly from the action \eqref{eq:genact2} as explained in footnote \ref{foot:stress}.}
\be \label{eq:stgen2}
\Tuu\mu\nu=\int_{\mathcal{W}}d^{p}\sigma\left(T^{\mu\nu}\widehat\delta^{(D)}(x)-\nabla_\rho\left(T^{\mu\nu\rho}\widehat\delta^{(D)}(x)\right)+\nabla_\lambda\nabla_\rho\left(T^{\mu\nu\rho\lambda}\widehat\delta^{(D)}(x)\right)\right) \ ,
\ee
where now $\widehat\delta^{(D)}(x)=\delta^{(p)}(x^{\alpha}-X^{\alpha})\widehat\delta(x)$. By adding such integration, it is clear that the tangential components $T^{\mu\nu\rho}\hdu\rho\lambda$ and $T^{\mu\nu\rho\lambda}\hdu\lambda\sigma$ can be removed via an integration by parts with appropriate boundary conditions. In particular, the stress tensor \eqref{eq:stgen2} is invariant under two independent transformations, $\delta_{\varepsilon_1}$ and $\delta_{\varepsilon_2}$, acting according to
\be \label{eq:transform}
\begin{split}
&\delta_{\varepsilon_1} \left(T^{\mu\nu\lambda}\hdu\lambda\rho\right)=\varepsilon^{\mu\nu\rho} \ , \quad \delta_{\varepsilon_1} T^{\mu\nu}=\overline\nabla_\rho\varepsilon^{\mu\nu\rho} \ , \quad \varepsilon^{\mu\nu\rho}\wt n_\rho |_{\partial\mathcal{W}}=0 \ , \\
&\delta_{\varepsilon_2} \left(T^{\mu\nu\rho\sigma}\hdu\sigma\lambda\right)=\varepsilon^{\mu\nu\rho\lambda} \ , \quad \delta_{\varepsilon_2}  T^{\mu\nu\rho}=\overline\nabla_\lambda\varepsilon^{\mu\nu\rho\lambda} \ , \quad \varepsilon^{\mu\nu\rho\lambda}\wt n_\lambda |_{\partial\mathcal{W}}=0 \ ,
\end{split}
\ee
where the coefficients $\varepsilon^{\mu\nu\rho}$ and $\varepsilon^{\mu\nu\rho\lambda}$ are symmetric in their first two indices and tangential in their last index. These transformation properties indicate that not all components of the stress tensor are physical. Performing the transformations \eqref{eq:transform} without imposing the boundary conditions on $\varepsilon^{\mu\nu\rho}$ and $\varepsilon^{\mu\nu\rho\lambda}$ gives rise to different physics on the edges of the surface.

With little effort, it is possible to identify the frame-invariant components. The following two combinations are invariant under $\delta_{\varepsilon_1}$
\be \label{eq:inv1}
\delta_{\varepsilon_1}\left(T^{\mu\nu}-\overline\nabla_\rho\left(T^{\mu\nu\lambda}\hdu\lambda\rho\right)\right)=0 \ , \quad \delta_{\varepsilon_1}\left(\pdu\lambda\rho T^{\mu\nu\lambda}\right)=0 \ ,
\ee
while the following combinations are invariant under $\delta_{\varepsilon_2}$
\be \label{eq:inv2}
\delta_{\varepsilon_2}\left(T^{\mu\nu\rho}-\overline\nabla_\sigma\left(T^{\mu\nu\rho\lambda}\hdu\lambda\sigma\right)\right)=0 \ , \quad \delta_{\varepsilon_2}\left(\pdu\lambda\sigma T^{\mu\nu\rho\lambda}\right)=0 \ .
\ee
Ref.~\cite{Vasilic:2007wp} had in fact identified a subset of \eqref{eq:inv1} when $T^{\mu\nu\rho\lambda}$ vanishes. The presence of a non-zero $T^{\mu\nu\rho\lambda}$ implies that only the last combination in \eqref{eq:inv2} is invariant under both symmetries by itself. Combining both transformations one arrives at the mutually independent invariant combinations
\be \label{eq:inv3}
\begin{split}
&\mathsf{T}^{\mu\nu}=T^{\mu\nu}-\overline\nabla_\lambda\left(\hdu\rho\lambda\left( T^{\mu\nu\rho}+\overline\nabla_\alpha\left(T^{\mu\nu\rho\sigma}\hdu\sigma\alpha\right)\right)\right) \ , \\
&\mathsf{T}^{\mu\nu\rho}=\pdu\rho\lambda\left(T^{\mu\nu\rho}-\overline\nabla_\sigma\left(T^{\mu\nu\rho\lambda}\hdu\lambda\sigma\right)\right) \ , \\
&\mathsf{T}^{\mu\nu\rho\lambda}=\pdu\sigma\lambda T^{\mu\nu\rho\sigma} \ .
\end{split}
\ee
The invariants \eqref{eq:inv3} can be used as a means for comparison between stress tensors in different frames. It can be observed from \eqref{eq:inv3} that there are three components in the first invariant $\mathsf{T}^{\mu\nu}$, one purely tangential and the other two with at least one transverse index. The same holds for the invariant $\mathsf{T}^{\mu\nu\rho}$, while $\mathsf{T}^{\mu\nu\rho\lambda}$ has six independent components. General covariance imposes four relations among these. The remaining independent components can be chosen to be, in the frame explained below, $\mathcal{T}^{\mu\nu}$, $\mathcal{D}^{\mu\nu\rho}$, $\mathcal{S}^{\mu\nu\rho}$ and $\mathcal{Q}^{\mu\nu\lambda\rho}$.

In \eqref{eq:stact3}, we obtained the stress tensor that followed from \eqref{eq:genact2} and by inspection it comes in a given frame, dictated by the coefficients $T^{\mu\nu\sigma}\hdu\sigma\rho=2\mathcal{D}^{\rho(\mu\nu)}$ and $T^{\mu\nu\rho\sigma}\hdu\sigma\lambda=-4\mathcal{Q}^{\sigma\rho(\mu\nu)}\hdu\sigma\lambda$. The fact that the stress tensor is obtained in a given frame from the action \eqref{eq:genact2} is not significant. While performing the variation \eqref{eq:varL2}, further integrations by parts could have been performed in order to remove the tangential components from $T^{\mu\nu\lambda}$ and $T^{\mu\nu\rho\sigma}$, at the expense of finding a non-zero inflow of stress from the surface to its edges. The natural frame \eqref{eq:stact2} is defined as the frame with vanishing inflow of stress from the surface to its edges. As a consequence of this ambiguity, the constraints found in \eqref{eq:bt} and \eqref{eq:pt} are not frame-independent but only apply in the frame \eqref{eq:stact2}. In order to make them applicable to any given frame, one may first move to a frame where $T^{\mu\nu\lambda}\hdu\lambda\rho$ and $T^{\mu\nu\rho\sigma}\hdu\sigma\lambda$ vanish and then add arbitrary $\delta_{\varepsilon_1}$ and $\delta_{\varepsilon_2}$ transformations.

The stress tensor \eqref{eq:sts+e} is affected by the same frame choices. In particular, the edge contribution to \eqref{eq:sts+e} can be written in a different frame using the analogous transformations to \eqref{eq:transform}. 


\subsection{Conserved currents and charges} \label{sec:charges}

With the covariant formulation presented here one can identify a set of conserved charges associated with symmetries of the background. The existence of an isometry implies the existence of a background Killing vector field $\tku\mu(x^{\alpha})$ that can be totally transverse. If $\tku\mu$ is the Killing vector field associated with an isometry then the action \eqref{eq:genactedge} must be invariant under variations along those Killing directions. This fact can be used in order to obtain a set of conserved surface currents. Setting $\xi^{\mu}=\beta \,\tku\mu$ for some constant $\beta$, the variation of the action \eqref{eq:genactedge} leads to
\be
\begin{split}
&\delta_{\textbf{k}}S[\Phi(\sigma),\Phi_{e}(\wt\sigma)]=\beta \int_{\mathcal{W}}d^{p}\sigma\left[B^{\mu\nu}\pdu\mu\rho\nabla_\rho\tkd\nu+B^{\mu}\tkd\mu+\overline\nabla_\lambda\left(\mathbb{T}^{\lambda\sigma}\tkd\sigma+\Sigma^{\lambda\sigma\alpha}\nabla_\alpha\tkd\sigma\right)\right]\\
&+\!\beta\!\int_{\partial\mathcal{W}}\!\!\!\!\!d^{p-1}\wt\sigma\sqrt{|h|}\left[\wt B^{\nu\rho}_e\psud\lambda\nu\nabla_\lambda\tkd\rho+\left(\wt B^{\nu}_e-\wt\nabla_\lambda\left(\wt n_\mu\hsud\lambda\nu\mathcal{D}^{\mu\nu\sigma}\right)\right)\tkd\nu+\wt\nabla_\lambda \left(\wt{\mathbb{T}}^{\lambda\sigma}\tkd\sigma+\wt\Sigma^{\lambda\sigma\alpha}\nabla_\alpha\tkd\sigma\right)\right]~,
\end{split}
\ee
where we have defined the effective surface stress tensor $\mathbb{T}^{\lambda\sigma}$ and the effective edge stress tensor $\wt{\mathbb{T}}^{\lambda\sigma}$ as
\be\label{eq:truestress}
\begin{split}
&\mathbb{T}^{\lambda\sigma}=\mathcal{T}^{\lambda\sigma}+{{\mathcal{P}}^{\mu}}_\nu\puu\nu\sigma\hud\lambda\mu-\hdu\mu\lambda\Pi^{\sigma\mu}-\hud\lambda\nu\overline\nabla_\mu\mathcal{D}^{\mu\nu\sigma}-{{\mathcal{S}^{\mu}}_\alpha}^{\sigma}\Kduu\mu\lambda\alpha-\mathcal{V}_\mu\huu\mu\lambda n^\sigma\delta_{n,1} \ ,\\
&\wt{\mathbb{T}}^{\lambda\sigma}=\wt{\mathcal{T}}^{\lambda\sigma}+{{\wt{\mathcal{P}}}^{\mu}}_{~\nu}\psuu\nu\sigma\hsud\lambda\mu-\hsdu\mu\lambda\wt{\Pi}^{\sigma\mu}-\hsud\lambda\nu\wt\nabla_\mu\wt{\mathcal{D}}^{\mu\nu\sigma} - \wt{\mathcal{S}}^{\mu}{_\alpha}{^\sigma}  \Ksduu\mu\lambda\alpha+\wt n_\mu\hsud\lambda\nu\mathcal{D}^{\mu\nu\sigma}-\wt{\mathcal{V}}_\mu\hsuu\mu\lambda \wt n^\sigma \ ,\\
\end{split}
\ee
as well as the tensors $\Sigma^{\lambda\sigma\alpha}$ and $\wt \Sigma^{\lambda\sigma\alpha}$ according to
\be\label{eq:trueSigmas}
\Sigma^{\lambda\sigma\alpha}=\mathcal{D}^{\lambda\alpha\sigma}+\mathcal{S}^{\lambda\sigma\alpha} \ , \quad \wt\Sigma^{\lambda\sigma\alpha}=\wt{\mathcal{D}}^{\lambda\alpha\sigma}+\wt{\mathcal{S}}^{\lambda\sigma\alpha} \ ,
\ee
which parametrise the non-trivial modification to the surface and edge currents due to the presence of dipole terms in the stress tensor which couple to the background Riemann tensor. The effective surface stress tensor $\mathbb{T}^{\lambda\sigma}$ is not symmetric, is tangential in its first index but has orthogonal components in its second index. The same holds for the effective edge stress tensor $\wt{\mathbb{T}}^{\lambda\sigma}$. The structures $\Sigma^{\lambda\sigma\alpha}$ and $\wt \Sigma^{\lambda\sigma\alpha}$ are also tangential in their first index.

Referring to  $\mathbb{T}^{\lambda\sigma}$ and $\wt{\mathbb{T}}^{\lambda\sigma}$ as effective surface and edge stress tensors, respectively, is justified since the equations of motion \eqref{eq:genshape} and \eqref{eq:genshapeedge} can be recast as stress conservation equations, such that
\be \label{eq:eom}
\begin{split}
&\overline\nabla_\lambda \mathbb{T}^{\lambda\sigma}=\Sigma^{\mu\lambda\rho}\Buddd\sigma\mu\rho\lambda+2\mathcal{Q}^{\mu\nu\lambda\rho}\nabla_\nu\Buddd\sigma\rho\mu\lambda  \ ,\\
&\wt\nabla_\lambda \wt{\mathbb{T}}^{\lambda\sigma}=\wt\Sigma^{\mu\lambda\rho}\Buddd\sigma\mu\rho\lambda+2\wt{\mathcal{Q}}^{\mu\nu\lambda\rho}\nabla_\nu\Buddd\sigma\rho\mu\lambda+\wt n_\lambda \mathbb{T}^{\lambda\sigma} \ .
\end{split}
\ee
On-shell, when the equations of motion \eqref{eq:eom} are satisfied, we have that
\be
\begin{split}
&B^{\mu\nu}\pdu\mu\rho=0 \ , \quad B^{\mu}=0 \ , \quad \wt B^{\nu[\sigma}_e\psud{\alpha]}\nu=-\wt n_\lambda\wt n_\mu \mathcal{D}^{\lambda\mu[\sigma}\wt{n}^{\alpha]}-\wt n_\lambda \psud\alpha\mu\psud\sigma\nu\mathcal{S}^{\lambda\nu\mu} \ ,\\
&\wt B^{\sigma}_e=\wt\nabla_\lambda\left(\wt n_\mu\hsud\lambda\nu\mathcal{D}^{\mu\nu\sigma}\right)-\wt n_\lambda\mathbb{T}^{\lambda\sigma} \ ,
\end{split}
\ee
where we have used the Killing equation $\nabla_{(\mu}\tkd{\nu)}=0$. Therefore, requiring the action to be invariant under such isometry, we find the surface and edge current conservation equations
\be \label{eq:cc}
\overline\nabla_\lambda t^{\lambda}_k=0 \ , \quad \wt\nabla_\mu \wt t^{\mu}_k-\wt n_\lambda T^{\lambda}_k=0 \ ,
\ee
for the purely tangential surface current $t^{\lambda}_k$ and edge current $\wt t^{\lambda}_k$ defined as
\be \label{eq:cc1}
t^{\lambda}_k=\mathbb{T}^{\lambda\sigma}\tkd\sigma+\Sigma^{\lambda\sigma\alpha}\nabla_\alpha\tkd\sigma \ , \quad \wt t^{\lambda}_k=\wt{\mathbb{T}}^{\lambda\sigma}\tkd\sigma+\wt\Sigma^{\lambda\sigma\alpha}\nabla_\alpha\tkd\sigma \ .
\ee
It is straightforward to see that Eqs.~\eqref{eq:cc} are satisfied for the currents \eqref{eq:cc1}. In order to do so, one must make use of the equations of motion \eqref{eq:eom}, the constraints \eqref{eq:st}-\eqref{eq:pt} and \eqref{eq:st1}, the Killing equation and the fact that for a Killing vector field one has that $\nabla_\nu\nabla_\mu\tkd\rho=\Bdddd\rho\mu\nu\lambda\tku\lambda$. Having identified the set of conserved currents, one may define a set of conserved charges. Assuming the topology of the surface to be $\mathbb{R}\times \mathcal{B}_{p-1}$ and that of the edge to be $\mathbb{R}\times \wt{\mathcal{B}}_{p-2}$, the conserved charge associated with a given Killing vector $\tku\mu$ is
\be \label{eq:chargesS}
\textbf{Q}_k=\int_{\mathcal{B}_{p-1}}d^{p-1}\sigma\sqrt{|\gamma|}~t^{\lambda}_k~ l_\lambda+\int_{\wt{\mathcal{B}}_{p-2}}d^{p-2}\wt\sigma\sqrt{|h|}~\wt t^{\lambda}_k ~l_\lambda \ ,
\ee
where $l_\lambda$ is a unit normalised timelike co-vector normal to $\mathcal{B}_{p-1}$ and $\wt{\mathcal{B}}_{p-2}$.

\subsubsection{Currents from the spacetime stress tensor}
The spacetime stress tensor can be used in order to provide an alternative derivation of the currents \eqref{eq:cc1}.
Given the spacetime stress tensor $\textbf{T}^{\mu\nu}$, whose components include the surface \eqref{eq:stact2} and the edge contributions \eqref{eq:sts+e}, one may straightforwardly construct a set of conserved currents $\textbf{T}^{\mu}_{(k)}$ given by
\be \label{eq:kcur}
\textbf{T}^{\mu}_{(k)}=\textbf{T}^{\mu\nu}\tkd\nu \ ,
\ee
which obviously satisfy $\nabla_\mu\textbf{T}^{\mu}_{(k)}=0$ due to the conservation equation \eqref{eq:stress} and the Killing equation. An appropriate integration of the currents $\eqref{eq:kcur}$ over a fixed time-slice yields a set of conserved charges, one for each $\tku\mu$. However, in order to understand how the currents \eqref{eq:kcur} can be written in terms of the different components of the stress tensor, i.e. as in \eqref{eq:cc1}, one needs to use a "Gaussian pillbox" to get rid of the delta functions. By explicit evaluation of the integral
\be
\int_{\mathcal{M}}d^{D}x\sqrt{|g|}~\nabla_\mu\left(\textbf{T}^{\mu\nu}\tkd\nu\right) f
\ee
for some arbitrary function $f(x^{\alpha})$ of compact support and using the specific form of the stress tensor \eqref{eq:stgenform}, with components \eqref{eq:stact2} and \eqref{eq:sts+e}, as well as the constraints \eqref{eq:st}-\eqref{eq:pt} and \eqref{eq:st1}, one obtains the conservation equations \eqref{eq:cc} for the currents \eqref{eq:cc1}, as expected.

\subsection{Spacetime stress tensor for interfaces}

One may turn the action \eqref{eq:intact} into an action over spacetime using the Theta and Dirac delta functions, such that
\be \label{eq:intactspace}
S[\Phi_{\text{int}}(x),\Phi(\sigma),\Phi_\text{ext}(x)]=\int_{\mathcal{M}}d^{D}x\left(\mathcal{L}[\Phi_{\text{int}}(x)] \Theta(x)+\mathcal{L}[\Phi(\sigma)]\widehat\delta(x) +\mathcal{L}[\Phi_\text{ext}(x)]\Theta(-x)\right)\ ~,
\ee
where $\Theta(x)=\Pi_\alpha\Theta(x^{\alpha}-X^{\alpha}(\sigma))$ and similarly for $\Theta(-x)$ with the opposite sign.  Using this form, under a Lagrangian variation one finds, as in \eqref{eq:stress}, that the equations of motion are given by
\be \label{eq:genintereq}
\nabla_\mu \textbf{T}^{\mu\nu}=0 \ ,
\ee
where the full spacetime stress tensor takes the form
\be
\textbf{T}^{\mu\nu}=\textbf{T}^{\mu\nu}_{\text{b}}+\textbf{T}^{\mu\nu}_{\text{s}} \ ,
\ee
where $\textbf{T}^{\mu\nu}_{\text{b}}$ is the bulk contribution while $\textbf{T}^{\mu\nu}_{\text{s}}$ is the surface contribution. Taking into account the specific couplings \eqref{eq:varintact}, the bulk stress tensor is given by
\be \label{eq:stbulk}
\textbf{T}^{\mu\nu}_{\text{b}}=T^{\mu\nu}_{\text{b}}\Theta(x) \ ,~~T^{\mu\nu}_{\text{b}}=T^{\mu\nu}_{\text{int}}-2\nabla_\lambda\nabla_\rho L^{\lambda\rho(\mu\nu)} \ .
\ee
Using the form \eqref{eq:stbulk} into \eqref{eq:genintereq}, one obtains two sets of equations
\be \label{eq:eomsinter}
\left(\nabla_\mu T^{\mu\nu}_{\text{b}}\right)\Theta(x)=0 \ , \quad \nabla_\mu \textbf{T}^{\mu\nu}_{\text{s}}=T^{\mu\nu}_{\text{b}}n_\mu\widehat\delta(x) \ ,
\ee
where we have used that $\nabla_\mu \Theta(x)=-n_\mu\widehat\delta(x)$. The first equation in \eqref{eq:eomsinter} gives rise to \eqref{eq:eqbulk} while the second equation gives rise to the diffeomorphism constraints and \eqref{eq:interfull}. The surface spacetime stress tensor now takes the form
\be\label{eq.interfaceinflow}
\begin{split}
\Tuu\mu\nu_{\text{s}}=&\left[\mathcal{T}^{\mu\nu}+\mathcal{B}^{\mu\nu}-\left(\mathcal{V}_\lambda n^{\lambda}n^\mu n^\nu +2\mathcal{V}_\lambda\huu\lambda{(\mu}n^{\nu)}\right)+2n_\lambda\nabla_\rho L^{\lambda\rho(\mu\nu)}\right]\widehat\delta(x)\\
&-\nabla_\rho\left[\left(2\mathcal{D}^{\rho(\mu\nu)}-\mathcal{D}^{\mu\nu\rho}-2n_\lambda L^{\rho\lambda(\mu\nu)}\right)\widehat\delta(x)\right]-4\nabla_\lambda\nabla_\rho\left[\mathcal{Q}^{\lambda\rho(\mu\nu)}\widehat\delta(x)\right] \ .
\end{split}
\ee
It can be observed that the bulk curvature moment $L^{\mu\nu\lambda\rho}$ yields an inflow of energy-momentum into the interface. Considering geometric bulk actions with derivatives of the Riemann tensor will also induce an inflow of surface curvature moments, beyond the inflow of monopole and dipole moments. If one takes the bulk action to be the Einstein-Hilbert action $\sqrt{|g|}R$ then there will not be any inflow of monopole energy-momentum to the interface but there will be a contribution to the dipole surface moment. This contribution is of the form $T^{\mu\nu\rho}=-2\huu\mu\nu n^\rho+2n^{(\mu}\huu{\nu)}\rho$. The second term can be removed by a frame-choice, while the first contribution is twice that which arises from a surface action of the form $\int_{\mathcal{W}}\sqrt{|\gamma|}d^p\sigma K^{\rho}n_\rho$. The two dipole moments (the one arising from the Einstein-Hilbert action and the one arising from the mean extrinsic curvature), therefore, differ by frame-choices.


\subsubsection{Conserved currents and charges}
Analogously to the case of surfaces studied above, one may obtain a set of conserved currents associated with symmetries  of the background by performing a diffeomorphism along a Killing direction such that $\xi^{\mu}=\beta \textbf{k}^\mu$ for a constant $\beta$, leading to
\be
\begin{split}
\delta_{\textbf{k}}S[\Phi_{\text{int}}(x),\Phi(\sigma)]=&\beta\int_{\mathcal{B}_{\text{int}}}d^{D}x\sqrt{|g|}\left(\mathring B^{\mu}\textbf{k}_\mu+\nabla_\mu\left(T^{\mu\nu}_{\text{b}}\textbf{k}_\nu\right)\right)\\
&+\beta \int_{\mathcal{W}}d^{p}\sigma\left[B^{\mu\nu}\pdu\mu\rho\nabla_\rho\tkd\nu+B^{\mu}\tkd\mu+\overline\nabla_\lambda\left(\mathbb{T}^{\lambda\sigma}\tkd\sigma+\Sigma^{\lambda\sigma\alpha}\nabla_\alpha\tkd\sigma\right)\right] \ .
\end{split}
\ee
When the bulk equations of motion are satisfied $\mathring B^{\mu}=0$ and hence for the bulk action to be invariant under the symmetry associated with $\tku\mu$ we must have
\be
\nabla_\mu \mathring t_k^\mu=0 \ ,\quad \text{for} \quad \mathring t_k^\mu=T^{\mu\nu}_{\text{b}}\tkd\nu \ ,
\ee
which indeed follows from the symmetry of the bulk stress tensor and the Killing equation. On the interface, in turn, when the equations of motion are satisfied we have $B^{\mu\nu}\pdu\mu\rho=0$ and
\be
B^{\mu}\tkd\mu=\overline\nabla_\lambda\left(\hud\lambda\mu\left(\mathring T^{\mu\nu}-\overline\nabla_\rho\left(\hud\rho\alpha\mathring T^{\alpha\nu\mu}\right)\right)\right)\tkd\nu -\left(\mathring T^{\mu\alpha\rho}+\frac{1}{2}\pud\rho\nu\mathring T^{\nu\alpha\mu}\right)\Buddd\sigma\alpha\mu\rho \tkd\sigma-n_\lambda T^{\lambda\sigma}_{\text{b}}\tkd\sigma \ .
\ee
Therefore, for the interface action to be invariant under this symmetry, we identify the interface conservation equation and current
\be \label{eq:curinter}
\overline\nabla_\mu t^{\mu}_k=n_\mu\mathring t^{\mu}_k \ , \quad \text{for} \quad t^{\mu}_k=\mathring{\mathbb{T}}^{\mu\nu}\tkd\nu+\mathring{\Sigma}^{\mu\sigma\alpha}\nabla_\alpha \tkd\sigma \ ,
\ee
where, using the definitions of $\mathbb{T}^{\mu\nu}$ and ${\Sigma}^{\mu\sigma\alpha}$ in \eqref{eq:truestress}--\eqref{eq:trueSigmas}, we have introduced
\be
\mathring{\mathbb{T}}^{\mu\nu}={\mathbb{T}}^{\mu\nu}+\hud\mu\lambda\left(\mathring T^{\lambda\nu}-\overline\nabla_\rho\left(\hud\rho\alpha\mathring T^{\alpha\nu\lambda}\right)\right) \ ,\quad \mathring{\Sigma}^{\mu\sigma\alpha}={\Sigma}^{\mu\sigma\alpha}+\hud\mu\nu\mathring T^{\nu\sigma\alpha} \ .
\ee
One can check that the current conservation equation \eqref{eq:curinter} is satisfied given the equations of motion \eqref{eq:interfull}. The corresponding conserved charges can be constructed as in \eqref{eq:chargesS} by appropriately integrating the currents
\be
\textbf{Q}_k=\int_{\text{B}_{\text{int}}}d^{D-1}x\sqrt{|g|}~\mathring t^{\lambda}_k~ l_\lambda+\int_{{\mathcal{B}}_{p-1}}d^{p-1}\sigma\sqrt{|\gamma|}~t^{\lambda}_k ~l_\lambda \ ,
\ee
where we assumed that $\mathcal{B}_{\text{int}}=\mathbb{R}\times\text{B}_{\text{int}}$ and $\mathcal{B}_{p}=\mathbb{R}\times\mathcal{B}_{p-1}$. Here, $l_\lambda$ is a unit normalised timelike co-vector normal to $\text{B}_{\text{int}}$ and ${\mathcal{B}}_{p-1}$.


\section{Two-derivative surface action with edges} \label{sec:2derivative}

In this section we employ classification methods analogous to those used in the context of hydrodynamics to constrain the main results of Sec.~\ref{sec:surfaces} in a derivative expansion, up to second order in derivatives.\footnote{We note that we are not assuming this derivative expansion to be a perturbative expansion.} To this end we  first construct the most general surface geometric bulk action of $\Gdd\mu\nu$ with at most two derivatives, including a quite general (but not exhaustive) edge contribution.
Spacetime covariance implies that the diffeomorphism constraints \eqref{eq:st}-\eqref{eq:bt}, as well as the tangential projection \eqref{eq:genshapetang}, must be automatically satisfied off-shell for all shape configurations, as explained at the end of Sec.~\ref{sec:constraints}. This implies that whenever a contribution is added to the action, it must be checked whether or not such contribution satisfies the tangential diffeomorphism constraints.

This construction has applications in many physical systems whose description includes embedded surface in a given spacetime. These include fluid membranes, entangling surfaces, spinning brane systems, membrane elasticity or effective theories of black holes.

The restriction to a two-derivative action implies that the corresponding stress tensor  should include at most two derivative terms, and hence must be of the form \eqref{eq:sts+e}. To count the number of derivatives we use a bookkeeping parameter $\varepsilon$. The components of the stress tensor are at most of order $\mathcal{O}(\varepsilon^2)$, and in particular, $T^{\mu\nu}\sim\mathcal{O}(\varepsilon^2)$, $T^{\mu\nu\rho}\sim\mathcal{O}(\varepsilon)$ and $T^{\mu\nu\lambda\rho}\sim\mathcal{O}(1)$. In turn, due to \eqref{eq:stact2}, this implies that
\be
\mathcal{T}^{\mu\nu}\sim\mathcal{P}^{\mu\nu}\sim\mathcal{B}^{\mu\nu}\sim\mathcal{O}(\varepsilon^2) \ , \quad \mathcal{D}^{\mu\nu\lambda}\sim\mathcal{S}^{\mu\nu\lambda}\sim\mathcal{O}(\varepsilon) \ ,  \quad \mathcal{Q}^{\mu\nu\lambda\rho}\sim\mathcal{O}(1) \ ,
\ee
and similarly for the edge components \eqref{eq:stactedge}. However, since $\mathcal{P}^{\mu\nu}$ and $\mathcal{B}^{\mu\nu}$ are related to the dipole and quadrupole terms via Eqns.~\eqref{eq:bt} and \eqref{eq:pt}, it is not necessary to be concerned with them here, as they will be determined by the remaining couplings. 

It is convenient to split the action into the sum of parity-even $S_+$, parity-odd $S_-$ and edge $S_e$ contributions according to
\be\label{eq:secderaction}
S[\Phi(\sigma),\Phi_{e}(\wt \sigma)]=S_+[\Phi(\sigma),\Phi_{e}(\wt \sigma)]+S_-[\Phi(\sigma),\Phi_{e}(\wt \sigma)] +S_e[\Phi_e(\wt\sigma)]\ ,
\ee
and first consider the parity-even sector, which does not depend on the dimension $p$ nor the codimension $n$ of the surface.


\subsection{Parity-even sector}
The procedure for identifying all the covariant two-derivate scalars consists on classifying the different independent contributions to the surface stress tensor $\mathcal{T}^{\mu\nu}$, to the dipole terms $\mathcal{D}^{\mu\nu\lambda}$ and $\mathcal{S}^{\mu\nu\lambda}$ as well as to the curvature quadrupole moment $\mathcal{Q}^{\mu\nu\lambda\rho}$, and similarly for the edge components.

\paragraph{Surface stress tensor.} All terms in an action which involve contractions with the induced metric will contribute to $\mathcal{T}^{\mu\nu}$. Here we are interested in the contributions which do not appear due to contractions with the other geometric tensors $\Kddu\mu\nu\rho$, $\oduu\mu\nu\rho$, $\Rdddd\mu\nu\lambda\rho$ and $\Bdddd\mu\nu\lambda\rho$, i.e., contributions which are purely intrinsic. This implies that such contributions can only be composed of combinations of the induced metric $\huu\mu\nu$, the internal rotation tensor and its derivatives.  

At order $\mathcal{O}(1)$, the only possible contribution to $\mathcal{T}^{\mu\nu}$ arises from the surface tension term 
\be
\alpha \int_{\mathcal{W}}d^p\sigma \, \sqrt{|\gamma|} \ ,
\ee
for some constant $\alpha$. 

At first order in derivatives, as in General Relativity, there is no covariant scalar that can be constructed from the induced metric $\hdd\mu\nu$ and internal rotation tensor ${{\rho_\mu}^{\nu}}{_\rho}$, which should be understood as the intrinsic Christoffel connection. At second order, the only scalar is the induced Ricci scalar but that is included in the coupling to the intrinsic Riemann tensor.

\paragraph{Bending moment.} Turning our attention to $\mathcal{D}^{\mu\nu\rho}$, it is noticeable that the symmetries and index structure implies that at one derivative level one must have
\be \label{eq:bendingmoment}
\mathcal{D}^{\mu\nu\rho}=\mathcal{Y}^{\mu\nu\lambda\sigma}\Kddu\lambda\sigma\rho \ ,
\ee
where $\mathcal{Y}^{\mu\nu\lambda\sigma}$ is the Young modulus of the surface, first introduced in the context of perturbations of black branes \cite{Armas:2011uf}, and has the symmetry $\mathcal{Y}^{(\mu\nu)(\lambda\sigma)}$, as a classical elasticity tensor. The above expression for the bending moment \eqref{eq:bendingmoment} takes the from of \emph{stress times strain}, where the extrinsic curvature $\Kddu\lambda\sigma\rho$ can be interpreted as strain.\footnote{In fact, when working with a foliation of surfaces, one has that $2\nud i\rho \Kddu\mu\nu\rho=-\hud\lambda\mu\hud\sigma\nu\mathcal{L}_{n^{i}}\hdd\lambda\sigma$, which makes it clear that the extrinsic curvature is a measure of the change in distances on the surface along transverse directions \cite{Armas:2011uf}.} 

From the bending moment \eqref{eq:bendingmoment}, one concludes that $\mathcal{Y}^{(\mu\nu)(\lambda\sigma)}\sim\mathcal{O}(1)$. Therefore, given the symmetries of $\mathcal{Y}^{\mu\nu\lambda\sigma}$, the most general form of the Young modulus is
\be
\mathcal{Y}^{\mu\nu\lambda\sigma}=2\lambda_1 \huu\mu\nu\huu\lambda\sigma+2\lambda_2\huu\mu{(\lambda}\huu{\sigma)}\nu \ ,
\ee
for some constants $\lambda_1$, $\lambda_2$, which turns out to have the symmetry $\mathcal{Y}^{\mu\nu\lambda\sigma}=\mathcal{Y}^{\lambda\sigma\mu\nu}$, as a classical elasticity tensor. It should also be noted that the form \eqref{eq:bendingmoment} immediately implies that the term $\mathcal{D}^{\mu\nu[\sigma}\Kddu\mu\nu{\alpha]}$ in \eqref{eq:st} vanishes.

\paragraph{Spin current.} Due to the symmetries of $\mathcal{S}^{\mu\nu\rho}$ and its index structure, one observes that the spin current must be of the general form
\be
\mathcal{S}^{\mu\nu\rho}=\mathcal{S}^{\mu\lambda}\oduu\lambda\nu\rho \ ,
\ee
for some spin tensor $\mathcal{S}^{\mu\lambda}$ of $\mathcal{O}(1)$. The only possible component of $\mathcal{S}^{\mu\lambda}$ would be, therefore
\be \label{eq:spineven}
\mathcal{S}^{\mu\lambda}=\vartheta_0\huu\mu\lambda \ ,
\ee
but the constraints fix $\vartheta_0=0$. To see this notice the first term in \eqref{eq:st} vanishes, as does also  the third term in that equation, as we will see below. The spin current conservation equation becomes $\pdu\lambda\sigma\pdu\rho\alpha\overline\nabla_\mu \mathcal{S}^{\mu\lambda\rho}=0$. However, it is easy to see that a term of the form given in Eqn.~\eqref{eq:spineven} does not satisfy this condition for arbitrary $\oduu\lambda\nu\rho$ and therefore must be discarded. This is an example of how diffeomorphism invariance along the surface leads to non-trivial constraints on the action.

\paragraph{Surface curvature moment.} The intrinsic Riemann tensor is of $\mathcal{O}(\varepsilon^2)$, which in turn implies that ${\mathcal{I}}^{\mu\nu\lambda\rho}\sim\mathcal{O}(1)$. The only possible choice is therefore 
\be \label{eq:classi}
{\mathcal{I}}^{\mu\nu\lambda\rho}=\alpha_1 \gamma^{\mu[\nu}\gamma^{\rho]\lambda} \ ,
\ee
where $\alpha_1$ is a constant. The contraction ${\mathcal{I}}^{\mu\nu\lambda\rho}\Rdddd\mu\lambda\nu\rho$ is proportional to the induced Ricci scalar $\mathcal{R}$. Due to the Gauss-Codazzi equation \eqref{eq:GC}, this term can be replaced by a linear combination of terms to the square of the extrinsic curvature and a term proportional to the background Riemann tensor. These two possibilities are equivalent, as it may be checked using the equations of motion and spacetime stress tensor obtained in App.~\ref{app:highdev}. For practical reasons we keep this term explicitly. 

\paragraph{Background curvature moment.} Similarly to the surface curvature moment, since the background Riemann tensor is of $\mathcal{O}(\varepsilon^2)$,  the curvature moment ${\mathcal{Q}_\mu}^{\nu\lambda\rho}$ must be of order $\mathcal{O}(1)$. Hence, its most general form, given its symmetries, is
\be \label{eq:classq}
{\mathcal{Q}_\mu}^{\nu\lambda\rho}=\alpha_2 \hud{[\nu}\mu\huu{\rho]}\lambda+\alpha_3\pud{[\nu}\mu\huu{\rho]}\lambda+\alpha_4\pud{[\nu}\mu\puu{\rho]\lambda} \ .
\ee
Since  ${\pdu\nu{[\sigma}\pdu\lambda{\alpha]}\Pi^{\nu\lambda}=0}$, this form of the background curvature moment satisfies the constraint in \eqref{eq:st}  for arbitrary constants $\alpha_2$, $\alpha_3$ and $\alpha_4$. Note also that the first term in \eqref{eq:classq} can be replaced by the term \eqref{eq:classi} when using the Gauss-Codazzi equation \eqref{eq:GC}. For presentation purposes, given \eqref{eq:classq} it is useful to define the three contractions
\be \label{eq:defR}
R_{||}=\hud{\lambda}\mu\huu{\rho}\nu\Buddd\mu\nu\lambda\rho \ , \quad R_{\angle}=\pud{\lambda}\mu\huu{\rho}\nu\Buddd\mu\nu\lambda\rho \ , \quad R_\perp=\pud{\lambda}\mu\puu{\rho}\nu\Buddd\mu\nu\lambda\rho \ .
\ee

\subsubsection{Parity-even action at second order in derivatives}
Taking these considerations into account, and the fact that there is no parity-even scalar built out of the outer curvature moment, we can now write down the most general parity-even two-derivative geometric action. This is given by
\be \label{eq:seven}
\begin{split}
S_+[\Phi(\sigma)]=\int_{\mathcal{W}}d^{p}\sigma\sqrt{|\gamma|}\Big(\alpha&+\lambda_1K^{\rho}K_{\rho}+\lambda_2\Kddu\mu\nu\rho \Kuud\mu\nu\rho+\alpha_1 \mathcal{R}+\alpha_2 R_{||}+\alpha_3 R_{\angle}+\alpha_4 R_{\perp}\Big) \ ,
\end{split}
\ee
and, as mentioned above, the term proportional to $\alpha_1$ is redundant. All the terms involved in this action were implicitly classified  earlier, in particular in the literature of conformal anomalies of two-dimensional submanifolds (see e.g. \cite{Schwimmer:2008yh}). The shape equation that arises from this action was considered recently in \cite{Fonda:2016ine} and agrees with the general form of \eqref{eq:genshapeperp}.\footnote{The authors of \cite{Fonda:2016ine} did not use the terms $\alpha_2 R_{||}+\alpha_3 R_{\angle}$ in the action but instead two other linear combinations, namely $\puu\mu\nu R_{\mu\nu}$ and $R$, which can be rewritten in terms of $\alpha_2 R_{||}+\alpha_3 R_{\angle}+\alpha_4 R_{\perp}$.} All the terms in \eqref{eq:seven} may appear in entanglement entropy functionals \cite{Camps:2013zua, Dong:2013qoa}. This concludes the parity-even sector of the surface action and hence we now turn to the parity-odd sector.


\subsection{Parity-odd sector}
In the parity-odd sector, the contributions to the action are either dependent on the dimension of the surface or on its codimension.

\paragraph{Codimension $n=1$.} In this case there is only one normal vector, $n^{\rho}$, to the surface. Since there is only one normal direction, the extrinsic twist potential vanishes and there are no couplings to the spin current. However, the bending moment may have a $\mathcal{O}(1)$ term. That is, besides \eqref{eq:bendingmoment} one can have a contribution of the form
\be \label{eq:bend0}
\mathcal{D}^{\mu\nu\rho}=\lambda_0\huu\mu\nu n^\rho\delta_{n,1} \ ,
\ee
which, in the context of General Relativity, leads to the known Gibbons-Hawking boundary term $n_\rho K^{\rho}$. This term is parity-odd, in the sense that, in the absence of any bulk (contrary to the case of General Relativity for which there is a preferred orientation of the normal vector) it is not invariant under reflection $n_\rho\to-n_\rho$.

\paragraph{Dimension $p=1$ and codimension $n=2$.} In the case of $p=1$, there is only one tangent vector $\partial_1 X^{\mu}=u^{\mu}\sqrt{|\gamma_{11}|}$, which is interpreted as the unnormalised velocity of the point-particle in the relativistic case. Also, when $n=2$, one may make use of the Levi-Civita tensor in the transverse space $\epsilon^{\mu\nu}_\perp$ in order to obtain new contributions. In this case the spin current can have an $\mathcal{O}(1)$ and an $\mathcal{O}(\varepsilon)$ contribution, such that
\be \label{eq:spin0}
\mathcal{S}^{\mu\nu\rho}=\vartheta_1u^{\mu}\epsilon^{\nu\rho}_\perp\delta_{p,1}\delta_{n,2}+\vartheta_2u^{\lambda}\omega_\lambda u^{\mu}\epsilon^{\nu\rho}_\perp\delta_{p,1}\delta_{n,2}  \ ,
\ee
where we have made used of the definition of the normal fundamental 1-form. 

The first of these terms is the usual coupling due to the particle's intrinsic spin and the second contribution is its square - a sub-leading spin-orbit effect. However, using the identity ${\epsilon^{\mu\nu\rho}u_\mu=\epsilon^{\nu\rho}_\perp}$, it is easily verifiable that only the term with coefficient $\vartheta_1$ satisfies $\pdu\lambda\sigma\pdu\rho\alpha\overline\nabla_\mu \mathcal{S}^{\mu\lambda\rho}=0$. The second term can satisfy the conservation equation if we assume that the velocity $u^{\mu}$ is aligned with a surface Killing vector field. While this possibility is interesting, as it allows to describe embedded fluids \cite{Armas:2013hsa, Armas:2013goa} or fluids with surfaces \cite{Armas:2015ssd, Armas:2016xxg}, assuming the existence of surface Killing vectors is beyond the scope of this paper.\footnote{If this assumption would be taken seriously, it would lead to many extra couplings besides the ones considered here. See \cite{Armas:2013hsa, Armas:2013goa} for examples.} Therefore we set $\vartheta_2=0$. 

Finally, we note that the equations of motion for a point-particle do not change if the codimension is increased. However, in order to write down such couplings, one must specialise to backgrounds with further rotation symmetries for which the rotation group is Abelian in a given transverse two-plane \cite{Armas:2014rva}. Here, we are performing an analysis that holds for any background spacetime, irrespectively of its symmetries. 

\paragraph{Dimension $p=2$ and codimension $n=2$.} In this case, the spacetime Levi-Civita tensor $\epsilon^{\mu\nu\lambda\rho}$ can be used to build new scalars by contracting it with Eq.~\eqref{eq:RV}. In particular, there are three apparent scalars
\be \label{eq:3terms}
2{\epsilon^{\lambda\mu}}_{\rho\alpha} \Kduu\lambda\nu\rho\Kddu\mu\nu\alpha \ , \quad \epsilon^{\alpha\sigma\kappa\tau}\hdd\mu\alpha\hud\nu\sigma\pud\lambda\kappa\pud\rho\tau\Buddd\mu\nu\lambda\rho \ , \quad \Omega \ .
\ee
It can be verified that the first two terms satisfy the constraints \eqref{eq:st}-\eqref{eq:pt} individually, however, for arbitrary extrinsic curvature and background Riemann tensor they do not satisfy individually the tangentiality condition \eqref{eq:genshapetang}, only a linear combination does. Hence, only this  combination provides a new contribution, as it will be proven in the next section.\footnote{This implies that we fully disagree with the counting of two independent parity-odd coefficients done in \cite{Cvitan:2015ana} in the context of conformal anomalies for two-dimensional submanifolds, as only one coefficient is allowed by diffeomorphism invariance along the surface.} This is a non-trivial consequence of the diffeomorphism constraints and shows that the naive expectation that any appropriate contraction of surface tensors yields a covariant contribution, is not correct.\footnote{Since ${\epsilon^{\lambda\mu}}_{\rho\alpha}$ is a pseudo-tensor, one might have expected that none of the terms in \eqref{eq:3terms} would be covariant. However, the outer curvature scalar $\Omega$ is.} 

Given \eqref{eq:3terms}, the bending moment and the background curvature moment will receive new contributions. In particular, the bending moment \eqref{eq:bendingmoment} admits the following generalisation
\be \label{eq:bendingmoment1}
\mathcal{D}^{\mu\nu\rho}={\mathcal{Y}^{\mu\nu\lambda\sigma\rho}}_\alpha \Kddu\lambda\sigma\alpha \ ,
\ee
where now the generalised Young modulus has the most general form
\be
{\mathcal{Y}^{\mu\nu\lambda\sigma\rho}}_\alpha=2\left(\lambda_1 \huu\mu\nu\huu\lambda\sigma+\lambda_2\huu\mu{(\lambda}\huu{\sigma)}\nu\right)\pud\rho \alpha +4\lambda_3\hud{(\lambda}\tau\huu{\sigma)}{(\nu}{\epsilon^{|\tau|\mu)\rho}}_\beta\pud\beta\alpha\delta_{p,2}\delta_{n,2} \ .
\ee
The coefficient $\lambda_3$ is a new purely geometric parity-odd contribution to the Young modulus of thin elastic surfaces.

This generalised version still has the symmetry ${\mathcal{Y}^{\mu\nu\lambda\sigma\rho}}_\alpha={\mathcal{Y}^{(\mu\nu)(\lambda\sigma)\rho}}_\alpha$ but no longer the symmetry ${\mathcal{Y}^{\mu\nu\lambda\sigma\rho}}_\alpha={\mathcal{Y}^{\lambda\sigma\mu\nu\rho}}_\alpha$ due to the parity-odd effects. It should not be surprising that the Young modulus must be generalised compared to its classical counterpart. What is surprising is that the usual classical definition \eqref{eq:bendingmoment}, first introduced in the context of codimension $n=1$ surfaces, is sufficient for capturing all parity-even effects for arbitrary codimension. In turn, the background curvature will now have the form
\be \label{eq:q0}
{\mathcal{Q}_\mu}^{\nu\lambda\rho}=\alpha_2 \hud{[\nu}\mu\huu{\rho]}\lambda+\alpha_3\pud{[\nu}\mu\huu{\rho]}\lambda+\alpha_4\pud{[\nu}\mu\puu{\rho]}\lambda+\lambda_3\epsilon^{\alpha\sigma\kappa\tau}\hdd\mu\alpha\hud\lambda\sigma\pud\nu\kappa\pud\rho\tau\delta_{p,2}\delta_{n,2} \ .
\ee
The linear combination of the first two terms in \eqref{eq:3terms} proportional to $\lambda_3$ is equivalent to the last term in $\Omega$ due to the Ricci-Voss equation \eqref{eq:RV}. However, similarly to the discussion above for the coefficient $\alpha_1$, we keep it explicitly for practical purposes. Therefore, the outer curvature moment can have one parity-odd contribution of the form
\be
{\mathcal{H}_\mu}^{\nu\lambda\rho}=\alpha_5\epsilon^{\beta_1\beta_2\beta_3\beta_4}\pdd{\beta_1}\mu\hud\nu{\beta_2}\pud\lambda{\beta_3}\hud\rho{\beta_4}\delta_{p,2}\delta_{n,2} \ ,
\ee
which leads to a contribution proportional to $\Omega$. As explained in Sec.~\ref{sec:geometry}, this term is topological. Therefore, it will not affect the surface dynamics, though it will certainly contribute to the on-shell value of the action as well as to the edge dynamics.

\subsubsection{Parity-odd action at second order in derivatives}

Given these considerations, we can write down the parity-odd sector of the two-derivative surface action
\be \label{eq:sodd}
\begin{split}
S_-[\Phi(\sigma)]=\int_{\mathcal{W}}d^{p}\sigma\sqrt{|\gamma|}&\Big(\lambda_0 n_\rho K^{\rho}\delta_{n,1}+2\vartheta_1u^{\mu}\omega_\mu\delta_{p,1}\delta_{n,2}\\
&+\left(\lambda_3 \left(2{\epsilon^{\lambda\mu}}_{\rho\alpha} \Kduu\lambda\nu\rho\Kddu\mu\nu\alpha+ R_-\right) +\alpha_5\Omega\right)\delta_{p,2}\delta_{n,2} \Big) \ ,
\end{split}
\ee
where we have defined $R_-=\epsilon^{\alpha\sigma\kappa\tau}\hdd\mu\alpha\hud\nu\sigma\pud\lambda\kappa\pud\rho\tau\Buddd\mu\nu\lambda\rho$. As mentioned earlier the contributions involving $\lambda_3$ and $\alpha_5$ are equivalent. All the terms present in \eqref{eq:sodd} had been enumerated in \cite{Armas:2013hsa} but the last two terms, in particular, had not been properly analysed. The second term and the last term also have a role to play in the context of entanglement entropy \cite{Castro:2014tta, Ali:2016cng}.


\subsection{Edge action}

We now consider an example of a quite thorough, though not exhaustive, edge action up to two derivatives. As explained in Sec.~\ref{sec:geometry}, there are other possible couplings between the edge and the surface geometry, which were not considered in \eqref{sec:edgegeomtr}. This  supervenes on the existence of the unit normal vector to the surface boundary $\wt n^\rho$, which can be used to introduce new contributions in the action. Nevertheless, without a full characterisation of all couplings, the possible couplings that appear are sufficient to exhibit the richness of the edge dynamics. We treat the parity-even and the parity-odd sectors of the edge action simultaneously.

\paragraph{Edge stress tensor.} As in the case of the surface action, there is only one purely intrinsic independent contribution to the surface stress tensor. This is the edge tension term 
\be
\chi\int_{\partial\mathcal{W}}d^{p-1}\, \wt\sigma\sqrt{|h|} \ .
\ee

\paragraph{Edge bending moment.} The bending moment exhibits quite a rich structure due to the existence of the normal vector $\wt n^\rho$ which represents a preferred direction. It is, in essence, a mixture of arbitrary codimension and codimension $n=1$ contributions. The edge bending moment can be written in the same form as in \eqref{eq:bendingmoment1} but with the additional contribution \eqref{eq:bend0}, that is\footnote{Note that this term proportional to $\wt\lambda_0$, contrary to $\lambda_0$, is not parity-odd because the presence of the surface and the natural outward-pointing direction of $\wt n^\rho$ breaks the reflection symmetry $\wt n^\rho\to-\wt n^\rho$.}
\be \label{eq:bendingmoment3}
\wt{\mathcal{D}}^{\mu\nu\rho}={\wt{\mathcal{Y}}^{\mu\nu\lambda\sigma\rho}}_{~~~~~ \quad \alpha} \Ksddu\lambda\sigma\alpha+\wt\lambda_0 \hsuu\mu\nu\wt n^{\rho} \ ,
\ee
where the generalised Young modulus takes the form
\be \label{eq:bendingmoment4}
\begin{split}
{\wt{\mathcal{Y}}^{\mu\nu\lambda\sigma\rho}}_{~~~~~ \quad \alpha}=&~2\left(\wt\lambda_1 \hsuu\mu\nu\hsuu\lambda\sigma+\wt\lambda_2\hsuu\mu{(\lambda}\hsuu{\sigma)}\nu\right)\pud\rho\alpha+4\wt\lambda_3\hsuu\nu{(\sigma}\epsilon_{||}^{\lambda)\mu}{\epsilon_{\perp}^{\rho}}_\alpha\delta_{p,3}\delta_{ n,1}\\
&+2\left(\wt\lambda_4 \hsuu\mu\nu\hsuu\lambda\sigma+\wt\lambda_5\hsuu\mu{(\lambda}\hsuu{\sigma)}\nu\right)\wt n^\rho\wt n_\alpha \ .
\end{split}
\ee
The existence of the preferred direction $\wt n^\rho$ appears to introduce further possible contributions to the Young modulus. However, it is easily seen that the first three terms above satisfy the condition $\wt{\mathcal{D}}^{\mu\lambda[\sigma}\Ksddu\mu\lambda{\alpha]}=0$ but last two do not, except in the case for which the boundary is codimension $n=1$ but we have excluded such possibility by only considering non-space-filling surfaces. Hence, we must set $\wt\lambda_4=\wt\lambda_5=0$. Furthermore, the term $\wt\lambda_0$ in \eqref{eq:bendingmoment3} does not satisfy the condition $\wt{\mathcal{D}}^{\mu\lambda[\sigma}\Ksddu\mu\lambda{\alpha]}=0$. However, explicit evaluation of \eqref{eq:st} leads to
\be \label{eq:edgebending}
\begin{split}
&\wt{\mathcal{D}}^{\mu\lambda[\sigma}\Ksddu\mu\lambda{\alpha]}-\wt n_\lambda\wt n_\mu \mathcal{D}^{\lambda\mu[\sigma}\wt{n}^{\alpha]}=0 \quad \\
&\Rightarrow~\left(\wt\lambda_0+2\lambda_1\right)\mathcal{K}^{[\alpha}\wt n^{\sigma]}-2\left(\lambda_1+\lambda_2\right)\wt n_\lambda \wt n_\rho \Kuuu\lambda\mu{[\sigma}\wt n^{\alpha]}=0 \ .
\end{split}
\ee
For this condition to be satisfied for arbitrary surface and edge extrinsic curvature one must set $\wt\lambda_0=-2\lambda_1$ and $\lambda_2=-\lambda_1$.\footnote{\label{foot:edge}Note that in the absence of non-trivial edges, the edge conditions \eqref{eq:b1} implied constraints on the extrinsic curvature and spin current at the edges, in particular, certain components must vanish. Here we are not requiring this though one could insist on such edge conditions. Instead, we allow for the surface extrinsic curvatures to be arbitrary at the edges and, by continuity, do not impose any restrictions on the edge components of the extrinsic curvature.} The presence of the surface bending moment introduces extra terms that violate the edge spin conservation equation \eqref{eq:st1}. Requiring consistency on the edges implies a constraint among the two surface response coefficients besides fixing $\wt \lambda_0$ in terms of a surface coefficient. This consistency condition implies that the surface cannot have arbitrary elastic response coefficients, in fact, the condition $\lambda_2=-\lambda_1$ implies that the linear combination involving extrinsic curvatures in \eqref{eq:seven} must be proportional to the linear combination $(\mathcal{R}-R_{||})$ due to the Gauss-Codazzi equation \eqref{eq:GC}. Furthermore, the presence of a term proportional to $\mathcal{R}$ requires a non-vanishing edge coefficient $\wt \lambda_0$ such that $\wt\lambda_0=-2\lambda_1$.\footnote{\label{foot:weyl}This can also be derived using the edge constraints involving $\mathcal{I}^{\mu\nu\lambda\rho}$ in App.~\ref{app:highdev}. We note that if $p=2$, then $\mathcal{R}$ is topological and the necessary term $\wt n_\rho\mathcal{K}^\rho=\wt{n}_\mu\epsilon^{\mu\nu}_{||}\rho_\mu/2$ is equal to its boundary contribution (see \eqref{eq:bricci}). This means that a term proportional to $\mathcal{R}$ for $p=2$ is not allowed by diffeomorphism invariance if the surface has edges.} This is the equivalent of the Gibbons-Hawking boundary term (with the appropriate coefficient) in General Relativity. In fact, if we keep the term proportional to $\alpha_1$ explicitly in \eqref{eq:seven} then one obtains that, using App.~\ref{app:highdev}, $\wt\lambda_0=-2\lambda_1-2\alpha_1$.

\paragraph{Edge spin current.} At the edge there is a similar contribution to the spin current, as for the surface, and an additional contribution, such that
\be \label{eq:spin1}
\wt{\mathcal{S}}^{\mu\nu\rho}=\wt\vartheta_1 \wt u^{\mu}\wt\epsilon_\perp^{\nu\rho}\delta_{p,2}\delta_{n,1}+\wt\vartheta_2\epsilon^{\alpha\lambda\nu\rho}\wt u_\alpha \wt n_\lambda \wt u^{\mu}\delta_{p,2}\delta_{n,2} \ .
\ee
The first term satisfies \eqref{eq:st1} individually but the second term does not. However, explicit evaluation of \eqref{eq:st} leads to
\be
\begin{split}
&\psud\sigma\lambda\psud\alpha\rho\wt \nabla_\mu \wt{\mathcal{S}}^{\mu\lambda\rho}-\wt n_\lambda\wt n_\mu \mathcal{D}^{\lambda\mu[\sigma}\wt{n}^{\alpha]}=0 \quad \\
&\Rightarrow~2(2\lambda_3+\wt\vartheta_2)\wt n_\mu\wt n^\nu {\epsilon^{\lambda\mu\beta}}_\rho\Kddu\nu\lambda\rho \pud{[\sigma}\beta \wt n^{\alpha]}=0 \ ,
\end{split}
\ee
which is satisfied if we set $\wt\vartheta_2=-2\lambda_3$.\footnote{Again, we are insisting on arbitrary surface spin current components on the edges. See footnote \ref{foot:edge}.} This is expected since the term proportional to $\Omega$ in \eqref{eq:sodd} is a total derivative $\overline\nabla_\mu(\epsilon^{\mu\nu}_{||}\omega_\nu)$ which leads to the edge contribution $\wt n_\mu\epsilon^{\mu\nu}_{||}\omega_\nu$. This contribution is proportional to the contribution induced by the term $\wt\vartheta_2$. In fact, if we keep the term $\alpha_5$ explicitly in \eqref{eq:sodd} we find, using App.~\ref{app:highdev}, that $\wt\vartheta_2=-2\lambda_3-2\alpha_5$. In essence, this means that the topological term $\Omega$ cannot be added to an action for a surface with non-trivial edges. This is yet another instance where it does not suffice to enumerate terms that can contribute to the action, as it is also necessary to check the constraints imposed by tangential diffeomorphism invariance. Furthermore, note that the term proportional to $\vartheta_1$ in \eqref{eq:sodd} satisfies the constraint $\wt n_\lambda \psud\alpha\mu\psud\sigma\nu\mathcal{S}^{\lambda\nu\mu}=0$ automatically. 

\paragraph{Edge background curvature moment.} The background curvature bending moment can have the following contributions
\be \label{eq:q1}
\begin{split}
{\wt{\mathcal{Q}}_\mu}^{~\nu\lambda\rho}=&~\wt\alpha_2 \hsud{[\nu}\mu\hsuu{\rho]}\lambda+\wt \alpha_3\psud{[\nu}\mu\hsuu{\rho]}\lambda+\wt\alpha_4\psud{[\nu}\mu\psuu{\rho]}\lambda+\wt\lambda_3\epsilon^{\alpha\sigma\kappa\tau}\hsdd\mu\alpha\hsud\lambda\sigma\psud\nu\kappa\psud\rho\tau\delta_{p,3}\delta_{n,1} \quad \\
&+\wt\alpha_5 \hsud{[\nu}\mu \wt n^{\rho]}\wt n^{\lambda}+\wt \alpha_6\psud{[\nu}\mu \wt n^{\rho]}\wt n^{\lambda} \ .
\end{split}
\ee
which are analogous to those in \eqref{eq:q0}, except for the last two contributions. In particular, the parity-odd term has the same coefficient as the last parity-odd term in \eqref{eq:bendingmoment4}, as in the surface case since such is necessary for that linear combination to satisfy the tangential projection of \eqref{eq:genshapeedge}. The last two terms do not satisfy \eqref{eq:st1} and hence we must set $\wt\alpha_5=\wt\alpha_6=0$.

\subsubsection{Edge action up to second order in derivatives}
One may also consider the analogous contributions to the surface and outer curvature moments as in the surface case. However, since they can always be exchanged by linear combinations of the remaining terms, we have not consider them here. Given the above considerations, the total edge action reads
\be \label{eq:sevenedge}
\begin{split}
S_e[\Phi_e(\wt\sigma)]=\int_{\partial\mathcal{W}}d^{p-1}\wt \sigma\sqrt{|h|}\Big(\chi&-2(\lambda_1+\alpha_1) \wt n_\rho \mathcal{K}^{\rho}+\wt\lambda_1\mathcal{K}^{\rho}\mathcal{K}_{\rho}+\wt \lambda_2\Ksddu\mu\nu\rho \Ksuud\mu\nu\rho+\wt\alpha_2 \wt{R}_{||}+\wt\alpha_3 \wt{R}_{\angle}\\
&+\wt\alpha_4 \wt{R}_{\perp}+2\wt\vartheta_1\wt u^{\mu}\varpi_\mu\delta_{p,2}\delta_{n,1}-2(\lambda_3+\alpha_5){\epsilon^{\alpha\lambda}}_{\nu\rho}\wt u_\alpha \wt n_\lambda \wt u^{\mu}\osduu\mu\nu\rho\delta_{p,2}\delta_{n,2} \\
&+\wt \lambda_3 \left(2{\epsilon^{\lambda\mu}}_{\rho\alpha} \Ksduu\lambda\nu\rho\Ksddu\mu\nu\alpha+ \wt{R}_-\right)\delta_{p,3}\delta_{n,1} \Big)~,
\end{split}
\ee
where we must set $\lambda_2=-\lambda_1$ in \eqref{eq:seven}. The scalars $\wt{R}_{||}$, $\wt{R}_{\angle}$, $\wt{R}_{\perp}$ and $\wt{R}_-$ are defined analogously to \eqref{eq:defR} and below \eqref{eq:sodd}.


\subsection{Constraints on the stress tensor} \label{eq:constraintsStress}

In this section we consider the constraints on the stress tensor that arise from requiring that the dynamics are determined by an action of the type \eqref{eq:genact}. 

We study the parity-even and parity-odd sectors of the action, including the edge contribution. The method employed is analogous to the classification methods, inspired by effective field theories, used in the context of hydrodynamics. In the present context, the results presented below yield the first example of constraining the stress tensor of entangling surfaces using a conservation equation. In particular, invariance of the action \eqref{eq:genact} under tangential diffeomorphisms implies that the constraints \eqref{eq:st}-\eqref{eq:bt} and the tangential projection of the equation of motion \eqref{eq:genshapetang} must be automatically satisfied. This implies that by classifying the tensor structures that can appear at the two-derivative level in the independent quantities $\mathcal{T}^{\mu\nu}$, $\mathcal{D}^{\mu\nu\rho}$, $\mathcal{S}^{\mu\nu\rho}$, $\mathcal{Q}^{\mu\nu\lambda\rho}$ and imposing invariance under tangential diffeomorphisms must yield the stress tensor as derived directly from \eqref{eq:seven}, \eqref{eq:sodd} and \eqref{eq:sevenedge}. It is shown that this is indeed the case.

\subsubsection{Constraints from the parity-even sector}

We now classify the contributions to the surface stress tensor $\mathcal{T}^{\mu\nu}$, which must be purely tangential and symmetric in the two indices. The most general form at the two-derivative level is given by
\be \label{eq:stclass}
\begin{split}
\mathcal{T}^{\mu\nu}=&~\alpha \huu\mu\nu+\beta_1 \huu\mu\nu K_\rho K^{\rho}+4\beta_2\Kuuu\mu\nu\rho K_\rho+\beta_3\huu\mu\nu \Kddu\alpha\beta\rho \Kuud\alpha\beta\rho+4\beta_4\Kudd{(\mu}\alpha\rho \Kuuu{\nu)}\alpha\rho \\
&+\beta_5\huu\mu\nu R_{||}+2\beta_6\huu\mu\beta\huu\nu\alpha \huu\lambda\sigma \Bdddd\beta\lambda\alpha\sigma+\beta_7\huu\mu\nu R_{\angle}+2\beta_8\huu\mu\beta\huu\nu\alpha 
 \puu\lambda\sigma \Bdddd\beta\lambda\alpha\sigma\\
&+\beta_9\huu\mu\nu R_{\perp}+\kappa_1\huu\mu\nu\mathcal{R}+2\kappa_2\mathcal{R}^{\mu\nu}+\kappa_3\huu\mu\nu \oduu\lambda\alpha\rho\oudd\lambda\alpha\rho+2\kappa_4\oudd\mu\lambda\rho\ouuu\nu\lambda\rho  \ .
\end{split}
\ee
Note that we have added a term proportional to $\mathcal{R}^{\mu\nu}$ and another to $\huu\mu\nu \mathcal{R}$, but these could be replaced by linear combinations of the remaining terms using the Gauss-Codazzi equation \eqref{eq:GC}. The reason for keeping them is just to demonstrate that one could also choose to work with $\kappa_1$ and $\kappa_2$ instead of the terms involving $\beta_5$ and $\beta_6$.

Given the general form of $\mathcal{T}^{\mu\nu}$, imposing the conservation equation \eqref{eq:genshapetang}, and using \eqref{eq:bendingmoment}, \eqref{eq:spineven} and \eqref{eq:classq}, a lengthy calculation reveals
\be \label{eq:eeven}
\begin{split}
&4(\beta_2+\lambda_1)\hud\sigma\nu\overline\nabla_\lambda\left(\Kuuu\lambda\nu\rho K_\rho\right)+2(\beta_1-\lambda_1)K^\rho\overline\nabla^\sigma K_\rho+2(\beta_3-\lambda_2)\Kuud\mu\nu\rho\overline\nabla^\sigma \Kddu\mu\nu\rho \\
&+4(\beta_4+\lambda_2)\huu\sigma\nu\overline\nabla_\lambda\left(\Kuud\mu\lambda\rho\Kddu\mu\nu\rho\right)+\huu\sigma\lambda\huu\mu\nu\huu\alpha\rho\left((\beta_5-\alpha_2)\nabla_\lambda \Bdddd\mu\alpha\nu\rho+2(\beta_6+\alpha_2)\nabla_\alpha\Bdddd\mu\rho\nu\lambda\right) \\
&+2(\beta_6+\alpha_2)\huu\sigma\alpha\left(\huu\mu\nu K^\rho \Bdddd\mu\rho\nu\alpha-\Kuuu\rho\nu\lambda\Bdddd\nu\alpha\lambda\rho\right)+2(\beta_6+2\beta_5-\alpha_2)\huu\mu\nu \Kuuu\rho\sigma\alpha \Bdddd\mu\rho\nu\alpha\\
&+\puu\mu\nu\huu\alpha\rho\huu\sigma\lambda\left((\beta_7-\alpha_3)\nabla_\lambda \Bdddd\mu\alpha\nu\rho+2(\beta_8+\alpha_3)\nabla_\alpha \Bdddd\mu\lambda\nu\rho\right)-2(\beta_7-\alpha_3)\Kuuu\sigma\mu\nu\huu\alpha\rho\Bdddd\mu\alpha\nu\rho\\
&+2(\beta_8+\alpha_3)\huu\sigma\alpha\left(\puu\mu\nu K^\rho\Bdddd\mu\alpha\nu\rho+2\Kuuu\rho\mu\nu\Bdddd\alpha\mu\nu\rho\right)+2(\beta_7+\beta_8)\puu\mu\nu\Kuuu\sigma\alpha\rho\Bdddd\mu\alpha\nu\rho \quad \\
&+(\beta_9-\alpha_4)\left(\puu\mu\nu\puu\lambda\rho\huu\sigma\alpha\nabla_\alpha \Bdddd\mu\lambda\nu\rho-4\puu\alpha\rho\Kuuu\sigma\mu\nu\Bdddd\mu\alpha\nu\rho\right)+(\kappa_1+\kappa_2)\overline\nabla^\sigma \mathcal{R} \quad \\
&+2(\kappa_3-\vartheta_0)\oduu\lambda\nu\rho\overline\nabla^\sigma\oudd\lambda\nu\rho+2(\kappa_4+\vartheta_0)\oudd\lambda\nu\rho\overline\nabla_\lambda \ouuu\sigma\nu\rho+2\kappa_4\ouuu\sigma\nu\rho\overline\nabla_\lambda\oudd\lambda\nu\rho=0 \ .
\end{split}
\ee
Note that we have made use of the equation \eqref{eq:fullshapeeq} to account for the terms appearing in $\mathcal{I}^{\mu\nu\lambda\rho}$, however, these yield vanishing contributions to \eqref{eq:eeven}. The conservation equation must be satisfied for arbitrary geometric tensors, and since each of these terms is independent\footnote{The term involving $(\kappa_1+\kappa_2)$ is not independent but we are treating it as so because one could have chosen to work with $\kappa_1,\kappa_2$ instead of $\beta_5,\beta_6$.}  there is the unique solution
\be
\begin{split}
&\beta_1=\lambda_1 \ , \quad \beta_2=-\lambda_1 \ , \quad \beta_3=\lambda_2 \ , \quad \beta_4=-\lambda_2 \ , \quad \beta_5=\alpha_2 \ , \quad \beta_6=-\alpha_2 \ ,\\
&\beta_7=\alpha_3 \ , \quad \beta_8=-\alpha_3 \ , \quad \beta_9=\alpha_4 \ , \quad \kappa_1=-\kappa_2 \ , \quad \kappa_3=\kappa_4=\vartheta_0=0 \ ,
\end{split}
\ee
with the surface tension $\alpha$ being unconstrained.  Indeed, one may check that once introducing this solution into the stress tensor \eqref{eq:stclass} one gets precisely what is obtained by direct variation of \eqref{eq:seven}. Note that the last term in \eqref{eq:eeven} is precisely the conservation equation for the spin current \eqref{eq:spineven} and forces $\kappa_3=\kappa_4=\vartheta_0=0$. As mentioned in the previous section, adding a term of the form \eqref{eq:spineven} is not allowed as it violates the off-shell diffeomorphism constraint \eqref{eq:st}. Furthermore, note that the term appearing in $\mathcal{I}^{\mu\nu\lambda\rho}$ is unconstrained, but it will be related to $\kappa_1$ via the edge constraints. This analysis shows that, as in hydrodynamics, these methods can successfully constrain the stress tensor of (entangling) surfaces.

\subsubsection{Constraints from the parity-odd sector}
For the parity-odd sector we proceed in a similar manner. First of all, we note that the bending moment, spin current and background curvature moment have non-trivial contributions given by \eqref{eq:bend0}, \eqref{eq:bendingmoment1}, \eqref{eq:spin0} and \eqref{eq:q0}. For the present purposes, however, we will not assume that the coefficient $\lambda_3$ appearing in the parity-odd contribution to the Young modulus \eqref{eq:bendingmoment1} is the same as the one appearing in the background curvature moment \eqref{eq:q0}. Therefore, we replace the coefficient $\lambda_3$ appearing in the Young modulus by another coefficient $\theta_3$, though we will at the end derive that we must have $\lambda_3=\theta_3$. In this case, the most general surface stress tensor can be written as \footnote{We could have added terms to $\mathcal{T}^{\mu\nu}$ of the form $\gamma_1\huu\mu\nu{\epsilon^{\lambda\sigma}}_{\rho\alpha}\Kduu\lambda\beta\alpha\Kddu\sigma\beta\rho$, $\gamma_2\huu\mu\nu R_-$ and $\gamma_3{\epsilon_{||}^{\alpha(\mu}} \Kduu\alpha{\nu)}\lambda K^\rho\epsilon^{\perp}_{\lambda\rho}$. However, ultimately, the conservation equation \eqref{eq:genshapetang} requires $\gamma_1=\gamma_2=\gamma_3=0$ and therefore we have avoided making this explicit for clarity of presentation.}

\be \label{eq:stclass1}
\begin{split}
\mathcal{T}^{\mu\nu}=&\left(\beta_{10}\huu\mu\nu K^\rho n_\rho+2\beta_{11}\Kuuu\mu\nu\rho n_\rho\right)\delta_{n,1}\\
&+\left(2\kappa_5\huu\mu\nu u^\lambda\omega_\lambda+\kappa_6\huu\mu\nu (u^\alpha\omega_\alpha)^2+2\kappa_7\hud{(\mu}\lambda\overline\nabla^{\nu)}\omega^\lambda\right)\delta_{p,1}\delta_{n,2} \\
&+\left(4\beta_{13}\epsilon^{\lambda\sigma\rho\alpha}\Kdud\lambda{(\mu}\rho\Kudd{\nu)}\sigma\alpha+2\beta_{14}\huu\alpha\sigma\puu\lambda\kappa\puu\rho\tau{\epsilon^{(\mu}}_{\sigma\kappa\tau}\hud{\nu)}\beta\Buddd{\beta}\alpha\lambda\rho\right)\delta_{p,2}\delta_{n,2} \ .
\end{split}
\ee
A similarly lengthy calculation as in the parity-even case, using \eqref{eq:genshapetang}, \eqref{eq:bend0}, \eqref{eq:bendingmoment1}, \eqref{eq:spin0} and \eqref{eq:q0} leads to 
\be \label{eq:longodd}
\begin{split}
&(\beta_{10}-\lambda_0)\overline\nabla^{\sigma}\left(K^\rho n_\rho\right)\delta_{n,1}+2(\beta_{11}+\lambda_0)\hud\sigma\alpha\overline\nabla_\lambda\left(\Kuuu\lambda\alpha\rho n_\rho\right)\delta_{n,1}\\
&+\left[2\kappa_5\overline\nabla^\sigma\left(u^{\mu}\omega_\mu\right)+\kappa_6\overline\nabla^\sigma(u^\alpha\omega_\alpha)^2\right]\delta_{p,1}\delta_{n,2}\\
&+\kappa_7\left(K_\alpha\overline\nabla^\sigma\omega^\alpha+\overline\nabla_\alpha\overline\nabla^\sigma\omega^\alpha+\Kduu\lambda\sigma\alpha\overline\nabla^\lambda\omega_\alpha+\hud\sigma\alpha\overline\nabla_\lambda\overline\nabla^\lambda\omega^\alpha\right)\delta_{p,1}\delta_{n,2}\\
&+\Big[(\beta_{14}-\lambda_3)\epsilon^{\lambda\alpha}_{||}\epsilon^{\beta\rho}_\perp\nabla_\lambda \Buddd\nu\alpha\beta\rho+4(\beta_{13}+\theta_3)\epsilon^{\lambda\alpha}_{||}\epsilon^{\beta\rho}_\perp\left(\Kdud\lambda\sigma\beta\overline\nabla_\mu \Kudd\mu\alpha\rho+\hud\sigma\nu\Kudd\mu\alpha\rho\overline\nabla_\mu\Kdud\lambda\nu\beta\right) \\
&+(4\theta_3 -2(\beta_{14}+\lambda_3))\epsilon^{\lambda\alpha}_{||}\epsilon^{\beta\rho}_\perp\hud\sigma\nu\Buddd\nu\alpha\mu\rho+2(\theta_3-\lambda_3)\epsilon^{\lambda\alpha}_{||}\epsilon^{\beta\rho}_\perp\Kdud\nu\sigma\beta\Buddd\nu\rho\alpha\lambda\\
&+(\beta_{14}-\lambda_3)\hud\sigma\kappa\overline\nabla_\lambda\left(\epsilon^{\kappa\alpha}_{||}\epsilon^{\beta\rho}_\perp\hud\lambda\nu\Buddd\nu\alpha\beta\rho\right)\Big]\delta_{p,2}\delta_{n,2}=0 \ ,
\end{split}
\ee
where we have made use of Eq.~\eqref{eq:fullshapeeq} in order to deal with the parity-odd contribution to $\mathcal{H}^{\mu\nu\lambda\rho}$. However, this contribution drops out from the conservation equation and will only be related to the other coefficients via an edge analysis.
Since all these terms are independent from each other, satisfying this equation implies that all terms must individually vanish, giving the unique solution
\be\label{eq:sol2}
\begin{split}
&\beta_{10}=\lambda_0 \ , \quad \beta_{11}=-\lambda_0 \ , \quad \kappa_5=\kappa_6=\kappa_7=0 \ ,\\
&\beta_{13}=-\theta_3 \ , \quad \beta_{14}=\lambda_3 \ , \quad \lambda_3=\theta_3 \ .
\end{split}
\ee
Note that the terms in the spin current proportional to $\vartheta_1$, $\vartheta_2$ do not appear in \eqref{eq:longodd} and so are unconstrained by the stress conservation equation. This is a consequence of the  surface dimensionality $p=1$. However, one must still impose the off-shell constraint \eqref{eq:st} which leads to $\vartheta_2=0$, as in the previous section, and $\vartheta_1$ is unconstrained, as expected for a spinning point-particle. This shows that the conservation equation \eqref{eq:genshapetang} is not sufficient in order to implement all diffeomorphism constraints. Furthermore, the result \eqref{eq:sol2} provides a formal derivation of the fact that only the linear combination of the first two terms in \eqref{eq:3terms} satisfies the requirements of tangential diffeomorphism invariance, as advertised in the previous section, since we must have $\lambda_3=\theta_3$. Finally, we note that the solution \eqref{eq:sol2} agrees with the surface stress tensor that is obtained by direct variation of \eqref{eq:sodd}.

\subsubsection{Constraints from the edges}
We now consider the constraints on the edge stress tensor. The edge stress tensor contains the analogous contributions to \eqref{eq:stclass} and \eqref{eq:stclass1}. The tangential projection of \eqref{eq:genshapeedge} leads to the same constraints as \eqref{eq:eeven} and \eqref{eq:sol2}. Here, we will therefore derive the constraints due to contributions that are a feature of the edge dynamics, considering the parity-even and parity-odd sectors simultaneously. We therefore consider the non-trivial contributions to the edge stress tensor\footnote{We could have added other terms such as those proportional to $\wt\gamma_1 \hsuu\mu\nu (\mathcal{K}^\rho\wt n_\rho)^2$ and $\wt\gamma_2\hsuu\mu\nu \Ksuud \alpha\beta\rho\wt n^\rho \Ksddu\alpha\beta\sigma \wt n_\sigma$ but a similar analysis would ultimately set $\wt\gamma_1=\wt\gamma_2=0$ and therefore we have avoided presenting these for the sake of clarity of presentation.}
\be \label{eq:stressedgec}
\wt{\mathcal{T}}^{\mu\nu}=\wt \beta_{15}\hsuu\mu\nu \wt n_\rho \mathcal{K}^{\rho}+2\wt\beta_{16}\Ksuud\mu\nu\rho\wt n^\rho+2\wt\kappa_1\hsuu\mu\nu{\epsilon^{\alpha\sigma}}_{\lambda\rho}\wt u_\alpha\wt n_\sigma \wt u^{\kappa}\osduu\kappa\lambda\rho\delta_{p,2}\delta_{n,2} \ .
\ee
We now use the stress tensors \eqref{eq:stclass1}, \eqref{eq:stclass1}, together with the surface bending moment \eqref{eq:bendingmoment1}, spin current \eqref{eq:spin0} and the constraints \eqref{eq:eeven}, \eqref{eq:sol2}, as well as the edge bending moment \eqref{eq:edgebending} and spin current \eqref{eq:spin1}, and introduce it into \eqref{eq:genshapeedge}. The tangential projection of \eqref{eq:genshapeedge} then yields 
\be \label{eq:tanedge}
\begin{split}
&(\wt\beta_{15}-\wt\lambda_0)\wt\nabla^\sigma(\wt n_\rho\mathcal{K}^\rho)+2(\wt\beta_{16}+\wt\lambda_0)\hsud\sigma\nu\wt\nabla_\mu\left(\Ksuud\mu\nu\rho\wt n^\rho\right)+(\wt\lambda_0+2\lambda_1+2\kappa_1)\mathcal{K}^\rho\wt\nabla^\sigma \wt n_\rho \\
&+2(\alpha_1-\kappa_1)\Ksuud\rho\sigma\nu\wt\nabla_\rho\wt n^\nu+2(\kappa_1-\alpha_1)\wt n_\lambda \hsuu\sigma\rho\hsuu\mu\nu \Buddd\lambda\mu\rho\nu \\
&+2(\lambda_1+\lambda_2)\hsuu\sigma\lambda\wt n_\mu\wt n_\nu\wt n_\alpha \Kuuu\mu\nu\rho \Kudd\alpha\lambda\rho+2\wt\kappa_1\wt\nabla^\sigma\left({\epsilon^{\alpha\nu}}_{\lambda\rho}\wt u_\alpha\wt n_\nu \wt u^{\mu}\osduu\mu\lambda\rho\right)\delta_{p,2}\delta_{n,2}=0 \ ,
\end{split}
\ee
which leads to the constraints
\be \label{eq:sol3}
\wt\beta_{15}=\wt\lambda_0 \ , \quad \wt\beta_{16}=-\wt\lambda_0 \ , \quad \wt\lambda_0=-2\lambda_1-2\kappa_1 \ , \quad \lambda_1=-\lambda_2 \ , \quad \wt\kappa_1=0 \ , \quad \kappa_1=\alpha_1 \ .
\ee
Note that these constraints capture the correct conditions found in the previous section for the terms involving the parity-even sector of the bending moment. As advertised, the boundary analysis relates the surface quadrupole moment with the stress tensor coefficient $\kappa_1$ and the presence of $\alpha_1$ also requires a non-trivial $\wt\lambda_0$. Note also that the terms involving $\lambda_3$, $\vartheta_1$, $\wt\vartheta_1$, $\wt\vartheta_2$ and $\alpha_5$ have dropped out from \eqref{eq:tanedge}. The term proportional to $\wt\vartheta_1$ is left unconstrained as in the surface case but in order to constraint the remaining terms we must impose the off-shell constraints
\be
\begin{split}
&\wt{\mathcal{D}}^{\mu\lambda[\sigma}\Ksddu\mu\lambda{\alpha]}-\wt n_\lambda\wt n_\mu \mathcal{D}^{\lambda\mu[\sigma}\wt{n}^{\alpha]}-\wt n_\lambda \psud\alpha\mu\psud\sigma\nu\mathcal{S}^{\lambda\nu\mu}\\
&-2\psud\alpha\tau\psud\sigma\beta\wt\nabla_\mu\left(\hsud\mu\lambda\wt n_\kappa\mathcal{H}^{\beta\kappa\tau\lambda}\right)-2\wt n_\lambda \puu\mu{\sigma}\pud\alpha\tau\overline\nabla_\kappa{\mathcal{H}_\mu}^{\kappa\tau\lambda}=0 \ ,
\end{split}
\ee
which leads to  $\wt\vartheta_2=-2\lambda_3-2\alpha_5$ as in the previous section, where we have used App.~\ref{app:highdev}. The form of \eqref{eq:stressedgec} with the constraints \eqref{eq:sol3} can be obtained directly from \eqref{eq:sevenedge}. The edge dynamics provide a very non-trivial example of the diffeomorphism constraints imposed by a well-defined variational principle.


\subsection{Constraints from Weyl invariance}\label{sec.weylsecondorder}

In this section we study the constrains imposed on actions of the form \eqref{eq:varL} and \eqref{sec:edgegeomtr} if the action is required to be invariant under infinitesimal Weyl rescalings, i.e. under the metric rescaling
\be
\delta_\omega g_{\mu\nu} = 2 \omega\, g_{\mu\nu} \ ,
\ee
where $\omega(x)$ is an arbitrary real function. The transformation properties of the various geometric quantities is given in App.~\ref{app:variations}. The variation of the action still takes the form of \eqref{eq:varL} and \eqref{sec:edgegeomtr} but with the variations being those associated with Weyl rescalings instead of Lagrangian variations. Under these variations the action varies according to
\be \label{eq:Weylvar}
\begin{split}
\delta_\omega S=&\int_{\mathcal{W}}d^p\sigma\sqrt{|\gamma|}\left[\left({\mathcal{T}^{\mu}}_\mu+{\mathcal{D}^{\mu\nu}}_\rho\Kddu\mu\nu\rho+\pdd\mu\nu\Pi^{\mu\nu}\right)\omega+{{\mathcal{D}^{\mu}}_\mu}^\rho\nabla_\rho \omega+4{\mathcal{Q}_\mu}^{(\nu\lambda)\mu}\nabla_\nu\nabla_\lambda\omega\right]\\
&\int_{\partial\mathcal{W}}d^{p-1}\wt\sigma\sqrt{|h|}\left[\left({\wt{\mathcal{T}}^{\mu}}_{~~\mu}+{\wt{\mathcal{D}}^{\mu\nu}}_{~~\rho}\Ksddu\mu\nu\rho+\psdd\mu\nu\wt{\Pi}^{\mu\nu}\right)\omega+{{{\wt{\mathcal{D}}_{\mu}}^{~~\mu\rho}}}\nabla_\rho \omega+4{\wt{\mathcal{Q}}_\mu}^{~~(\nu\lambda)\mu}\nabla_\nu\nabla_\lambda\omega\right]~~,
\end{split}
\ee
where we have used the diffeomorphism constraints \eqref{eq:st}-\eqref{eq:pt} and \eqref{eq:st1} and also ignored variations with respect to the intrinsic and outer curvatures, since they are not necessary. For the action to be invariant under Weyl rescalings each of the terms above must vanish independently, therefore a total of six conditions must be satisfied.  We will first analyse the case of a two-derivative surface action without edges and then later consider the inclusion of the edges.

\subsubsection{Weyl invariance for the edgeless surface} 
In the case of a surface without edges the first line in \eqref{eq:Weylvar} leads to three independent conditions. The last term in \eqref{eq:Weylvar} implies that 
\be \label{eq:W1}
\begin{split}
&{\mathcal{Q}_\mu}^{(\nu\lambda)\mu}=-\frac{1}{2}\left(\alpha_2(p-1)+\alpha_3\frac{n}{2}\right)\huu\lambda\nu-\frac{1}{2}\left(\alpha_4(n-1)+\alpha_3\frac{p}{2}\right)\puu\lambda\nu=0~~\\
&\Rightarrow~\alpha_2=-\alpha_3\frac{n}{2(p-1)}~~,~~\alpha_4=-\alpha_3\frac{p}{2(n-1)}~~,
\end{split}
\ee
where we have used \eqref{eq:q0}. Note that the term $\lambda_3$ in \eqref{eq:q0} drops out from this equation. In turn, the second condition in \eqref{eq:Weylvar} leads to
\be \label{eq:W2}
\begin{split}
&{{\mathcal{D}^{\mu}}_\mu}^\rho=2(\lambda_1p+\lambda_2)K^\rho+\lambda_0 p n^\rho=0\\
&\Rightarrow~\lambda_1=-\frac{\lambda_2}{p}~~,~~\lambda_0=0~~,
\end{split}
\ee
for which we have used \eqref{eq:bend0} and \eqref{eq:bendingmoment1} and where again $\lambda_3$ in \eqref{eq:bendingmoment1} does not play a role. Finally, the requirement that the first term in \eqref{eq:Weylvar} vanishes, using \eqref{eq:stclass} and \eqref{eq:stclass1}, leads to
\be \label{eq:W3}
\alpha p+(p-2)\left(\lambda_1K^\rho K_\rho+\lambda_2\Kddu\mu\nu\rho\Kuud\mu\nu\rho+\alpha_2R_{||}+\alpha_3R_{\angle}+\alpha_4R_{\perp}\right)=0~~.
\ee
This condition, together with \eqref{eq:W1} and \eqref{eq:W2} leads to $p=2$ and $\alpha=0$. These results imply that $\lambda_3$ (or equivalently $\alpha_5$) as well as $\vartheta_1$ are free coefficients. Using the conformal tensors defined in Sec.~\ref{sec:geometry}, the two-derivative Weyl-invariant surface action takes the form
\be\label{eq.Weylaction}
\begin{split}
S[\Phi(\sigma)] = \int_{\mathcal{W}}d^{2}\sigma\sqrt{|\gamma|} & \left(\lambda_2 \, C_{\mu\nu\rho} C^{\mu\nu\rho}+ \frac{n(n+1)}{2}\alpha_4 \widehat W_{\mu\nu}{^{\mu\nu}} \right) \delta_{p,2} \\
+ \int_{\mathcal{W}}d^{p}\sigma\sqrt{|\gamma|} & \,\bigg( \left[\lambda_3 \left(2{\epsilon^{\lambda\mu}}_{\rho\alpha} C_\lambda{^{\nu\rho}} C_{\mu\nu}{^\alpha} + R_-\right) +\alpha_5 \, \Omega\right]\delta_{p,2}\delta_{n,2}\\
& \, + 2\vartheta_1u^{\mu}\omega_\mu\delta_{p,1}\delta_{n,2} \bigg) \ ,
\end{split}
\ee
with $C_{\mu\nu}{^\rho}$ the conformation tensor defined in Eqn.~\eqref{eq:conformation}.

\subsubsection{Edge contribution}
If the surface has edges, the analysis changes considerably due to the non-trivial diffeomorphism constraints obtained in \eqref{eq:st1}, which relate surface coefficients to edge coefficients. If the surface has dimensionality $p=1$ then the only non-trivial contribution to the full Weyl-invariant action is the last term in \eqref{eq.Weylaction}, as the edge is point-like. For $p=2$ there is a non-trivial contribution in \eqref{eq.Weylaction} due to the extrinsic curvatures. As shown in Sec.~\ref{sec:2derivative}, diffeomorphism invariance at the edges requires $\lambda_1=-\lambda_2$, which is incompatible with \eqref{eq:W2} for $p=2$. Therefore we must set $\lambda_2=0$.   Similarly, in the presence of edges, the contributions due to $\lambda_3$ (or $\alpha_5$) must be balanced by an equivalent edge term as $\lambda_3$ is topological. In other words, we must set $\lambda_3=\alpha_5=0$. For a two-dimensional surface, the edges are one-dimensional lines and, analogously to the surface, the edge Weyl constraints in \eqref{eq:Weylvar} do not impose any restrictions on $\wt{\vartheta}_1$. In this case, the full Weyl-invariant action is given by
\be \label{eq:edgeanom}
\begin{split}
S[\Phi(\sigma),\Phi_e(\wt\sigma)] = & \int_{\mathcal{W}}d^{2}\sigma\sqrt{|\gamma|}   \frac{n(n+1)}{2}\alpha_4 \widehat W_{\mu\nu}{^{\mu\nu}}  + 2\int_{\partial\mathcal{W}}d^{1}\wt \sigma\sqrt{|h|} \wt\vartheta_1\wt u^{\mu}\varpi_\mu\delta_{n,1}~~.
\end{split}
\ee
This is rather significantly different from the edgeless case. If the dimensionality of the surface is $p=3$, then there is no contribution from the surface action, however, the last three edge conditions in \eqref{eq:Weylvar} lead to the equivalent results as in the case of surface actions with $p=2$. In particular, we have that $\chi=\wt \alpha_2=\wt \alpha_3=0$ and the Weyl-invariant action is given by
\be\label{eq:p3edges}
\begin{split}
S[\Phi_e(\wt\sigma)] =  \int_{\partial\mathcal{W}}d^{2}\wt \sigma\sqrt{|h|} & \bigg( \wt \lambda_2\left( \Ksddu\mu\nu\rho \Ksuud\mu\nu\rho - \frac{1}{2} \mathcal{K}^{\rho}\mathcal{K}_{\rho} \right) +\wt\alpha_4 \wt{R}_{\perp}  \\
&+\wt \lambda_3 \left(2{\epsilon^{\lambda\mu}}_{\rho\alpha} \Ksduu\lambda\nu\rho\Ksddu\mu\nu\alpha+ \wt{R}_-\right)\delta_{n,1} \bigg) \ .
\end{split}
\ee
The first term above is the square of the edge conformation tensor (defined analogously to \eqref{eq:conformation}) while the second term coincides with the trace of the pull-back of the Weyl tensor onto the edge. For $p\ge4$ there are no non-trivial two-derivative Weyl-invariant contributions.

\subsubsection{Weyl anomalies}
The above results are pertinent in the context of Weyl anomalies for two-dimensional submanifolds. In particular, certain classes of CFTs with vacuum energy $W[\Gdd\mu\nu,X^\mu]$ living on embedding surfaces have conformal anomalies, denoted by $\mathcal{A}$, which are defined as
\be\label{eq.weylanomaly}
\delta_\omega W = \int d^Dx \sqrt{|g|}\, {\mathcal A} \, \omega \ .
\ee
Using the formulation of the action \eqref{eq:genact2} in terms of the spacetime stress tensor (with $S$ replaced by $W$), a Weyl transformation simply yields\footnote{Formally, the quantities $ \mathbf{T}_s{}^\mu_{~\mu}$ and $\mathbf{T}_e{}^\mu_{~\mu}$ are one-point functions since they are obtained via the variation of the vacuum energy $W$ with respect to $\Gdd\mu\nu$.}
\be
\cA =  \mathbf{T}_s{}^\mu_{~\mu} + \mathbf{T}_e{}^\mu_{~\mu}~~,
\ee
where the trace of the surface spacetime stress tensor is given by
\be
\mathbf{T}_s{}^\mu_{~\mu} =  \left( \cT^\mu_{~~\mu}  + \cD^{\mu\lambda\alpha} K_{\mu\lambda\alpha}  + \perp_{\lambda\sigma} \Pi^{\lambda\sigma} \right)\widehat \delta(x) + \nabla_\rho \left(  \cD^\mu{_\mu}{^\rho}  \widehat \delta (x)\right) - 4 \nabla_\lambda\nabla_\rho \left(  \cQ_\mu{^{\mu(\lambda\rho)}} \widehat \delta (x) \right) ~~,
\ee
and similarly for the edge contribution. Anomalies can be classified into three different types \cite{Deser:1993yx}. In particular, type B anomalies are anomalies composed of local terms which are Weyl invariant. All the terms in \eqref{eq.Weylaction} are locally Weyl invariant and hence are possible conformal anomalies for two-dimensional submanifolds. The first line corresponds to the Graham-Witten anomalies found in \cite{Graham:1999pm} but the term corresponding to $\Omega$, which is present in $D=4$, seems not to have received attention in the literature. 

If the submanifold has edges and assuming diffeomorphism invariance, the usual anomaly proportional to the square of the conformation tensor must vanish and we are left with \eqref{eq:edgeanom}.\footnote{Here we are insisting that the diffeomorphism constraints are satisfied for arbitrary extrinsic curvature components of both surface and edge geometry instead of imposing ad-hoc boundary conditions on them. See footnote \ref{foot:edge}.} This includes a new possible edge anomaly in $D=3$ which is type B. It is also worth noting that for $p=2$ there is only one type A surface anomaly given by (see e.g. \cite{Schwimmer:2008yh})
\be
\int_{\mathcal{W}}d^2\sigma\sqrt{|\gamma|}\mathcal{R}~~.
\ee
However, as we have seen in Sec.~\ref{sec:2derivative} (see footnote \ref{foot:weyl}), in the presence of edges and assuming diffeomorphism invariance, this term must vanish. Therefore, there are no type A anomalies for submanifolds with boundaries. In turn, this implies that the weak c-theorem of \cite{Jensen:2015swa} for CFTs coupled to defects simply does not apply if the defect has edges.


\section{Actions for DCFTs and BCFTs} \label{sec:DCFTs}

In previous sections we focused on the consequences of spacetime diffeomorphism invariance on embedded surfaces where the embedding map $X^\mu(\sigma^a)$ was kept fixed. This gave rise to a rich set of equations of motion and constraints that such a surface must satisfy. In this last section of the paper we relax the requirement that the embedding map is kept fixed, and in particular we will consider the transformation properties
\be\label{eq:variationsMM}
\delta_\xi\Gdd\mu\nu=2\nabla_{(\mu}\xi_{\nu)} \ , \quad \delta_X X^{\mu}=-\xi^{\mu} \ ,
\ee 
i.e. the change in the local coordinates is compensated by the change on the embedding map such that the surface is unmoved. 

This new set of variations, \eqref{eq:variationsMM}, is of special interest to particular cases of conformal field theories coupled to defects (surfaces) with edges (DCFTs) and boundary conformal field theories (BCFTs). When coupling defects to CFTs one wishes that the defect is consistently coupled regardless of its shape. In turn, this implies that there is no dynamics associated with a shape equation. Instead, the variational principle must render the shape equation trivial, which implies the simultaneously displacement of the embedding map under a diffeomorphism. This is precisely what Eqn.~\eqref{eq:variationsMM} ensures.

 A further generalisation can be taken into account, by letting actions have more dynamics than those dictated by pure geometry, in particular with the existence of couplings to a scalar field and $q$-form gauge fields. These fields are not background fields to which the defect couples to, as when coupling probes to supergravity actions \cite{Armas:2016mes}, such that the variation of the action with respect to them produces a source term in the equations of motion. Instead, the fields are dynamic and variation of the defect action with respect to these fields yields the corresponding equation of motion. 
We use the spacetime formulation of the variational principle to deal with these extra fields and identify new contributions to these Ward identities, which arise due to the spacetime formulation of the variational principle and were previously overlooked (see Eqs.\eqref{eq:dbp}).

Some of these results were considered in \cite{Billo:2016cpy}. Here, we provide a different method and interpretation, using a spacetime formulation, and also include further degrees of generality, e.g. the Ward identities are derived for defects in curved space and include the existence of non-trivial defect edges, whose detailed analysis is given in App.~\ref{app:DCFTedge}.

\subsection{Variational principle}

Consider a quantum field theory living on a manifold $\mathcal{M}$ with boundary or defect $\mathcal{W}$ of dimensionality $p$. This quantum field theory is characterised by a vacuum energy functional of the form
\be \label{eq:vacuum}
W[\Gdd{\mu\nu},X^{\mu}]=\log \int [D\Psi]\exp\left(-S[\Gdd{\mu\nu},X^{\mu},\Psi,\partial\Psi,\cdots]\right) \ ,
\ee
where $\Psi$ denotes a collection of fields, while $S$ denotes the action functional which can depend on the fields $\Psi$ and its derivatives. We require the action $S$ to be invariant under diffeomorphisms when all fields are allowed to vary accordingly, namely
\be \label{eq:actCFT}
S[\Gdd{\mu\nu},X^{\mu},\Psi,\partial\Psi,\cdots]=S[\Gdd{\mu\nu}+\delta_\xi\Gdd{\mu\nu} ,X^{\mu}+\delta_X X^{\mu},\Psi+\delta_\xi \Psi,\partial\Psi+\delta_\xi\partial\Psi,\cdots] \ ,
\ee
where the metric and embedding map have the transformation properties given by Eqn.~\eqref{eq:variationsMM}.

The variation of the collection of fields $\Psi$ under diffeomorphisms is not necessary to be given since we assume that diffeomorphism invariance holds on-shell, i.e. when the equations of motion for the fields $\Psi$ are satisfied. The variation of $X^{\mu}$ is defined differently than in \cite{Billo:2016cpy}, in particular, it is defined as the total variation of $S$ with respect to $X^\mu$, instead of just being the variation restricted to a particular set of fields.

As in Sec.~\ref{sec:stresstensor}, we can recast the action in terms of a spacetime action and define its total variation as the sum of a Lagrangian variation and a variation of the embedding map, such that 
\be \label{eq:vvv}
\delta W[\Gdd\mu\nu,X^{\mu}]=\int_{\mathcal{M}}\sqrt{|g|}d ^Dx~\left(\frac{1}{2}\langle\textbf{T}^{\mu\nu}\rangle\delta_\xi\Gdd\mu\nu-\langle\textbf{D}_\mu\rangle\delta_X X^{\mu}\right) \ ,
\ee
where $\langle\textbf{D}_{\mu}\rangle$ is the one-point function referred to as the \emph{displacement operator}, as it is the operator dual to an infinitesimal displacement of the surface and defined as
\be
\langle\textbf{D}_{\mu}\rangle=-\frac{1}{\sqrt{|g|}}\frac{\delta W}{\delta_X X^{\mu}} \ .
\ee
Introducing the variations for background diffeomorphisms leads to the conservation equation
\be \label{eq:force}
\nabla_\mu\langle\textbf{T}^{\mu\nu}\rangle=\langle\textbf{D}^{\nu}\rangle \ ,
\ee
where $\langle\textbf{T}^{\mu\nu}\rangle$ is the one-point function defined as in \eqref{eq:varL2} but using the vacuum energy, i.e. by trading $S\to W$. It takes the general form
\be
\langle\Tuu\mu\nu\rangle=\langle\Tuu\mu\nu_{\text{b}}\rangle+\langle\Tuu\mu\nu_{\text{s}}\rangle+\langle\Tuu\mu\nu_{\text{e}}\rangle \ ,
\ee
with $\Tuu\mu\nu_{\text{b}}=T^{\mu\nu}_{\text{b}}$ for DCFTs and $\Tuu\mu\nu_{\text{b}}=T^{\mu\nu}_{\text{b}}\Theta(x)$ for BCFTs and the remaining two contributions being the surface and edge contributions respectively, which  for the case of BCFT's may contain additional contributions due to an inflow of energy-momentum from the bulk part of the action as in \eqref{eq.interfaceinflow}. 

 It is clear from \eqref{eq:force} that the displacement operator has the role of a force term in the stress tensor conservation equation. Its role is to force or to displace the surface in such a way that the orthogonal components of \eqref{eq:force} are trivially satisfied. The displacement operator has an expansion analogous to the stress tensor, that is
\be
\langle\textbf{D}^{\mu}\rangle=\langle\textbf{D}^{\mu}_{\text{s}}\rangle+\langle\textbf{D}^{\mu}_{\text{e}}\rangle \ ,
\ee
where both the surface and edge contributions to the displacement operator can be written as\footnote{One could expect that the displacement operator, analogously to the stress tensor, would have an expansion in terms of derivatives of $\widehat\delta(x)$. This is however not the case. Derivatives of the embedding map appearing in all geometric structures are tangential since orthogonal derivatives are not well-defined. Therefore, any variation of the embedding map will at most involve tangential derivatives. These tangential derivatives can be integrated out, as for the case of the stress tensor, by making a frame choice. It turns out that the frame choice that yields the same equations of motion and boundary conditions as for the stress tensor is the one for which all tangential derivatives of the variation are integrated out.}
\be \label{eq:displacement}
\textbf{D}^{\mu}_{\text{s}}=D^\mu \widehat\delta(x) \ , \quad \textbf{D}^{\mu}_{\text{e}}=D^\mu \widehat\delta_e(x) \ .
\ee
This definition agrees with the definition of displacement operator introduced in \cite{McAvity:1993ue, Jensen:2015swa} and it accounts for the full displacement of the surface as it appears in the conservation equation \eqref{eq:force}.\footnote{In full generality, the surface and edge displacement operators \eqref{eq:displacement} should be defined as integrals over the surface and edges, respectively, as for the spacetime stress tensor (see footnote \ref{foot:stress}).} We will determine the Ward identities for CFTs with vacuum energy \eqref{eq:vacuum} and in the process uncover the stress tensor and the displacement operator.

\subsection{Ward identities for DCFTs and BCFTs}
In order to obtain the Ward identities, consider a set of bulk operators $O_i$ and define the collection $\boldsymbol{\chi}$ of bulk operators as
\be
\boldsymbol{\chi}=O_1(x_1,z_1)\cdots O_n(x_n,z_n) \ .
\ee
It follows from \eqref{eq:force} that for any correlation function the Ward identity for diffeomorphism invariance can be written as
\be\label{eq:ward11}
\delta_\xi\left\langle\boldsymbol{\chi}\right\rangle+\int_{\mathcal{M}}\sqrt{|g|}d ^Dx~\left(\left\langle-\nabla_\mu\textbf{T}^{\mu\nu}\boldsymbol{\chi}\right\rangle+\left\langle\textbf{D}_\nu\boldsymbol{\chi}\right\rangle\right)\xi_\nu=0 \ .
\ee
Hence, in the absence of external operators or when the correlator is invariant under the symmetry, one must have
\be \label{eq:ward1}
\left\langle\nabla_\mu\textbf{T}^{\mu\nu}\boldsymbol{\chi}\right\rangle=\left\langle\textbf{D}^\nu\boldsymbol{\chi}\right\rangle \ .
\ee
While \eqref{eq:ward1} does capture all constraints from diffeomorphism invariance, we require that the defect action, including its edges, is invariant under tangential diffeomorphisms that do not displace the surface. From \eqref{eq:ward1}, this implies the stricter results\footnote{The authors of \cite{Jensen:2015swa} claim to have derived surface reparametrisation invariance in Eq. (B3) of \cite{Jensen:2015swa}. This however is not something that can be derived but something that must be imposed on abstract actions of the form \eqref{eq:vvv}. In particular, they seem to have forgotten the change in the argument in $\Gdd\mu\nu(X)$ in (B2) of \cite{Jensen:2015swa}. If such had been taken into account, a linear combination of the two Ward identities in \eqref{eq:ward2} would have been obtained.}
\be \label{eq:ward2}
\left\langle\hud\sigma\nu\nabla_\mu\textbf{T}^{\mu\nu}\boldsymbol{\chi}\right\rangle=0 \ , \quad \left\langle\hud\sigma\nu\textbf{D}^\nu\boldsymbol{\chi}\right\rangle=0 \ .
\ee
The first condition above states that the invariance of the action under tangential diffeomorphism while the second condition states that the defect action must be reparametrisation invariant. As we will see below, for specific classes of DCFT actions, these two Ward identities will lead to different constraints. Finally, for a Weyl transformation in which $\delta_\omega \Gdd\mu\nu=2\omega \Gdd\mu\nu$ and $\delta_\omega X^{\mu}=0$ one obtains the Ward identity
\be
\delta_\omega\left\langle\boldsymbol{\chi}\right\rangle+\int_{\mathcal{M}}\sqrt{|g|}d ^Dx~\omega\left\langle{\textbf{T}^{\mu}}_{\mu}\boldsymbol{\chi}\right\rangle=0 \ .
\ee
Hence, in the absence of external operators, up to possible conformal anomalies, this leads to $\left\langle{\textbf{T}^{\mu}}_{\mu}\boldsymbol{\chi}\right\rangle=0$. This concludes the general study of Ward identities which holds for any DCFT or BCFT. However, when written in this abstract manner, its usefulness is questionable. Below, we focus on a large class of DCFTs/BCFTs and write down the Ward identities in a more practical manner. We also give examples in which further local symmetries are imposed, leading to further Ward identities. The inclusion of edges in full generality is considered in App.~\ref{app:DCFTedge}.

\subsection{Ward identities for a class of DCFT actions} \label{eq:WardDCFT}
We now consider a class of DCFT actions and write down their Ward identities. We focus on the case in which the defect couples to the different geometric tensors introduced in Sec.~\ref{sec:surfaces} but we ignore couplings to curvature tensors for simplicity. As we have noted above, the vacuum energy is a functional of $\Gdd\mu\nu$ and $X^\mu$ but the action can be a function of non-geometric fields. If the fields are background fields such as a scalar field $\phi(x)$ or some vector field $\phi^\mu(x)$ then the couplings to the geometric fields introduced in \eqref{eq:varL} are sufficient. However, if the fields are purely surface fields such as a vector field $\phi^i(X)$ or $\phi^a(X)$ that only have support on $\mathcal{W}$, then one may define their respective push-forwards onto the background spacetime as $\phi^i\nud\mu i$ or $\phi^a\eud\mu a$. Hence their variations will need to include variations with respect to $\nud\mu i$ and $\eud\mu a$ which were not accounted for in \eqref{eq:varL}.\footnote{This means that in practice, we need to work with both spacetime and surface indices though, as we shall see, we can always push-forward the required structures in order to have a purely spacetime description. In the gauge formulation of the variational principle, the problem is reversed and it would be necessary to introduce couplings to $\nud\mu i$ and $\eud\mu a$ in order to deal with background fields leading again to the necessity of dealing with both spacetime and surface indices.}

Consider a general variation of a correlator such that\footnote{A subclass of this class of DCFTs was studied in \cite{Billo:2016cpy} but using a gauge formulation of the variational principle and without considering the possibility of non-trivial edges.}
\be \label{eq:varDCFT}
\begin{split}
&\delta\left\langle\boldsymbol{\chi}\right\rangle+\frac{1}{2}\int_{\mathcal{M}}d^Dx\sqrt{|g|}\left\langle T^{\mu\nu}_{\text{b}}\boldsymbol{\chi}\right\rangle\delta \Gdd\mu\nu \\
&+\int_{\mathcal{W}}d^p\sigma\sqrt{|\gamma|} \left\langle \Big(-\frac{1}{2}\mathcal{T}_{\mu\nu} \, \delta_\xi\huu\mu\nu-{\mathcal{P}_{\mu}}^{\nu} \, \delta_\xi \pud\mu\nu+\frac{1}{2}\mathcal{B}^{\mu\nu} \, \delta_\xi\! \pdd\mu\nu +{\mathcal{D}^{\mu\nu}}_{\rho} \, \delta_\xi\Kddu\mu\nu\rho \right.\\
&~~~~~~~~~~~~~~~~~~~~~~~~~\left.+{\mathcal{S}^{\mu}}_{\lambda\rho}\, \wt \delta_\xi\oduu\mu\lambda\rho+\mathcal{D}_\mu\delta X^\mu+{\mathcal{V}^{\mu}}_\nu\delta {n_{\mu}}^{\nu}+{\mathcal{C}^{\mu}}_\nu\delta {e_{\mu}}^{\nu}\Big)\boldsymbol{\chi}\right\rangle\\
&\int_{\partial\mathcal{W}}d^{p-1}\wt\sigma\sqrt{|h|}\left\langle\Big(-\frac{1}{2}\wt{\mathcal{T}}_{\mu\nu}\delta_\xi \hsuu\mu\nu-{\wt{\mathcal{P}}_{\mu}}^{~\nu}\delta_\xi \psud\mu\nu+\frac{1}{2}\wt{\mathcal{B}}^{\mu\nu}\delta_\xi \psdd\mu\nu+{\wt{\mathcal{D}}^{\mu\nu}}_{ \quad \rho}\delta_\xi\Ksddu\mu\nu\rho \right. \\
&~~~~~~~~~~~~~~~~~~~~~~~~~\left.+{\wt{\mathcal{S}}^\mu}_{~\lambda\rho}\delta_\xi \osduu\mu\lambda\rho+\wt{\mathcal{D}}_\mu\delta \wt{X}^\mu+{\wt{\mathcal{V}}^{\mu}}_{ ~~ \nu}\delta {\wt{n}_{\mu}}^{~\nu}+{\wt{\mathcal{C}}^{\mu}}_{ ~~ \nu}\delta {\wt{e}_{\mu}}^{~\nu}\Big)\boldsymbol{\chi}\right\rangle=0 \ ,
\end{split}
\ee
where we have defined the variations $\delta {e_{\mu}}^{\nu}=\edu\mu a\delta \eud\nu a$ and $\delta {n_{\mu}}^{\nu}=\ndu\mu i\delta \nud\nu i$ and analogously for the edge variation. These are the required structures to deal with fields with support on $\mathcal{W}$ and $\partial\mathcal{W}$. We have also introduced the dual operators to these variations in the manner
\be
\mathcal{D}_{\mu}=\frac{1}{\sqrt{|\gamma|}}\frac{\delta \mathcal{L}}{\delta X^{\mu}}\bigg|_{\text{g fixed}} \ ,~~{\mathcal{V}^{\mu}}_\nu=\frac{1}{\sqrt{|\gamma|}}\nud\mu i\frac{\delta\mathcal{L}}{ \delta \nud\nu i} \ ,~~{\mathcal{C}^{\mu}}_\nu=\frac{1}{\sqrt{|\gamma|}}\eud\mu a\frac{\delta\mathcal{L}}{ \delta \eud\nu a} \ ,
\ee
and similarly for the edge action. Here ${\mathcal{V}^{\mu}}_\nu$ is transverse in its first index while ${\mathcal{C}^{\mu}}_\nu$ is tangential in its first index and similarly for the edge terms. These variations and their conjugate operators mark the difference between the analysis presented in this section and that of Sec.~\ref{sec:surfaces}. These three variations were introduced in order to deal with non-geometric fields such as gauge fields and scalar fields. In particular, $\mathcal{D}_{\mu}$ is defined as the variation of the action with respect to variations of the embedding map keeping the intrinsic and extrinsic geometry fixed. Therefore, it should not be confused with $\textbf{D}_\mu$, which takes into account variations of the geometric fields as well. Similarly, the variations with respect to $\delta {e_{\mu}}^{\nu}$ and $\delta {n_{\mu}}^{\nu}$ should be understood as couplings between the purely surface non-geometric fields and background non-geometric fields or the surface geometry, except in the case $p=1$ for which the coupling to ${e_{\mu}}^{\nu}$ is essentially a coupling to the point particle velocity $u^{\mu}$ or in the case of a BCFT where the coupling to ${n_{\mu}}^{\nu}$ is essentially a coupling to the normal co-vector $n_\mu$. In these particular cases, their variations also include direct couplings between the geometric fields. 

\paragraph{Ward identity for tangential diffeomorphisms.} We now wish to express the Ward identities \eqref{eq:ward2} for the class of DCFTs whose action varies according to \eqref{eq:varDCFT}. We begin by considering the first of the identities in \eqref{eq:ward2} and perform a tangential Lagrangian variation that does not displace the mapping functions, i.e. 
\be
\delta_\xi\Gdd\mu\nu=2\nabla_{(\mu}\xi_{\nu)}^{||} \ , \quad \delta_\xi X^\mu=0 \ ,~~ \xi_{\nu}^{||}=\hud\nu\mu\xi^\mu \ .
\ee
Under this restricted variation, the variations associated with the operators $\mathcal{D}_\mu$ and ${\mathcal{C}^{\mu}}_\nu$ vanish, and similarly for the corresponding edge operators, however ${\mathcal{V}^{\mu}}_\nu$ and ${\wt{\mathcal{V}}^{\mu}}_{~~ \nu}$ do play a role. Under the tangential diffeomorphism, one finds
\be \label{eq:difftang}
\begin{split}
\delta_{\xi^{||}}\left\langle\boldsymbol{\chi}\right\rangle&-\int_{\mathcal{M}}\sqrt{|g|}d^{D}x\left\langle\nabla_{\mu}T^{\mu\nu}_{\text{b}}\boldsymbol{\chi}\right\rangle\xi_\nu^{||}+\int_{\mathcal{W}}\sqrt{|\gamma|}d^p\sigma\left(\left\langle B^{\mu\nu}\pud\lambda\mu\boldsymbol{\chi}\right\rangle\nabla_\lambda\xi_\nu^{||}+\left\langle B^{\nu}\boldsymbol{\chi}\right\rangle \xi_\nu^{||}\right)\\
&+\int_{\partial \mathcal{W}}\sqrt{|h|}d^{p-1}\wt{\sigma}~\wt{n}_{\mu}\left(\left\langle\wt{B}^{\mu\rho\nu}\psud\lambda\rho\boldsymbol{\chi}\right\rangle\nabla_\lambda\xi_\nu^{||}+\left\langle\wt{B}^{\mu\nu}\boldsymbol{\chi}\right\rangle\xi_\nu^{||}\right)=0 \ .
\end{split}
\ee
The quantities $B^{\mu\nu}, B^{\nu}, \wt{B}^{\mu\rho\nu}, \wt{B}^{\mu\nu}$ were introduced in Sec.~\ref{sec:shapeequations}. However, now there is a new contribution to $B^{\mu\nu}$ and $\wt{n}_{\mu}\wt{B}^{\mu\rho\nu}$. In particular, 
\be
{\mathcal{P}^{\alpha}}_\sigma\puu\sigma\nu \hud\mu\alpha+\mathcal{B}^{\mu\nu}-\mathcal{D}^{\alpha\lambda\nu}\Kddu\alpha\lambda\mu+\mathcal{S}^{\alpha\lambda\mu}\Kdud\alpha\nu\lambda-\pdu\lambda\nu\pdu\rho\mu\overline\nabla_\alpha \mathcal{S}^{\alpha\lambda\rho}-\pud{(\mu}\lambda\mathcal{V}^{\nu)\lambda}-{\mathcal{V}^{\nu}}_\lambda\huu\lambda\mu =0\ ,
\ee
while $\wt{n}_{\mu}\wt{B}^{\mu\rho\nu}$ is a linear combination of the three equations \eqref{eq:st1} with the analogous addition of ${\wt{\mathcal{V}}^{\mu}}_{ ~~ \nu}$ as for $B^{\mu\nu}$. Contrary to Sec.~\ref{sec:shapeequations}, in the presence of a defect one cannot conclude that $B^{\mu\nu}, B^{\nu}, \wt{B}^{\mu\rho\nu}, \wt{B}^{\mu\nu}$ must vanish independently. This is because on-shell the divergence of the bulk stress tensor can be decomposed as
\be \label{eq:tbexpansion}
\nabla_{\mu}T^{\mu\nu}_{\text{b}}=\mathcal{E}^{\nu}\widehat{\delta}(x)-\nabla_\mu\left(\mathcal{E}^{\lambda\nu}\pud\mu\lambda\widehat{\delta}(x)\right)+\wt{\mathcal{E}}^{\nu}\widehat{\delta}_e(x)-\nabla_\mu\left(
\wt{\mathcal{E}}^{\lambda\nu}\psud\mu\lambda\widehat{\delta}_e(x)\right)+ \cdots \ ,
\ee
where the dots correspond to higher order terms that we are neglecting in this analysis. Introducing this into \eqref{eq:difftang}, one finds, in the absence of correlators, the constraints
\be \label{eq:conditionsdefect}
B^{\mu}\hud\nu\mu=\mathcal{E}^{\mu}\hud\nu\mu \ , \quad \mathcal{E}^{\lambda\nu}\pud\mu\lambda=B^{\lambda\nu}\pud\mu\lambda \ , \quad \wt{\mathcal{E}}^{\mu}\hsud\nu\mu=\wt{n}_\mu\wt{B}^{\mu\nu} \ , \quad \wt{\mathcal{E}}^{\lambda\nu}\psud\mu\lambda=\wt{n}_\alpha\wt{B}^{\alpha\lambda\nu}\psud\mu\lambda \ .
\ee
This implies that a priori, without knowing the specific on-shell value of \eqref{eq:tbexpansion}, one cannot determine the components of stress ${\mathcal{P}^{\alpha}}_\sigma$ and $\mathcal{B}^{\mu\nu}$. As we shall see, this is in contrast with the case of BCFTs. However, one can write general constraints imposed by the Ward identity. From the second condition in \eqref{eq:conditionsdefect}, we are lead to three conditions 
\be \label{eq:dbp}
\begin{split}
&\mathcal{D}^{\mu\lambda[\sigma}\Kddu\mu\lambda{\alpha]}+\pdu\lambda\sigma\pdu\rho\alpha\overline\nabla_\mu \mathcal{S}^{\mu\lambda\rho}=\mathcal{E}^{\mu\nu}\pud{[\sigma}\mu\pud{\alpha]}\nu \ , \\
&\mathcal{B}^{\alpha\sigma}-\mathcal{D}^{\mu\lambda(\sigma}\Kddu\mu\lambda{\alpha)}-\pud{(\sigma}\nu\mathcal{V}^{\alpha)\nu}=\mathcal{E}^{\mu\nu}\pud{(\sigma}\mu\pud{\alpha)}\nu \ ,\\
&{\mathcal{P}^{\sigma\alpha}}-\mathcal{S}^{\mu\alpha\lambda}\Kdud\mu\sigma\lambda-{\mathcal{V}^{\alpha}}_\mu\huu\mu\sigma=\mathcal{E}^{\mu\nu}\pdu\mu\alpha\hud\alpha\nu \ ,
\end{split}
\ee
which are modifications of \eqref{eq:st}-\eqref{eq:pt} due to the presence of the bulk stress tensor.\footnote{We note that these constraints are altogether lacking in the analysis of \cite{Billo:2016cpy}.} Finally, the first condition \eqref{eq:conditionsdefect} yields
\be \label{eq:firstreq1}
\begin{split}
\hud\nu\sigma\overline\nabla_\lambda\mathbb{T}^{\lambda\sigma}-\hud\nu\sigma\Sigma^{\mu\lambda\rho}\Buddd\sigma\mu\rho\lambda=\mathcal{E}^{\mu}\hud\nu\mu \ ,
\end{split}
\ee
where $\mathbb{T}^{\lambda\sigma}$ is a modification of the one introduced in \eqref{eq:truestress}, namely,
\be \label{eq:truestress2}
\mathbb{T}^{\lambda\sigma}=\mathcal{T}^{\lambda\sigma}+{{\mathcal{P}}^{\mu}}_\nu\puu\nu\sigma\hud\lambda\mu-\hud\lambda\nu\overline\nabla_\mu\mathcal{D}^{\mu\nu\sigma}-{{\mathcal{S}^{\mu}}_\alpha}^{\sigma}\Kduu\mu\lambda\alpha-{\mathcal{V}^{\sigma}}_\nu\huu\nu\lambda \ ,
\ee
while $\Sigma^{\mu\lambda\rho}$ was defined in \eqref{eq:truestress}, and similarly for the two last edge conditions in \eqref{eq:conditionsdefect} as we shall see in App.~\ref{app:DCFTedge}.

\paragraph{Ward identity for tangential displacements of $X^{\mu}$.} We now study the second Ward identity in \eqref{eq:ward2} obtained by displacing the mapping functions according to
\be
\delta_X X^\mu=-\xi^\mu_{||} \ .
\ee
This corresponds to a reparametrisation of the surface coordinates. Under this displacement, and contrary to the case of a BCFT, there is no contribution from the bulk action and bulk operators. Using the variation formulae of App.~\ref{app:variations} we find 
\be \label{eq:surfaceinv}
\begin{split}
&\left\langle\left(\mathcal{D}^{\nu}\hud\sigma\nu+\huu\alpha\sigma\left({\mathbb{S}^{\mu}}_{\nu}+{\mathbb{B}^{\mu}}_{\nu}+{\mathbb{N}^{\mu}}_\nu-{\mathbb{P}^{\mu}}_\nu-{\mathcal{C}^{\mu}}_\nu-{\mathcal{V}^{\mu}}_\nu\right)\Gamma^{\nu}_{\mu\alpha}\right)\boldsymbol{\chi}\right\rangle\\
&+\langle\left(\overline\nabla_\lambda\left({\mathbb{N}^{\lambda}}_\nu-{\mathbb{P}^{\lambda}}_\nu-{\mathcal{C}^{\lambda}}_\nu\right)\huu\sigma\nu\boldsymbol{\chi}\right\rangle \\
&=\left\langle\left(\hud\sigma\nu\overline\nabla_\lambda\left(\mathbb{T}^{\lambda\nu}-{\mathcal{E}^{\mu}}_\rho\hud\lambda\mu\puu\rho\nu-\hud\lambda\rho\mathcal{V}^{\mu\rho}\pud\nu\mu\right)-\hud\sigma\nu\Sigma^{\mu\lambda\rho}\Buddd\nu\mu\rho\lambda\right)\boldsymbol{\chi}\right\rangle
\end{split}
\ee
where we have used the last condition in \eqref{eq:dbp} and, for convenience, defined
\be \label{eq:defs}
\begin{split}
&{\mathbb{N}^{\mu}}_\nu={{\mathcal{T}}^{\mu}}_{\nu}-\frac{\mathcal{L}}{\sqrt{|\gamma|}}\hud\mu\nu+2{\mathcal{D}^{\lambda\mu}}_\rho \Kddu\nu\lambda\rho+{\mathcal{S}^{\mu}}_{\lambda\rho}\oduu\nu\lambda\rho \ ,~~{\mathbb{P}^{\mu}}_\nu={\mathcal{P}^{\mu}}_\nu +{\mathcal{S}^{\alpha}}_{\lambda\nu}\Kduu\alpha\mu\lambda  \ ,\\
&\mathbb{B}^{\lambda\rho}=\left(\mathcal{B}^{\lambda\rho}-\mathcal{D}^{\mu\nu(\lambda}\Kddu\mu\nu{\rho)}\right) \ , \quad \mathbb{S}^{\alpha\beta}=\left(\mathcal{D}^{\mu\nu[\lambda}\Kddu\mu\nu{\rho]}+\overline\nabla_\mu\mathcal{S}^{\mu\lambda\rho}+\mathcal{V}^{[\lambda\rho]}\right)\pud\alpha\lambda\pud\beta\rho \ .
\end{split}
\ee
The l.h.s. of Eq.~\eqref{eq:surfaceinv} is not manifestly covariant thought its combination must be as the r.h.s. is. This is a feature of working with variations of the embedding map. 

In the absence of couplings to non-geometric fields $\mathcal{D}^\mu={\mathcal{C}^{\mu}}_\nu={\mathcal{V}^{\mu}}_\nu=0$, this Ward identity reduces to the Ward identity for surface reparametrisations in the pure geometric setting of Sec.~\ref{sec:surfaces}. However, in such case, it does not follow that ${\mathbb{S}^{\mu}}_{\nu},{\mathbb{B}^{\mu}}_{\nu},{\mathbb{P}^{\mu}}_\nu$ and ${\mathbb{N}^{\mu}}_{\nu}$  vanish. In order to recover the correct equations of motion, including the shape equation, from variations of the embedding map in the pure geometric setting, one must implement the diffeomorphism constraints \eqref{eq:st}-\eqref{eq:pt} or equivalently \eqref{eq:dbp} with $\mathcal{E}^{\mu\nu}=0$. This sets ${\mathbb{S}^{\mu}}_{\nu}={\mathbb{B}^{\mu}}_{\nu}={\mathbb{P}^{\mu}}_\nu=0$. Once introducing this into \eqref{eq:surfaceinv}, we note that one of the terms involving ${\mathbb{N}^{\mu}}_{\nu}$ still contributes with non-covariant terms. Since the action obeys the diffeomorphism constraints and the last two lines in  \eqref{eq:surfaceinv} are manifestly covariant,  it follows that requiring \eqref{eq:surfaceinv} to be covariant implies that the first line must vanish, leading to ${\mathbb{N}^{\mu}}_{\nu}=0$. This is, however, an extra requirement beyond those obtained in \eqref{eq:dbp} and it is satisfied for all pure geometric actions that we consider. It can also be obtained by requiring the spacetime and gauge variational principles to be equivalent (see App.~\ref{app:highdev}).

\paragraph{Ward identity for diffeomorphism invariance.} We now turn to the Ward identity for diffeomorphism invariance \eqref{eq:ward11}. Using App.~\ref{app:variations}, a diffeomorphism that changes the metric and the embedding map simultaneously as in \eqref{eq:variationsMM} leads to the relation
\be
\begin{split}
&\delta_{\xi}\left\langle\boldsymbol{\chi}\right\rangle-\int_{\mathcal{M}}\sqrt{|g|}d^D x\left\langle\nabla_\mu T^{\mu\nu}_{\text{b}}\boldsymbol{\chi}\right\rangle\xi_\nu+\int_{\mathcal{W}}\sqrt{|\gamma|}d^p\sigma\left\langle\left(B^{\mu\nu}\pud\lambda\nu\nabla_\lambda\xi_\mu-\mathcal{D}_\lambda\xi^{\lambda}\right)\boldsymbol{\chi}\right\rangle\\
&+\int_{\mathcal{W}}\sqrt{|\gamma|}d^p\sigma\left\langle\left(\left({\mathbb{N}^{\mu}}_{\nu}-{\mathbb{P}^{\mu}}_{\nu}-{\mathcal{C}^{\mu}}_{\nu}\right)\partial_\mu\xi^\nu+\left({\mathcal{V}^{\mu}}_\nu-{\mathbb{B}^{\mu}}_{\nu}-{\mathbb{S}^{\mu}}_{\nu}\right)\Gamma^{\nu}_{\mu\alpha}\xi^\alpha\right)\boldsymbol{\chi}\right\rangle\\
&-\int_{\mathcal{W}}\sqrt{|\gamma|}d^p\sigma\left\langle\overline\nabla_\lambda\left({\mathcal{E}^{\mu}}_\nu\hud\lambda\mu\puu\nu\sigma+\hud\lambda\nu\mathcal{V}^{\mu\nu}\pud\sigma\mu\right)\xi_\sigma\boldsymbol{\chi}\right\rangle=0 \ .
\end{split}
\ee
This Ward identity can be turned into an unintegrated expression as in \eqref{eq:ward1} with a specific spacetime stress tensor and displacement operator that will be given below. One may demand the action to be invariant under further local symmetries. For instance, using a gauge variational principle, as in \cite{Billo:2016cpy}, requiring invariance under rotations of the normal vectors sets $\mathbb{S}^{\mu\nu}=0$.

\paragraph{Ward identity for Weyl transformations.}

Lastly, consider the Ward identity obtained from Weyl invariance $\delta_\omega \Gdd \mu\nu = 2 \omega \Gdd\mu\nu$ with $\delta_\omega \eud\mu a=0$ and $\delta_\omega n_\mu{^i}=\omega n_\mu{^i}$. The variation \eqref{eq:varDCFT} becomes
\be 
\begin{split}
&\delta_\omega\left\langle\boldsymbol{\chi}\right\rangle+ \int_{\mathcal{M}}d^Dx\sqrt{|g|}\left\langle T_\text{b}{^{\mu}{_\mu}}\boldsymbol{\chi}\right\rangle \omega  \\
&\quad +\int_{\mathcal{W}}d^p\sigma\sqrt{|\gamma|}\left\langle\left( \left( \mathcal{T}^{\mu}{_\mu}  +\mathcal{B}^\mu{_\mu}  -{\mathcal{V}^{\mu}}_\mu  \right) \omega- \mathcal{D}^{\mu}{_{\mu}}{^\alpha} \nabla_\alpha \omega \right)\boldsymbol{\chi}\right\rangle\\
& \quad + \int_{\partial\mathcal{W}}d^{p-1}\wt\sigma\sqrt{|h|}\left\langle\left( \left(\wt{\mathcal{T}}^{\mu}{_\mu}  + \wt{\mathcal{B}}^\mu{_\mu} - {\wt{\mathcal{V}}^{\mu}}{_\mu}  \right) \omega - {\wt{\mathcal{D}}^\mu{_\mu}{^\alpha}}  \nabla_\alpha \omega \right)\boldsymbol{\chi}\right\rangle=0 \ ,
\end{split}
\ee
Notice that the trace ${\mathcal{T}^{\mu}}_\mu$ only runs over the tangential indices while the traces ${\mathcal{V}^{\mu}}_\mu$ and ${\mathcal{B}^{\mu}}_\mu$ only run over the transverse directions. An equivalent statement holds for the edge tensors. This expression can be compared to the analysis performed in \cite{Billo:2016cpy}. Using the second identity in \eqref{eq:dbp} and ignoring edge contributions one finds
\be 
\begin{split}
\delta_\omega\left\langle\boldsymbol{\chi}\right\rangle&+ \int_{\mathcal{M}}d^Dx\sqrt{|g|}\left\langle T_\text{b}{^{\mu}{_\mu}}\boldsymbol{\chi}\right\rangle \omega\\ 
&+\int_{\mathcal{W}}d^p\sigma\sqrt{|\gamma|}\left\langle\left( \left( \mathcal{T}^{\mu}{_\mu} +{\mathcal{D}^{\mu\nu}}_\rho\Kddu\mu\nu\rho +\mathcal{E}^{\mu\nu}\pdd\mu\nu \right) \omega - \mathcal{D}^{\mu}{_{\mu}}{^\alpha} \nabla_\alpha \omega \right)\boldsymbol{\chi}\right\rangle=0\ ,
\end{split}
\ee
which agrees with Eq.(5.14) of \cite{Billo:2016cpy} only if $\mathcal{E}^{\mu\nu}\pdd\mu\nu=-{\mathcal{V}^\mu}_\mu$, or alternatively if the constraint \eqref{eq:bt} is satisfied. However, this is not an a priori requirement.

\subsection{Ward identities for a class of BCFT actions}
We now consider a similar class of BCFTs and write explicitly the Ward identities. This case is comparatively different than the case of DCFTs. The class of BCFTs studied here is broader than that in \cite{McAvity:1993ue}, in particular, it includes couplings to arbitrary non-geometric fields. Consider the variation of a correlator under some unspecified symmetry. As previously, this variation can be written as
\be \label{eq:varBCFT}
\begin{split}
&\delta\left\langle\boldsymbol{\chi}\right\rangle+\frac{1}{2}\int_{\mathcal{M}}d^Dx\sqrt{|g|}\Theta(x)\left\langle T^{\mu\nu}_{\text{b}}\boldsymbol{\chi}\right\rangle\delta \Gdd\mu\nu \\
&+\int_{\mathcal{W}}d^p\sigma\sqrt{|\gamma|}\left\langle\Big(-\frac{1}{2}\mathcal{T}_{\mu\nu} \, \delta_\xi \huu\mu\nu+\frac{1}{2}\mathcal{B}^{\mu\nu} \, \delta_\xi\! \pdd\mu\nu +{\mathcal{D}^{\mu\nu}}_{\rho} \, \delta_\xi\Kddu\mu\nu\rho+\mathcal{D}_\mu\delta X^\mu \right. \\
&~~~~~~~~~~~~~~~~~~~~~~~~~~~\left. +{\mathcal{V}_{\mu}}\delta {n^{\mu}}+{\mathcal{C}^{\mu}}_\nu\delta {e_{\mu}}^{\nu}\Big)\boldsymbol{\chi}\right\rangle=0 \ .\\
\end{split}
\ee
As there is only one normal vector to the CFT boundary $n^\mu$ there is no spin current, contrary to \eqref{eq:varDCFT}. Furthermore, note the appearance of the function $\Theta(x)$ in the bulk contribution. We now consider the Ward identities associated with this class of BCFTs.

\paragraph{Ward identity for tangential diffeomorphisms.} Under a tangential diffeomorphism that does not displace the mapping functions, the first Ward identity in \eqref{eq:ward2} reads
\be \label{eq:difftang1}
\begin{split}
\delta_{\xi^{||}}\left\langle\boldsymbol{\chi}\right\rangle&-\int_{\mathcal{M}}\sqrt{|g|}d^{D}x\Theta(x)\left\langle\nabla_{\mu}T^{\mu\nu}_{\text{b}}\boldsymbol{\chi}\right\rangle\xi_\nu^{||}\\
&+\int_{\mathcal{W}}\sqrt{|\gamma|}d^p\sigma\left(\left\langle B^{\mu\nu}\pud\lambda\mu\boldsymbol{\chi}\right\rangle\nabla_\lambda\xi_\nu^{||}+\left\langle B^{\nu}\boldsymbol{\chi}\right\rangle \xi_\nu^{||}\right)=0 \ .
\end{split}
\ee
Contrary to the case of a DCFT, $\nabla_{\mu}T^{\mu\nu}_{\text{b}}$ does not have an expansion in terms of derivatives of $\widehat{\delta}(x)$, therefore, in the absence of external operators one is lead to the results obtained for an interface in Sec.~\ref{sec:interfaces}. In particular, from \eqref{eq:dbp} we have that
\be \label{eq:cBCFT}
\begin{split}
\mathcal{B}^{\alpha\sigma}=\mathcal{D}^{\mu\lambda(\sigma}\Kddu\mu\lambda{\alpha)}+\mathcal{V}_\mu n^\mu \puu\alpha\sigma \ , \quad \mathcal{V}_\mu\hud\mu\alpha=0 \ ,
\end{split}
\ee
and from \eqref{eq:interfull} that
\be \label{eq:wardB1}
\hud\nu\sigma\nabla_{\mu}T^{\mu\sigma}_{\text{b}}=0 \ , \quad \hud\nu\sigma\overline\nabla_\lambda\mathbb{T}^{\lambda\sigma}-\hud\nu\sigma\Sigma^{\mu\lambda\rho}\Buddd\sigma\mu\rho\lambda=n_\lambda T^{\lambda\sigma}_{\text{b}}\hud\nu\sigma \ ,
\ee
with $\mathbb{T}^{\lambda\sigma}$ and $\Sigma^{\mu\lambda\rho}$ defined in \eqref{eq:truestress}. It is worth noting that, as in Sec.~\ref{sec:interfaces}, there may be contributions to the surface operators such as $\mathcal{T}^{\mu\nu}$ due to an inflow from the bulk.

\paragraph{Ward identity for tangential displacements of $X^{\mu}$.} We now consider the second Ward identity in \eqref{eq:ward2} by displacing tangentially the mapping functions. One obtains
\be
\begin{split}
& \left\langle \left(\mathcal{D}^\nu\hud\sigma\nu-\left({\mathcal{C}^{\mu}}_{\nu}-{\mathbb{N}^{\mu}}_{\nu}\right)\Gamma^{\nu}_{\mu\lambda}\huu\lambda\sigma-\nabla_\lambda\left(\mathcal{C}^{\mu\nu}-\mathbb{N}^{\mu\nu}\right)\hud\sigma\nu\right)\boldsymbol{\chi}
\right\rangle\\
& = \left\langle\left(\hud\sigma\nu\overline\nabla_\lambda\mathbb{T}^{\lambda\nu}-\hud\sigma\nu\Sigma^{\mu\lambda\rho}\Buddd\nu\mu\rho\lambda-n_\lambda T^{\lambda\nu}_{\text{b}}\hud\sigma\nu\right)\boldsymbol{\chi}\right\rangle=0 \ ,
\end{split}
\ee
where $\mathbb{N}^{\mu\nu}$ was given in \eqref{eq:defs} but now with vanishing spin current and the last equality follows due to \eqref{eq:wardB1}. The quite simpler form of this Ward identity compared to a DCFT is due to the fact that the constraints \eqref{eq:cBCFT} are generally valid. Furthermore note that the l.h.s. is not manifestly covariant but this is just an artefact of working with variations of the embedding map.

\paragraph{Ward identity for diffeomorphism invariance.} Turning to the Ward identity for diffeomorphism invariance \eqref{eq:ward11} we find 
\be
\begin{split}
&\delta_{\xi}\left\langle\boldsymbol{\chi}\right\rangle-\int_{\mathcal{M}}\sqrt{|g|}\Theta(x)d^D x\left\langle\nabla_\mu T^{\mu\nu}_{\text{b}}\boldsymbol{\chi}\right\rangle\xi_\nu\\
&-\int_{\mathcal{M}}\sqrt{|g|}d^Dx\widehat{\delta}(x)\left\langle\left(\mathcal{D}^\nu\hud\sigma\nu-\nabla_\lambda\left(\mathcal{C}^{\mu\nu}-\mathbb{N}^{\mu\nu}\right)\hud\sigma\nu-\left({\mathcal{C}^{\mu}}_{\nu}-{\mathbb{N}^{\mu}}_{\nu}\right)\Gamma^{\nu}_{\mu\lambda}\huu\lambda\sigma\right)\boldsymbol{\chi}\right\rangle\xi_\sigma=0 \ ,
\end{split}
\ee
which, in the absence of external operators leads to the two separate requirements 
\be
\left(\nabla_\mu T^{\mu\nu}_{\text{b}}\right)\Theta(x)=0 \ , \quad \left(\mathcal{D}^\sigma-\nabla_\lambda\left(\mathcal{C}^{\mu\sigma}-\mathbb{N}^{\mu\sigma}\right)-\left({\mathcal{C}^{\mu}}_{\nu}-{\mathbb{N}^{\mu}}_{\nu}\right)\Gamma^{\nu}_{\mu\lambda}\Guu\lambda\sigma\right)\widehat{\delta}(x)=0 \ .
\ee
The last condition, in particular, states that the non-manifestly covariant combination of terms must vanish for arbitrary spacetime directions.

\paragraph{Ward identity for Weyl transformations.} The Ward identity for Weyl transformations is just a simple addition of the bulk term to the Weyl transformation for codimension $n=1$ surfaces studied in Sec.~\ref{sec.weylsecondorder}. It reads
\be
\begin{split}
&\delta_\omega\left\langle\boldsymbol{\chi}\right\rangle+\int_{\mathcal{M}}d^Dx\sqrt{|g|}\Theta(x)\left\langle T_{\text{b}}{^\mu}{_\mu} \omega \boldsymbol{\chi}\right\rangle \\
&\qquad +\int_{\mathcal{W}}d^p\sigma\sqrt{|\gamma|}\left\langle \left( \left(\mathcal{T}^\mu{_\mu} + {\mathcal{D}^{\mu\nu}}_\rho\Kddu\mu\nu\rho  \right) \omega  - {\mathcal{D}^\mu{_\mu}}{^\alpha} \nabla_\alpha \omega  \right) \boldsymbol{\chi} \right\rangle=0 \ ,
\end{split}
\ee
where we have used \eqref{eq:cBCFT}.

\subsection{Spacetime stress tensor and displacement operator}
From the variational principle \eqref{eq:varDCFT}, one can easily extract the spacetime stress tensor and the displacement operator by making use of the delta function $\widehat{\delta}(x)$. In particular we obtain the stress tensor for the bulk and defect\footnote{For the special case $\mathbb{S}^{\mu\nu}=\mathbb{B}^{\mu\nu}=\mathbb{P}^{\mu\nu}=0$, this stress tensor is different than the ad-hoc stress tensor introduced in Eq.(5.21) of \cite{Billo:2016cpy} . In particular the authors of \cite{Billo:2016cpy} missed the dipole contribution $2\mathcal{S}^{(\mu\nu)\rho}$ in \eqref{eq:fullstress}. In turn, this lead them to postulate a Ward identity in the form of Eq.(5.22) of \cite{Billo:2016cpy} which is not generally valid neither physically meaningful.}
\be \label{eq:fullstress}
\begin{split}
\Tuu\mu\nu=&T^{\mu\nu}_{\text{b}}+\left[\mathcal{T}^{\mu\nu}+2 \, {\mathcal{P}^{\lambda}}{_\rho}\puu\rho{(\mu}\hdu\lambda{\nu)}+\mathcal{B}^{\mu\nu}-\pud{(\mu}\lambda\mathcal{V}^{\nu)\lambda}-2\huu\lambda{(\mu}{\mathcal{V}^{\nu)}}_\lambda\right]\widehat\delta(x) \\
&-\nabla_\rho\left[\left(2\mathcal{D}^{\rho(\mu\nu)}-\mathcal{D}^{\mu\nu\rho}+2\mathcal{S}^{(\mu\nu)\rho}\right)\widehat\delta(x)\right] \ .
\end{split}
\ee
In turn, for a BCFT, the stress tensor on the boundary is \eqref{eq.interfaceinflow}. This provides a formal derivation of the contact terms in the stress tensor for a DCFT that was in general lacking in the literature.

The displacement operator can be derived by making an arbitrary variation of the embedding map. In general, there will be tangential derivatives of the variation of the embedding map but these can always be integrated out. Performing this variation leads to the displacement operator
\be
\begin{split}
\textbf{D}^{\sigma}=&\Big[\left(\overline\nabla_\lambda\mathbb{T}^{\lambda\sigma}-\Sigma^{\mu\lambda\rho}\Buddd\sigma\mu\rho\lambda\right)-\overline\nabla_\lambda\left({\mathcal{E}^{\mu}}_\nu\hud\lambda\mu\puu\nu\sigma+\hud\lambda\nu\mathcal{V}^{\mu\nu}\pud\sigma\mu\right)\\
&-\left(\mathcal{D}^{\sigma}+g^{\alpha\sigma}\left({\mathbb{S}^{\mu}}_{\nu}+{\mathbb{B}^{\mu}}_{\nu}+{\mathbb{N}^{\mu}}_\nu-{\mathbb{P}^{\mu}}_\nu-{\mathcal{C}^{\mu}}_\nu-{\mathcal{V}^{\mu}}_\nu\right)\Gamma^{\nu}_{\mu\alpha}\right)\\
&-\left(\overline\nabla_\lambda\left({\mathbb{N}^{\lambda}}_\nu-{\mathbb{P}^{\lambda}}_\nu-{\mathcal{C}^{\lambda}}_\nu\right)\Guu\sigma\nu\right)\Big]\widehat{\delta}(x) \ ,\end{split}
\ee
while for a BCFT, the displacement operator takes a simpler form
\be
\begin{split}
\textbf{D}^{\sigma}=&\Big[\left(\overline\nabla_\lambda\mathbb{T}^{\lambda\sigma}-\Sigma^{\mu\lambda\rho}\Buddd\sigma\mu\rho\lambda-n_\lambda T^{\lambda\sigma}_{\text{b}}\right)\\
&-\left(\mathcal{D}^\sigma-\nabla_\lambda\left(\mathcal{C}^{\mu\sigma}-\mathbb{N}^{\mu\sigma}\right)-\left({\mathcal{C}^{\mu}}_{\nu}-{\mathbb{N}^{\mu}}_{\nu}\right)\Gamma^{\nu}_{\mu\lambda}\Guu\lambda\sigma\right)\Big]\widehat{\delta}(x) \ .
\end{split}
\ee
The form of the stress tensor and displacement operator provided here satisfy the Ward identity \eqref{eq:ward1}, given the constraints \eqref{eq:dbp}.




\section{Discussion} \label{sec:discussion}
In this paper we have introduced a new type of variational principle for (entangling) surfaces, interfaces and BCFTs/DCFTs. This variational principle is based on Carter's spacetime formulation of surface actions and the concept of Lagrangian variations \cite{Carter:1993wy, Battye:1995hv, Carter:1997pb}, for which the background coordinates are displaced but the embedding map is kept fixed, as in the case of (entangling) surfaces. In the case of actions for BCFTs/DCFTs, it is required to consider both Lagrangian variations and variations of the embedding map simultaneously, since one wishes to couple defects/boundaries to CFTs regardless of their shape. This led us to extend the variational calculus within this spacetime formulation  in two different directions: on the one hand, we give explicit variational formulae for Lagrangian variations of many geometric structures of interest while on the other, we provide variational formulae for variations of the embedding map of the same geometric structures.\footnote{As far as we are aware, variations of the embedding map within this spacetime formulation had not been carried out earlier.} In particular, Lagrangian variations are always manifestly spacetime covariant and only require defining geometric tensors on a single surface, rather than working with a foliation of such surfaces, even if just in a local neighbourhood. 

The variational principle introduced here encompasses all diffeomorphism constraints on surface/interface actions, as show in Sec.~\ref{sec:shapeequations}, in particular Eqs.\eqref{eq:st}-\eqref{eq:genshapeperp} and Eqs.~\eqref{eq:st1}-\eqref{eq:genshapeedge}. One of these constraints is the shape equation itself, describing the non-trivial dynamics of these surfaces. Others include a component of tangential diffeomorphism invariance \eqref{eq:genshapetang} and invariance under local rotations of the transverse background coordinates \eqref{eq:st}.\footnote{In a gauge formulation of the variational principle these two correspond to surface reparametrisation invariance and invariance under infinitesimal rotations of the normal vectors (see App.~\ref{app:highdev}). However, we have not investigated whether this statement also holds when taking into account non-trivial edges.} An additional set of constraints (Eqs.~\eqref{eq:bt}-\eqref{eq:pt}) describes the relations between certain components of the spacetime stress tensor associated with a given surface/interface, while others (Eq.~\eqref{eq:cnew}) are related to invariance under local Lorentz transformations on the surface and other local transformations. This last set of constraints is not captured within a gauge formulation of the variational principle, as show in App.~\ref{app:highdev}. A subset of these constraints allows to explicitly relate this spacetime formulation of the action with the multipole expansion of the stress tensor introduced in \cite{Vasilic:2007wp}.

These diffeomorphism constraints  impose restrictions on the type of terms that can contribute to surface actions. The naive expectation that, analogous to spacetime actions, contractions of tensor fields yield covariant scalars does not hold in the case of surface actions, as the diffeomorphism constraints impose stronger restrictions. These restrictions become intrinsically more complex when the surface has edges, such as in the case of an open string worldsheet with a point-particle attached to both its ends. Based on an enumeration of response coefficients, we analysed a two-derivative action with non-trivial edges in Sec.~\ref{sec:2derivative}. It was seen that requiring such action to be diffeomorphism invariant for all shape configurations (i.e. without imposing ad-hoc edge conditions that restrict the edge dynamics) imposed relations between surface and edge response coefficients. In particular, if the surface has edges then an arbitrary combination of extrinsic curvature invariants is not allowed. Diffeomorphism invariance at the edges requires that a particular combination of extrinsic curvatures to be equal to the intrinsic Ricci scalar plus a tangential contraction with the background Riemann tensor. In turn, these considerations implied that the presence of edges restricts considerably possible conformal anomalies for two-dimensional submanifolds, as seen in Sec.~\ref{sec.weylsecondorder}. It would be interesting to investigate whether these constraints also impose any restrictions on actions in the context of renormalised entanglement entropy \cite{Taylor:2016aoi}.

Analogous to classification procedures inspired by effective field theory methods employed in the context of hydrodynamics to constraint constitutive relations, it was shown that similar classification procedures can be employed to constraint the stress tensor of (entangling) surfaces at a given order in derivatives. This leads to non-trivial constraints among the response coefficients and the coefficients appearing in the surface/edge stress tensor. It would be interesting to extend this work by including couplings to background Killing vectors and worldvolume Killing vector fields. This direction has been pursued to a certain extent in e.g. \cite{Armas:2013goa, Armas:2015ssd, Armas:2016xxg} but in light of this new variational principle it would be interesting to revisit these constructions and to push them to higher levels of generality, such as including arbitrary couplings to derivatives of the Killing vector fields.

In the context of BCFTs/DCFTs, this new variational principle leads to several constraints among the bulk stress tensor and the spacetime surface/edge stress tensor as shown in Sec.~\ref{sec:DCFTs}. Some of these constraints had been previously overlooked in the literature (see Eq.~\eqref{eq:dbp}). A formal derivation of the contact terms in the spacetime stress tensor and a formal derivation of the displacement operator in curved spacetimes was given. Ad-hoc constructions of this spacetime stress tensor were present in earlier literature and turn out not to be fully correct. Partly, the reason for this supervenes on the usage of a gauge formulation of the variational principle. While this form is more suitable to deal with  variations of intrinsic quantities, it is less suitable to deal with the variations of background quantities and the existence of a bulk, for which a split between normal and tangential components is meaningless unless a foliation of surfaces is introduced. The correct form of the spacetime stress tensor and displacement operator can now be used to evaluate correlation functions in a broad class of BCFT/DCFT actions and to constrain CFT data as in \cite{Billo:2016cpy}. 

The main purpose of this work was to develop a spacetime framework to deal with a broad range of systems, however, we have only dealt with actions or vacuum energy functionals that can be treated as functionals of geometric fields only, namely, $\Gdd\mu\nu(x)$, $X^\mu(\sigma)$ and $\wt X^\mu(\sigma)$. It would be interesting to apply the same methods to actions that are functionals of other background fields such as vector fields, gauge fields and scalar fields. This extension is pertinent in the context of black holes and the coupling of probe branes to supergravity \cite{Armas:2016mes} as well as in the context of entangling surfaces in theories with gauge or scalar fields (see e.g. \cite{Azeyanagi:2015uoa, Caceres:2017lbr}). The existence of symmetries, such as gauge symmetries, will impose further restrictions on the type of couplings that can appear on surface actions. Furthermore, we have stream-lined a method for obtaining the surface diffeomorphism constraints in full generality for an arbitrary number of derivatives in terms of frame-invariant tensors in App.~\ref{app:highdev}. It would be interesting to obtain these constraints explicitly at higher orders following this procedure. We plan on addressing some of these problems in future work.


\section*{Acknowledgements}
Part of this work was based on the course \textbf{Effective theories for the dynamics of interfaces: from biophysical membranes to black holes} given at KUL in February 2017 by JA. JA would like to thank Thomas Van Riet for the invitation as well as the students that participated in the course for their critical questions. We would like to thank Tatsuo Azeyanagi, Matthias Blau, Joan Camps and Jemal Guven for useful discussions and we thank Joan Camps and Niels Obers for comments on a earlier draft of this manuscript. We also thank Bjarke T. Nielsen for checking some of the equations in this paper. The work of JA is supported by the ERC Starting Grant 335146 HoloBHC. JA would like to thank Geoffrey Comp\`{e}re for his support. JT is supported by the Advanced ARC project ``Holography, Gauge Theories and Quantum Gravity'' and by the Belgian Fonds National de la Recherche Scientifique FNRS (convention IISN 4.4503.15).
\appendix

\section{Variational calculus for submanifolds}\label{app:variations}
In this appendix we review the variational calculus of submanifolds. We focus on Lagrangian variations that were studied in some detail in \cite{Carter:1993wy, Carter:1997pb} and we generalise it to several geometric structures. The variations of the same geometric quantities under changes of the embedding map is also given and, as far as we aware, it is the first time that they are properly addressed in a spacetime formulation. Furthermore, we also present the transformation properties of many geometric structures under Weyl rescalings. 


\subsection{Lagrangian variations} 
For a given submanifold for which one has chosen the adapted frame where the background metric evaluated at the submanifold can be written as $\Gdd\mu\nu=\hdd\mu\nu+\pdd\mu\nu$, the tangent and normal vectors obey the conditions\footnote{One may also work with a time-like transverse space by replacing $\delta^{ij}\to\eta^{ij}$. See footnote \ref{foot:O}.}
\be \label{eq:conditions}
\euu\mu a \edu\mu b=\gamma^{ab} \ , \quad \nuu \mu i \ndu \mu j=\delta^{ij} \ , \quad \eud\mu a \ndu \mu i=0 \ .
\ee
When performing a variation such that the tangent and normal vectors change according to
\be
\eud\mu a\to {\widehat {e}^{\mu}}_ a= \eud\mu a+\delta \eud\mu a \ , \quad \nud\mu i\to {\widehat {n}^{\mu}}_ i= \nud\mu i+\delta \nud\mu i \ ,
\ee
the new tangent and normal vectors ${\widehat {e}^{\mu}}_ a$ and ${\widehat {n}^{\mu}}_ i$ must continue to obey \eqref{eq:conditions}. For a general variation where the background metric may also vary, one obtains from \eqref{eq:conditions} the relations
\be 
\delta \nuu \mu i\ndd \mu i=-\frac{1}{2}\nuu\mu i\nud\nu i\delta \Gdd\mu\nu \ , \quad \delta \nuu\mu i \edd\mu a=-\nuu\mu i\eud\nu a\delta \Gdd\mu\nu-\ndu\mu i\delta\eud\mu a \ ,
\ee
and hence in general one finds
\be\label{eq:varnupL}
\begin{split}
&\delta \nuu \mu i=-\frac{1}{2}\puu\mu\lambda\nuu\rho i\delta\Gdd\lambda\rho+{\omega^{i}}_j\nuu\mu j-\left(\nuu\lambda i\eud\nu a\delta \Gdd\lambda\nu+\ndu\lambda i\delta\eud\lambda a\right)\euu\mu a \ ,\\
&\delta \ndu \mu i=\frac{1}{2}\pud\lambda\mu\nuu\rho i\delta \Gdd\lambda\rho+{\omega^{i}}_j\ndu\mu j-\ndu\lambda i\edu\mu a\delta \eud\lambda a \ ,
\end{split}
\ee
for some anti-symmetric matrix ${\omega^{i}}_j$ in O($n$). The term ${\omega^{i}}_j\nuu\mu j$ expresses the freedom of rotating the normal vectors and still satisfying \eqref{eq:conditions}. 

We now focus on a specific class of variations, which we refer to as Lagrangian variations and denote by $\delta_\xi$. In this case, $\delta_\xi \Gdd\mu\nu=2\nabla_{(\mu}\xi_{\nu)}$, ${\omega^{i}}_j=0$ and the embedding functions remain fixed $\delta_\xi X^{\mu}=0$ so that $\delta_\xi \eud\mu a=0$. Using the fact that
\be \label{eq:varinduced}
\delta_\xi\gamma_{ab}=\eud\mu a\eud \mu b\delta_\xi\Gdd\mu\nu \ , \quad \delta_\xi \gamma^{ab}=-\gamma^{ac}\gamma^{db}\delta_\xi \gamma_{cd} \ ,
\ee
one obtains the variational formulae 
\be\label{eq:mvar}
\delta_\xi \huu\mu\nu=-\huu\mu\lambda\huu\nu\rho\delta_\xi\Gdd\lambda\rho \ , \quad \delta_\xi \pdd\mu\nu=\pud\lambda\mu\pud\rho\nu\delta_\xi\Gdd\lambda\rho \ , \quad \delta_\xi\pud\mu\nu=-\huu\mu\lambda\pud\rho\nu\delta_\xi\Gdd\lambda\rho \ ,
\ee
and also $\delta_\xi\hud\mu\nu=-\delta_\xi\pud\mu\nu$. The metric variations \eqref{eq:mvar} form a complete set of Lagrangian variations of $\Gdd\mu\nu$ as they account for the different tangential and orthogonal projections of $\delta_\xi\Gdd\mu\nu$. For completeness, we also have that
\be
\begin{split}
&\delta_\xi\hdd\mu\nu=-\hdd\alpha\mu\hdd\beta\nu\delta_\xi\huu\alpha\beta-2\hdd\lambda{(\mu}\pdu{\nu)}\rho\delta_\xi\pud\lambda\rho ~~, \\
&\delta_\xi\puu\mu\nu=-\puu\mu\alpha\puu\nu\rho\delta_\xi\pdd\alpha\rho+2\puu\lambda{(\mu}\hud{\nu)}\alpha\delta_\xi\pud\alpha\lambda \ .
\end{split}
\ee
Given this, one may now evaluate the variations of the extrinsic curvature, leading to
\be
\delta_\xi \Kddu\mu\nu\rho=\left[2\pud\lambda{\mu}\Kduu{\nu}\alpha\rho-\huu\lambda\rho\Kddu\mu\nu\alpha\right]\delta_\xi\Gdd\lambda\alpha+\pud\rho\sigma\hud\lambda\mu\hud\alpha\nu\delta_\xi\Gamma^\sigma_{\lambda\alpha} \ ,
\ee
where
\be
\delta_\xi\Gamma^\sigma_{\mu\alpha}=\frac{1}{2}\Guu\sigma\nu\left(\nabla_\alpha\delta_\xi\Gdd\mu\nu+\nabla_\mu\delta_\xi\Gdd\alpha\nu-\nabla_\nu\delta_\xi\Gdd\mu\alpha\right)=\nabla_{(\mu}\nabla_{\alpha)}\xi^\sigma-\Buddd\sigma{(\mu}{\alpha)}\rho\xi^\rho \ .
\ee
The Lagrangian variation of the normal vectors is also of interest. Using \eqref{eq:varnupL}, we find that 
\be \label{eq:varnupLL}
\delta_\xi \nud\mu i=-\nud\rho i\nabla_\rho \xi^\mu-\nud\rho i\huu\mu\lambda\nabla_\lambda \xi_\rho-\nud{[\rho}i\puu{\lambda]}\mu\nabla_\lambda\xi_\rho \ .
\ee
In turn, for the external rotation tensor we obtain
\be \label{eq:varext}
\begin{split}
\delta_\xi \oduu\mu\nu\rho=&\left[2\oduu\mu\alpha{[\nu}\huu{\rho]}\lambda+\pud\lambda\mu\ouuu\alpha\nu\rho+\oduu\mu\alpha{[\nu}\puu{\rho]}\lambda-\Kduu\mu\lambda{[\nu}\puu{\rho]}\alpha\right]\delta_\xi \Gdd\lambda\alpha +\pud{[\nu}\sigma\puu{\rho]}\alpha\hud\kappa\mu\delta_\xi\Gamma^{\sigma}_{\kappa\alpha} \ .
\end{split}
\ee
It is useful to write down the contractions
\be \label{eq:defsvards}
\begin{split}
&{\mathcal{D}^{\mu\nu}}_{\rho}\delta_\xi \Kddu\mu\nu\rho={\mathcal{D}^{\mu\nu}}_{\rho}\delta_\xi\Gamma^\rho_{\mu\nu} \ ,\\
&{\mathcal{S}^{\mu}}_{\nu\rho}\delta_\xi \oduu\mu\nu\rho={{\mathcal{S}^{\mu}}_\lambda}^{\sigma}\oduu\mu\alpha\nu\delta_\xi\pdd\sigma\alpha+{{{\mathcal{S}^{\mu}}}_{\lambda}}^{\sigma}\Kddu\mu\alpha\lambda\delta_\xi\pud\alpha\sigma+\mathcal{S}^{\mu\lambda\rho}\left(\Buddd\sigma\mu\lambda\rho\xi_\sigma+\overline\nabla_\mu\nabla_\rho\xi_\lambda\right) \ ,
\end{split}
\ee
and to define the variation $\wt \delta_\xi$ 
\be \label{eq:contracspin}
{\mathcal{S}^{\mu}}_{\nu\rho}\wt\delta_\xi \oduu\mu\nu\rho={\mathcal{S}^{\mu}}_{\nu\rho}\delta_\xi \oduu\mu\nu\rho-{{\mathcal{S}^{\mu}}_\lambda}^{\sigma}\oduu\mu\alpha\lambda\delta_\xi\pdd\sigma\alpha-{{{\mathcal{S}^{\mu}}}_{\lambda}}^{\sigma}\Kddu\mu\alpha\lambda\delta_\xi\pud\alpha\sigma={{\mathcal{S}^{\mu}}_{\nu}}^{\rho}\delta_\xi\Gamma^{\nu}_{\mu\rho} \ ,
\ee
where we have extracted the transverse components of the metric variations from the variation of the external rotation tensor. The reason for this, is that the transverse components of the metric variations are included in the definitions of $\mathcal{P}^{\mu\nu}$ and $\mathcal{B}^{\mu\nu}$ in \eqref{eq:def1a}. In order to consider variations of the intrinsic Riemann tensor, it is useful to obtain the variation of the internal rotation tensor, which reads
\be \label{eq:varint}
\delta_\xi{{\rho_\mu}^{\nu}}{_\rho}=\left(\huu\nu\alpha\hud\sigma\rho\Kddu\mu\sigma\beta+{\rho_\mu}^{\nu\alpha}\pud\beta\rho+\pud\beta\mu{\rho^{\alpha\nu}}_\rho\right)\delta_\xi \Gdd\alpha\beta+\hud\nu\sigma\hud\lambda\rho\hud\alpha\mu\delta_\xi\Gamma^{\sigma}_{\alpha\lambda} \ .
\ee
As for the variation of the external rotation tensor, we define the variation 
\be
\wt \delta_\xi{{\rho_\mu}^{\nu}}{_\rho}=\hud\nu\sigma\hud\lambda\rho\hud\alpha\mu\delta_\xi\Gamma^{\sigma}_{\alpha\lambda} \ ,
\ee
where we have stripped away the transverse components of the metric variations. Since the variation of the intrinsic rotation tensor is purely tangential, the effect on the spacetime stress tensor is to add dipole terms that can be removed by a choice of frame. It is also useful to consider the variation of the background Riemann tensor, Ricci tensor and Ricci scalar
\be
\begin{split}
&\delta_\xi\Buddd\mu\lambda\nu\rho=\nabla_\nu\delta_\xi\Gamma^{\mu}_{\lambda\rho}-\nabla_\rho\delta_\xi\Gamma^{\mu}_{\lambda\nu} \ , \quad \delta_\xi R_{\mu\nu}=\nabla_\lambda\delta_\xi\Gamma^{\lambda}_{\mu\nu}-\nabla_\nu\delta_\xi\Gamma^{\lambda}_{\mu\lambda} \ ,\\
&\delta_\xi R=\nabla_\mu\left(g^{\nu\lambda}\delta_\xi\Gamma_{\nu\lambda}^{\mu}\right)-\nabla^{\mu}\delta_\xi\Gamma_{\mu\lambda}^{\lambda}-R^{\mu\nu}\delta g_{\mu\nu} \ .
\end{split}
\ee
The Lagrangian variations of the intrinsic Riemann tensor are also of interest. A lengthy calculation reveals that
\be \label{eq:varrint}
\delta_\xi \Ruddd\mu\nu\kappa\lambda=2\hud\mu\sigma\hud\tau\nu\hud\pi{[\lambda}\overline\nabla_{\kappa]}\delta_\xi {{\rho_\pi}^{\sigma}}_\tau+\left(\delta \hud\mu\nu ~\text{terms}\right) \ ,
\ee
for which the last terms can be removed, since they are already included in the transverse monopole components of the stress tensor. Using the variation \eqref{eq:varint} into \eqref{eq:varrint} we remove the additional transverse components of the metric variations but keep the tangential components of the metric variations. Therefore, we define the independent variation
\be
\begin{split}
\wt \delta_\xi \Ruddd\mu\nu\kappa\lambda=&2\huu\mu\alpha\Kddu\nu{[\kappa}\rho\Kdud{\lambda]}\beta\rho\delta_\xi \Gdd\alpha\beta+2\huu\mu\alpha\Kddu\nu{[\lambda}\beta\overline\nabla_{\kappa]}\delta_\xi \Gdd\alpha\beta \\
&+ 2\hud\mu\sigma\hud\tau\nu\hud\pi{[\lambda}\overline\nabla_{\kappa]}\left(\hud\sigma\alpha\hud\rho\tau\hud\beta\pi\delta_\xi\Gamma^{\alpha}_{\beta\rho}\right) \ .
\end{split}
\ee
In the core of this paper we the definition of $\mathcal{T}_{\mu\nu}$ as the tangential components of the monopole part of the stress tensor coincided with that which is obtained by direct variation with respect to $\huu\mu\nu$. However, the variation of the intrinsic Riemann tensor gives new contributions to these tangential components. It is convenient to keep the definition of $\mathcal{T}_{\mu\nu}$ as that which is obtained by direct variation. Therefore, additional tangential components will appear in the spacetime stress tensor, as it will be seen in App.~\ref{app:highdev}. 

Similarly, it is also useful to consider Lagrangian variations of the outer curvature, which take the form
\be
\delta_\xi{\Omega^{\mu}}_{\nu\kappa\lambda}=2\pud\mu\sigma\pud\tau\nu\hud\pi{[\lambda}\overline\nabla_{\kappa]}\delta_\xi \odud\pi\sigma\tau+\left(\delta \hud\mu\nu ~\text{terms}\right)+\left(\delta \pud\mu\nu ~\text{terms}\right) \ .
\ee
Removing the transverse components of the metric variations and using \eqref{eq:varext}, the shifted variation takes  form
\be \label{eq:varouter}
\begin{split}
\wt\delta_\xi{\Omega^{\mu}}_{\nu\kappa\lambda}=&2\pud\mu\sigma\pdd\tau\nu\Kduu{[\lambda}\beta{[\sigma}\Kduu{\kappa]}{|\alpha|}{\tau]}\delta_\xi \Gdd\alpha\beta-2\pud\alpha\rho\Kduu{[\lambda}\beta{[\mu}\pud{\rho]}{|\nu|}\overline\nabla_{\kappa]}\delta_\xi\Gdd\alpha\beta\\
&+2\pud\mu\sigma\pdd\tau\nu\hud\pi{[\lambda}\overline\nabla_{\kappa]}\left(\pud{[\sigma}\beta\puu{\tau]}\alpha\hud\theta\pi\delta_\xi\Gamma^{\beta}_{\theta\alpha}\right) \ .
\end{split}
\ee
Analogously to variations of the intrinsic Riemann tensor, variations of the outer curvature will also induce new contributions to the tangential components of the monopole part of the stress tensor.

Finally, we note that variations of the edge geometry take the same form as for the surface geometry but with the surface projectors and tensor structures replaced by the edge projectors and tensor structures.

\subsection{Variations of the embedding map}
In order to obtain the equations of motion for a given surface action, one may perform an infinitesimal variation of the embedding map according to
\be
X^{\mu}(\sigma)\to X^{\mu}(\sigma)-\xi^{\mu}(\sigma) \ ,
\ee
though in general it will not lead to manifestly covariant variations. Under this transformation the background metric evaluated at the surface $\Gdd\mu\nu(X)$ and the tangent vectors vary as
\be \label{eq:varemb}
\delta_X \Gdd\mu\nu=-\xi^\alpha\partial_\alpha \Gdd\mu\nu \ , \quad \delta_X \eud\mu a=-\partial_a\xi^{\mu} \ ,
\ee
and $\omega_{ij}=0$ in \eqref{eq:varnupL}. It is easier to work with a gauge formulation to calculate these variations and then transcribe them into a spacetime formulation. Under an arbitrary variation of the induced metric one finds
\be
\delta \gamma_{ab}=\eud \mu a \eud\nu b\delta \Gdd\mu\nu+2\edd\mu{(a}\delta \eud\mu{b)} \ ,
\ee
and, hence, specialising to variations of the embedding map \eqref{eq:varemb} one gets
\be
\delta_X \gamma_{ab}=-2\eud\mu a\eud \nu b\nabla_{(\mu}\xi_{\nu)} \ , \quad \delta_X\huu\mu\nu=2\huu\mu\lambda\huu\nu\rho\nabla_{(\lambda}\xi_{\rho)}+2\euu a{(\mu}\delta_X\eud{\nu)}a \ .
\ee
Clearly, for a diffeomorphism for which the mapping functions are allowed to vary one finds, using \eqref{eq:varinduced}, that $\delta_\xi \gamma_{ab}+\delta_X \gamma_{ab}=0$. In turn, for variations of the normal vectors we find
\be \label{eq:varnup}
\delta_X \nud\mu i=\nud\rho i\nabla_\rho \xi^\mu+\ndd\lambda i\huu\mu\alpha\nabla_\alpha\xi^\lambda-\nud\rho i\partial_\rho \xi^\mu-\widehat{\omega}_{ij}\nuu\mu j \ ,
\ee
where ${\widehat{\omega}_j}^{~i}$ is the anti-symmetric matrix ${\widehat{\omega}_j}^{~i}=\nuu\mu i\nud\alpha j\xi^\lambda \partial_{[\alpha}\Gdd{\mu]}\lambda$. Using the corresponding Lagrangian variation \eqref{eq:varnupLL}, one obtains
\be
\delta_\xi \nud\mu i+\delta_X \nud\mu i=-\nud\rho i\partial_\rho \xi^\mu+\dot{\omega}_{ij}\nuu\mu j \ ,
\ee
for the anti-symmetric matrix ${\dot{\omega}_j}^{~i}=\nud\alpha j \ndu{[\mu} i\partial_{\alpha]}\xi^\mu$. Therefore, under diffeomorphisms that also displace the embedding map, the normal vectors $\nud\mu i$ transform as vectors, up to a transverse rotation.\footnote{For completeness, one also has that $\delta_X \ndu\mu i=\Gdd\mu\nu\delta_X\nuu\nu i-\nuu\nu i\xi^\alpha\partial_\alpha\Gdd\mu\nu$ and that $\delta_X \edu\mu a=\Gdd\mu\nu\delta_X \euu\nu a-\euu\nu a\xi^{\alpha}\Gdd\mu\nu$.}

For the extrinsic curvature we first introduce its alternative definition
\be
\Kddu ab \rho=\eud\mu a\eud\nu b\Kddu\mu\nu\rho=\nnb_a \eud\rho b \ ,
\ee
where $\nnb_a$ is the covariant derivative compatible with both metrics $\Gdd\mu\nu$ and $\hdd ab$ and acts on all indices $\mu, a, i$. The action of this covariant derivative is better expressed via the Weingarten decomposition
\be
\begin{split}
&\nnb_a \eud\rho b=\partial_a \eud\mu b-\gamma^{c}_{ab}\eud\mu c+\Gamma^{\mu}_{\lambda\alpha}\eud\lambda a\eud\alpha b=\nud\mu i\Kddu abi \ ,\\
&\nnb_a \nud\mu i=\partial_a \nud\mu i+\Gamma^{\mu}_{\lambda\alpha}\eud\lambda a \nud\alpha i+\oddu aij\nud\mu j=-\euu\mu b\Kddd abi \ ,
\end{split}
\ee
where $\gamma^{c}_{ab}$ is the Christoffel connection built from $\gamma_{ab}$. The external rotation tensor has been expressed as $\oddu aij=\eud\mu a\ndd\lambda i\ndu\rho j\oduu\mu\lambda\rho$. Using this, we find that for a general variation of the extrinsic curvature one gets
\be
\delta {K_{ab}}^{i}=\ndu\mu i\nnb_{(a}\delta \eud\mu{b)}+\ndu\mu i\eud\lambda a\eud\alpha b\delta \Gamma^{\mu}_{\lambda\alpha}+\ndu\mu i\delta \eud\lambda {(a}\eud\alpha{b)}\Gamma^{\mu}_{\lambda\alpha}+\nud\mu j{K_{ab}}^{j}\delta \ndu\mu i \ .
\ee
Hence, under a variation of the embedding map one obtains
\be \label{eq:varKx0}
\delta_X {K_{ab}}^{i}=-\ndu\mu i\nnb_{(a}\nnb_{b)}\xi^{\mu}+\ndu\mu i\xi^\lambda\eud\alpha{(a}\eud\sigma{b)}\Buddd\mu\alpha\sigma\lambda+{K_{ab}}^{j}{\widehat{\omega}_j}^{~i} \ ,
\ee
where, in deriving this, we have used that $\delta_X\Gamma^{\mu}_{\nu\lambda}=-\xi^{\alpha}\partial_\alpha\Gamma^{\mu}_{\nu\lambda}$. One may also evaluate $\delta_\xi  {K_{ab}}^{i}$ which leads to the result $\delta_\xi  {K_{ab}}^{i}+\delta_X {K_{ab}}^{i}={K_{ab}}^{j}{\dot{\omega}^i}_{~j}$. This means that under diffeomorphisms for which the mapping functions are allowed to vary we find that the extrinsic curvature ${K_{ab}}^{i}$ is invariant up to a normal rotation. Converting the result \eqref{eq:varKx0} into a spacetime formulation one obtains
\be \label{eq:varKx}
\begin{split}
\delta_X {K_{\mu\nu}}^{\rho}=&-\hdu\mu\sigma\hdu\nu\alpha\pud\rho\beta\nabla_{(\alpha}\nabla_{\sigma)}\xi^{\beta}-\pud\rho\sigma\Kddu\mu\nu\alpha\nabla_\alpha\xi^\sigma+\pud\rho\beta\hud\alpha\mu\hud\sigma\nu\Buddd\beta{(\alpha}{\sigma)}\lambda\xi^\lambda\\
&+\Kddu\mu\nu\lambda{\widehat{\omega}_\lambda}^{~\rho}+\Kddu\mu\nu \lambda \ndu\lambda i\delta_X\nud\rho i+2\Kddu\lambda{(\mu}\rho\eud\lambda b\delta_X \edu{\nu)}b \ .
\end{split}
\ee
Using the Lagrangian variation in \eqref{eq:defsvards} one finds that
\be
\delta_\xi {K_{\mu\nu}}^{\rho}+\delta_X {K_{\mu\nu}}^{\rho}=-\pud\rho\sigma\Kddu\mu\nu\alpha\nabla_\alpha\xi^\sigma+\Kddu\mu\nu\lambda{\widehat{\omega}_\lambda}^{~\rho}+\Kddu\mu\nu \lambda \ndu\lambda i\delta_X\nud\rho i+2\Kddu\lambda{(\mu}\rho\eud\lambda b\delta_X \edu{\nu)}b \ ,
\ee
where we have ignored terms that vanish when contracted with ${\mathcal{D}^{\mu\nu}}_\rho$. The second term in this variation leads to the invariance of the action under rotations of the normal vectors, contributing to \eqref{eq:st}, as in the gauge formulation of the variational principle where $\delta_\xi  {K_{ab}}^{i}+\delta_X {K_{ab}}^{i}={K_{ab}}^{j}{\dot{\omega}^i}_{~j}$. The other three terms in the variation mark the departure between the spacetime and the gauge version of the variational principle. In particular, it implies that spacetime variations carry more information than the corresponding gauge variations. The last two terms, in fact, will contribute to new constraints as shown in App.~\ref{app:comp}. The first term is responsible for the diffeomorphism constraints \eqref{eq:st}-\eqref{eq:pt}.

By the same token, we consider a general variation of the external rotation tensor
\be
\delta \oduu aij=\ndu\mu{[i}\nnb_a\delta \nuu{i]}{\mu}+\ndu\mu{[i}\nuu{j]}\nu\Gamma^{\mu}_{\lambda\nu}\delta\eud\lambda a+\ndu\mu{[i}\nuu{j]}\nu\eud\lambda a\delta \Gamma^{\mu}_{\lambda\nu} +\euu\mu b\Kddu ab{[i}\delta \ndu\mu{j]} \ ,
\ee
which, when specialised to variations of the embedding map leads to
\be
\delta_X \oduu aij=2\Kduu ab{[i}\nud{j]}\rho\nnb_b \xi^\rho+\ndu\mu{[i}\nuu{j]}\nu\xi^\lambda \eud\alpha a\Buddd\mu\nu\alpha\lambda+\nnb_a\widehat{\omega}^{ij} \ .
\ee
Evaluating the corresponding Lagrangian variation one finds $\delta_\xi \oduu aij+\delta_X \oduu aij=\nnb_a\dot{\omega}^{ij}$, while translating the variation into the spacetime formulation and using the shifted variation \eqref{eq:contracspin} one obtains
\be
\begin{split}
\wt \delta_X \oduu \mu\lambda\rho=&2\Kduu\mu\nu{[\lambda}\pud{\rho]}\sigma\nabla_\nu\xi^\sigma-\hud\alpha\mu\pud\rho\sigma\puu\nu\lambda\Buddd\sigma\nu\alpha\beta\xi^\beta+\pud\lambda\alpha\pud\rho\sigma\overline\nabla_\mu \widehat{\omega}^{\alpha\sigma} \\
&+\oduu\nu\lambda\rho\eud\nu a\delta_X \edu\mu a+\Kduu\mu\nu{[\lambda}\pud{\rho]}\alpha\edu\nu a\delta_X\eud\alpha a \ .
\end{split}
\ee
Given this and \eqref{eq:defsvards} one can compute the total variation under a diffeomorphism that shifts the embedding map
\be \label{eq:varextfull}
\begin{split}
\wt \delta_\xi \oduu \mu\lambda\rho+\wt \delta_X \oduu \mu\lambda\rho=&\hdu\mu \alpha\puu\sigma{[\lambda}\puu{\rho]}\tau \nabla_\alpha \nabla_\tau\xi_\sigma+2\Kduu\mu\nu{[\lambda}\pud{\rho]}\sigma\nabla_\nu\xi^\sigma\\
&+\pud\lambda\alpha\pud\rho\sigma\overline\nabla_\mu \widehat{\omega}^{\alpha\sigma}+\oduu\nu\lambda\rho\eud\nu a\delta_X \edu\mu a+\Kduu\mu\nu{[\lambda}\pud{\rho]}\alpha\edu\nu a\delta_X\eud\alpha a \ ,
\end{split}
\ee
where we have used that $\nud\mu j\delta_X\ndd\mu i$ is symmetric in the indices $i,j$ since under a variation of the embedding map $\omega_{ij}=0$ in \eqref{eq:varnupL}.
Again, we clearly see that the spacetime version of these variations carries more information than the corresponding gauge variation for which $\delta_\xi \oduu aij+\delta_X \oduu aij=\nnb_a\dot{\omega}^{ij}$. The last two terms in \eqref{eq:varextfull} will contribute to new constraints as shown in App.~\ref{app:comp}. The third term contributes to the invariance of the action under rotations of the normal vectors while the first two contribute to the constraints \eqref{eq:st}-\eqref{eq:pt}.

The same type of variations can be performed for the background Riemann tensor. In particular, for the background Ricci scalar, one simply obtains $\delta_X R=-\xi^\alpha\nabla_\alpha R$. Also, consider the contraction with the background Ricci tensor $\puu\lambda\alpha R_{\lambda\alpha}$ so that
\be
\begin{split}
\delta_X\left(\puu\lambda\alpha R_{\lambda\alpha}\right)&=2\nuu\alpha iR_{\lambda\alpha}\delta_X \nud\lambda i-\puu\lambda\alpha\xi^{\sigma}\partial_\sigma R_{\lambda\alpha} \\
&=-\puu\lambda\alpha\xi^\sigma\nabla_\sigma R_{\lambda\alpha}+2\pud\alpha\rho\huu\lambda\sigma R_{\lambda\alpha}\overline\nabla_\sigma \xi^\rho \ ,
\end{split}
\ee
where we have used \eqref{eq:varnup} in the second line. For specific examples, such as these contractions with the background Riemann tensor, one readily obtains, via variations of the embedding map, that there are not contributions to the constraints \eqref{eq:st}-\eqref{eq:pt}. Recasting the variational principle of Sec.~\ref{sec:shapeequations} in terms of variations of the embedding map leads to the same final results but without giving the explicit couplings the abstract results are not manifestly covariant. Nevertheless, the resulting equations of motion are equivalent, as one may explicitly check for all terms in  Sec.~\ref{sec:2derivative}. Equivalent variational formulae is obtained for the edge geometry.


\subsection{Weyl transformations} 

In this part of the appendix we provide the Weyl transformations needed in the bulk of the paper. Weyl transformations are defined according to
\be
\Gdd \mu\nu \to e^{2\omega} \Gdd \mu\nu \ ,
\ee
and are inherited by $\hdd \mu\nu$ and $\pdd \mu\nu$. In particular, the surface measure transforms accordingly as ${\sqrt{|\gamma|} \to e^{p\, \omega} \sqrt{|\gamma|}}$. The background curvature tensors transform as
\bse
\bal
\Buddd \pi\theta\mu\nu & \to \Buddd \pi\theta\mu\nu + \Gdu \nu\pi \nabla_\theta\, \nabla_\mu\, \omega - \Gdu \mu\pi \nabla_\theta\, \nabla_\mu\, \omega + \Gdd \theta\mu \nabla_\nu\, \nabla^\pi\, \omega - \Gdd \theta\nu \nabla_\mu\, \nabla^\pi\,  \omega \nonumber\\
& \quad + \Gdu \mu\pi \nabla_\theta\, \omega \nabla_\nu\, \omega - \Gdu \nu\pi \nabla_\theta\, \omega \nabla_\mu\, \omega + \Gdd \theta\nu \nabla_\mu\, \omega \nabla^\pi\, \omega - \Gdd \theta\mu \nabla_\nu\, \omega \nabla^\pi\, \omega \nonumber\\[2mm]
& \quad + \left( \Gdd \theta\mu \Gdu \nu\pi - \Gdd \theta\nu \Gdu \mu\pi \right) \nabla^\rho\, \omega \nabla_\rho\, \omega \ , \\
\Bdd \mu\nu & \to \Bdd \mu\nu - \Gdd \mu\nu \DAlembert \omega - (D-2) \left( \nabla_\mu\, \nabla_\nu\, \omega - \nabla_\mu\, \omega \nabla_\nu\, \omega + \Gdd \mu\nu \nabla^\pi\, \omega \nabla_\pi\, \omega \right) \  , \\
R & \to e^{-2\omega}\left( R - 2(D-1) \DAlembert \omega - (D-1)(D-2)\nabla^\pi\, \omega \nabla_\pi\, \omega \right) \ ,
\end{align}
\ese
whereas the background Schouten tensor \eqref{eq.bgSchouten} transforms as
\be
\Pdd \mu\nu \to \Pdd \mu\nu - \nabla_\mu\, \nabla_\nu\, \omega + \nabla_\mu\, \omega \nabla_\nu\, \omega - \frac{1}{2} \Gdd \mu\nu \nabla^\pi\, \omega \nabla_\pi\, \omega \ .
\ee
The former expressions imply the invariance of the Weyl tensor defined in \eqref{eq.bgWeyl}, that is
\be
W^\pi{_{\theta\mu\nu}} \to W^\pi{_{\theta\mu\nu}} \ .
\ee
In turn, the second fundamental form \eqref{eq:extc} and its trace \eqref{eq.Ktrace} transform as
\bse
\bal
\Kddu \mu\nu\pi & \to \Kddu \mu\nu\pi - \hdd \mu\nu \puu \theta\pi \!\nabla_\theta\, \omega \ , \\
\Ku \pi & \to e^{-2\omega} \left( \Ku \pi - p \puu \theta\pi \!\nabla_\theta\, \omega \right) \ ,
\end{align}
\ese
which imply the invariance of the conformation tensor \eqref{eq:conformation}, that is
\be\label{eq.weylconformation}
\Cddu \mu\nu\pi \to \Cddu \mu\nu\pi \ .
\ee
The transformation of the  internal curvature tensors can be deduced from the former expressions by means of the Gauss-Codazzi equation \eqref{eq:GC}
\bse
\bal
\Ruddd \pi\theta\mu\nu & \to \Ruddd \pi\theta\mu\nu + \left( \hdd \mu\theta \hdu \nu\pi - \hdd \nu\theta \hdu \mu\pi \right) \huu \rho\sigma \nabla_\rho\, \omega  \nabla_\sigma\, \omega \nonumber\\
& \quad +  \left( \hdu \mu\rho \hdu \nu\pi \hdu \theta\upsilon - \hdu \mu\pi \hdu \nu\rho \hdu \theta\upsilon + \hdd \mu\theta  \hdu \nu\upsilon \huu \rho\pi - \hdd \nu\theta \hdu \mu\upsilon \huu \rho\pi \right) \left( \nabla_\upsilon\,\nabla_\rho\, \omega - \nabla_\upsilon\, \omega \nabla_\rho\, \omega \right) \nonumber\\
& \quad - \left(  \Kduu \mu\pi\upsilon \hdd \nu\theta - \Kduu \nu\pi\upsilon \hdd \mu\theta + \Kddu \nu\theta\upsilon \hdu \mu\pi - \Kddu \mu\theta\upsilon \hdu \nu\pi \right) \nabla_\upsilon\, \omega \ , \\
\Rdd \mu\nu & \to \Rdd \mu\nu - \hdd \mu\nu \huu \rho\sigma \nabla_\rho\,  \nabla_\sigma\, \omega - (p-2) \hdu \mu\rho \hdu \nu\sigma \left( \nabla_\rho\, \nabla_\sigma\, \omega - \nabla_\rho\, \omega \nabla_\sigma\, \omega \right) \nonumber\\
& \quad - (p-2) \left( \hdd \mu\nu \huu \rho\sigma \nabla_\rho\, \omega  \nabla_\sigma\, \omega + \Kddu \mu\nu\pi \nabla_\pi\, \omega \right) - \hdd \mu\nu \Ku \pi \nabla_\pi\, \omega \ , \\
{\cal R} & \to e^{-2\omega} \left( {\cal R} - 2(p-1)\huu \rho\sigma \nabla_\rho\,  \nabla_\sigma\, \omega - (p-1)(p-2) \huu \rho\sigma \nabla_\rho\, \omega  \nabla_\sigma\, \omega - 2(p-1) \Ku \pi \nabla_\pi\, \omega \right) \ ,
\end{align}
\ese
whereas the Ricci-Voss equation \eqref{eq:RV} implies the invariance of the outer curvature tensor \eqref{eq:oint}
\be
\Ouddd \pi\theta\mu\nu \to \Ouddd \pi\theta\mu\nu \ .
\ee
This could have also been derived by noticing that the external rotation tensor \eqref{eq:extt} is Weyl invariant
\be
{{\omega_\mu}^{\nu}}{_\rho} \to {{\omega_\mu}^{\nu}}{_\rho} \ .
\ee
On the other hand, the internal rotation tensor \eqref{eq:intt} has the transformation property
\be
{{\rho_\mu}^{\nu}}{_\rho} \to {{\rho_\mu}^{\nu}}{_\rho}+2\gamma^{\nu}{_{(\mu}}\overline\nabla_{\rho)}\omega-\gamma_{\mu\rho}\overline\nabla^{\nu}\omega \ .
\ee
For the particular Riemann curvature contractions defined in Eqn.~\eqref{eq:defR} and appearing in the parity-even action \eqref{eq:seven} we find the following transformation properties
\be
\begin{split}
R_{||} & \to e^{-2\omega}\left[ R_{||} - 2(p-1)\gamma^{\rho\beta} \left( \nabla_\rho \nabla_\beta \omega - \nabla_\rho \omega \nabla_\beta \omega \right) + p(1-p) \nabla^\alpha \omega \nabla_\alpha \omega \right] \ , \\
R_{\perp} & \to e^{-2\omega}\left[ R_{\perp} - 2(n-1)\perp^{\rho\beta} \left( \nabla_\rho \nabla_\beta \omega - \nabla_\rho \omega \nabla_\beta \omega \right) + n(1-n) \nabla^\alpha \omega \nabla_\alpha \omega \right] \ , \\
R_{\angle} & \to e^{-2\omega}\left[ R_{\angle} -  \left(n \gamma^{\rho\beta} + p \perp^{\rho\beta} \right) \left( \nabla_\rho \nabla_\beta \omega - \nabla_\rho \omega \nabla_\beta \omega \right) -p \, n \nabla^\alpha \omega \nabla_\alpha \omega \right] \ ,
\end{split}
\ee
whereas the transformations of the tensors contributing to the parity-odd part of the action given in Eqn.~\eqref{eq:sodd} are
\be
\begin{split}
\epsilon_{\mu\nu\rho\sigma} & \to e^{4\omega}\epsilon_{\mu\nu\rho\sigma} ~~ , ~~ \omega_\mu  \to \omega_\mu ~~, ~~n_\rho  \to e^\omega n_\rho ~~,~~ u^\mu \to e^{-\omega} e^\mu_1 \ ,\\
R_- & \to e^{-2\omega} R_- ~~,~~\Omega \to e^{-2\omega} \Omega ~~,~~\epsilon^{\lambda\mu}{_{\rho\alpha}} K_\lambda{^{\nu\alpha}} K_{\mu\nu}{^\rho}  \to e^{-2\omega} \epsilon^{\lambda\mu}{_{\rho\alpha}} K_\lambda{^{\nu\alpha}} K_{\mu\nu}{^\rho} ~~.
\end{split}
\ee


\section{Details on the variational principle for actions} \label{app:highdev}
In this section we give further details on the variational principle used in the core of this paper and provide the generalised constraints and equations of motion, which include the couplings to the intrinsic Riemann tensor and the outer curvature. We also provide a brief comparison between the spacetime and gauge variational principles. At the end of this appendix we give further details on the Ward identities and displacement operator for CFTs coupled to defects with edges, along with specific instructive examples.

\subsection{Framework for variations at any derivative order}
We consider a purely geometric action that takes the form \eqref{eq:genact} for arbitrary geometric fields $\Phi(\sigma)$. The Lagrangian variation of any explicit term that appears in such action can be organised as
\be \label{eq:varfull}
\delta_\xi S[\Phi(\sigma)]=\frac{1}{2}\int_{\mathcal{W}}d^{p}\sigma\sqrt{|\gamma|}\left(T^{\mu\nu}\delta_\xi\Gdd\mu\nu+T^{\mu\nu\rho}\nabla_\rho \delta_\xi g_{\mu\nu}+T^{\mu\nu\rho\lambda}\nabla_\rho\nabla_\lambda \delta_\xi g_{\mu\nu}+\cdots\right)  \ ,
\ee
where the \emph{dots} represent higher multipole terms $T^{\mu\nu\mu_1\mu_2\cdots\mu_l}$, with $l$ being the highest order multipole moment, that can appear when terms involving more than two derivatives are taken into account. It is clear that, with the aid of the delta function, the spacetime stress tensor that arises from \eqref{eq:varfull} takes the general form
\be
\textbf{T}^{\mu\nu}=T^{\mu\nu}\widehat\delta(x)+\sum_{i=1}^{l}(-1)^{l} \nabla_{\mu_{1}}\cdots\nabla_{\mu_{i}} \left(T^{\mu\nu\mu_1\cdots\mu_l}\widehat\delta(x)\right) \ ,
\ee
and hence is in agreement with that of \eqref{eq:stgen}. Using that for a Lagrangian variation one has $\delta_\xi \Gdd\mu\nu=2\nabla_{(\mu}\xi_{\nu)}$, one may obtain the shape equation and diffeomorphism constraints from \eqref{eq:varfull}. In general, this is quite involved and here we provide a method that can be applied to higher orders, though we explicit use it only up to $l=1$. First, we decompose the $\rho$ index in the second term in \eqref{eq:varfull} in tangential and transverse parts and integrate the tangential component by parts in order to find
\be
\begin{split}
\delta_\xi S[\Phi(\sigma)]=&\frac{1}{2}\int_{\mathcal{W}}d^{p}\sigma\sqrt{|\gamma|}\left(\left(T^{\mu\nu}-\overline\nabla_\lambda\left(\hud\lambda\rho T^{\mu\nu\rho}\right)\right)\delta_\xi\Gdd\mu\nu+T^{\mu\nu\lambda}\pud\rho\lambda\nabla_\rho \delta_\xi g_{\mu\nu}\right)\\
&+\frac{1}{2}\int_{\partial\mathcal{W}}d^{p-1}\wt\sigma\sqrt{|h|}\wt n_\rho T^{\mu\nu\rho}\delta_\xi\Gdd\mu\nu \ ,
\end{split}
\ee
It may be observed that now each of the terms in the variation is frame-invariant, as it can be compared with \eqref{eq:inv1}, except for the boundary term, as earlier advertised. Therefore, the constraints and equations of motion that will be derived from this variation will necessarily be frame-invariant in the surface but not on the edges. We introduce the quantities $\widehat{\mathsf{T}}^{\mu\nu}$ and $\widehat{\mathsf{T}}^{\mu\nu\rho}$ to denote these two invariants such that
\be
\widehat{\mathsf{T}}^{\mu\nu}=T^{\mu\nu}-\overline\nabla_\lambda\left(\hud\lambda\rho T^{\mu\nu\rho}\right) \ , \quad \widehat{\mathsf{T}}^{\mu\nu\rho}=T^{\mu\nu\lambda}\pud\rho\lambda \ .
\ee
Making use of $\delta_\xi \Gdd\mu\nu=2\nabla_{(\mu}\xi_{\nu)}$ and the Riemann identity $\nabla_{[\mu}\nabla_{\nu]}\xi_\lambda=\Buddd\alpha\lambda\nu\mu \xi_\alpha$ into \eqref{eq:varfull}, one finds the two sets of constraints 
\be
\widehat{\mathsf{T}}^{\alpha\nu(\rho}\pud{\mu)}\alpha=0 \ , \quad \pud\rho\mu\left(\widehat{\mathsf{T}}^{\mu\nu}-\overline\nabla_\lambda\left(\hud\lambda\alpha\widehat{\mathsf{T}}^{\alpha\nu\mu}\right)\right)=0 \ ,
\ee
and the equation of motion
\be
\overline\nabla_\lambda\left(\hud\lambda\mu\left(\widehat{\mathsf{T}}^{\mu\nu}-\overline\nabla_\rho\left(\hud\rho\alpha\widehat{\mathsf{T}}^{\alpha\nu\mu}\right)\right)\right)=\left(\widehat{\mathsf{T}}^{\mu\alpha\rho}+\frac{1}{2}\pud\rho\sigma\widehat{\mathsf{T}}^{\sigma\alpha\mu}\right)\Buddd\nu\alpha\mu\rho \ .
\ee
Keeping track of the boundary terms one arrives at the edge constraint
\be
\left(\wt n_\rho T^{\mu\nu\rho}\psud\alpha\mu+\wt n_\sigma \widehat{\mathsf{T}}^{\sigma\nu\alpha}\right)|_{\partial\mathcal{W}}=0 \ ,
\ee
and equation of motion on the edge
\be
\wt\nabla_\lambda\left(\hsud\lambda\mu\wt n_\rho T^{\mu\nu\rho}\right)-\wt n_\mu\left(\widehat{\mathsf{T}}^{\mu\nu}-\overline\nabla_\lambda\left(\hud\lambda\alpha \widehat{\mathsf{T}}^{\alpha\nu\mu}\right)\right)|_{\partial\mathcal{W}}=0 \ .
\ee
One finds perfect agreement of this form of the constraints and equations with those obtained in Sec.~\ref{sec:shapeequations} when no quadrupole moments are present. These equations had been derived in \cite{Vasilic:2007wp} via a different method but here we have provided a cleaner, quicker, spacetime covariant and more physical derivation of these equations (i.e. in terms of frame-invariant quantities). We leave the problem of obtaining the higher order equations of motion in full generality to future work. Below, we derive these equations for a specific form of the spacetime stress tensor.

\subsection{Generalised diffeomorphism constraints and shape equation}
We now provide the constraints and equations of motion for the action \eqref{eq:varL} including the couplings to the intrinsic Riemann tensor and outer curvature. Using the variations \eqref{eq:varrint} and \eqref{eq:varouter}, we find the spin conservation equation
\be \label{eq:spinfull}
\mathcal{D}^{\mu\lambda[\sigma}\Kddu\mu\lambda{\alpha]}+\pdu\lambda\sigma\pdu\rho\alpha\overline\nabla_\mu \mathcal{S}^{\mu\lambda\rho}+\pdu\nu{[\sigma}\pdu\lambda{\alpha]}\Pi^{\nu\lambda}-2\pud\alpha\beta\pud\sigma\rho\overline\nabla_\theta\left(\puu\nu\beta\pud\rho\tau\hud\theta\pi\overline\nabla_\kappa{\mathcal{H}_\nu}^{\kappa\tau\pi}\right)=0 \ .
\ee
The coupling to the intrinsic and outer curvatures introduce further modifications of this equation. The orthogonal components of the monopole stress tensor take the form
\be 
\mathcal{B}^{\alpha\sigma}=\mathcal{D}^{\mu\lambda(\sigma}\Kddu\mu\lambda{\alpha)}+\pdu\nu{(\sigma}\pdu\lambda{\alpha)}\Pi^{\nu\lambda}+4\mathcal{I}^{\mu\kappa\nu\lambda}\Kddu\lambda\nu{(\sigma}\Kddu\kappa\mu{\alpha)}-4\pud{(\alpha}\nu\pud{\sigma)}\beta\overline\nabla_\kappa\left(\mathcal{H}^{\mu\kappa\nu\lambda}\Kdud\lambda\beta\mu\right) \ ,
\ee
while the mixed tangential-transverse components read
\be
\begin{split}
{\mathcal{P}^{\mu}}_\nu\puu\nu\alpha \hdu\mu\sigma=&\mathcal{S}^{\mu\alpha\lambda}\Kdud\mu\sigma\lambda+\hdu\nu\sigma\pdu\lambda\alpha\Pi^{\nu\lambda}+2\hud\sigma\beta\pud\alpha\rho\overline\nabla_\kappa\left(\mathcal{I}^{\beta\kappa\nu\lambda}\Kddu\lambda\nu\rho\right)\\
&+4\hud{(\sigma}\nu\hud{\beta)}\lambda\Kddu\mu\beta\alpha\overline\nabla_\kappa\mathcal{I}^{\mu\kappa\nu\lambda}-4\pud\alpha\theta\hud\sigma\beta\overline\nabla_\kappa\left(\mathcal{H}^{\mu\kappa(\theta|\lambda|}\Kdud\lambda{\beta)}\mu\right) \quad \\
&-2\puu\alpha\nu\Kdud\pi\sigma\tau\overline\nabla_\kappa{\mathcal{H}_\nu}^{\kappa\tau\pi} \ .
\end{split}
\ee
In turn, given the mixed tangential-transverse components, the surface equations of motion can be written as
\be \label{eq:fullshapeeq}
\begin{split}
&\overline\nabla_\lambda\left(\mathcal{T}^{\lambda\sigma}+\hdu\mu\lambda\left(\pdu\nu\sigma\Pi^{\mu\nu}-\Pi^{\sigma\mu}\right)-\hud\lambda\nu\overline\nabla_\mu\mathcal{D}^{\mu\nu\sigma}-2{{\mathcal{S}^{\mu}}_\alpha}^{\sigma}\Kduu\mu\lambda\alpha\right) \\
&-4\overline\nabla_\lambda\left(\hud\lambda\alpha\overline\nabla_\beta\left(\huu\mu{(\alpha}\hud{\sigma)}\lambda\hud\beta\nu\overline\nabla_\kappa{\mathcal{I}_\mu}^{\kappa\nu\lambda}\right)-\hud{(\lambda}\nu\hud{\beta)}\alpha\Kddu\mu\beta\sigma\overline\nabla_\kappa\mathcal{I}^{\mu\kappa\nu\alpha}\right)\\
&-2\nabla_\lambda\left(\hud\lambda\beta\overline\nabla_\theta\left(\puu\alpha\beta\pud\sigma\tau\hud\theta\pi\overline\nabla_\kappa{\mathcal{H}_\alpha}^{\kappa\tau\pi}\right)+\puu\sigma\nu\Kdud\pi\lambda\tau\overline\nabla_\kappa{\mathcal{H}_\nu}^{\kappa\tau\pi}\right)\\
&=\left(\mathcal{S}^{\mu\lambda\rho}-\mathcal{D}^{\mu\lambda\rho}\right)\Buddd\sigma\mu\lambda\rho+2\mathcal{Q}^{\mu\nu\lambda\rho}\nabla_\nu\Buddd\sigma\rho\mu\lambda+4\puu\lambda\beta\pud\alpha\tau\hud\theta\pi\overline\nabla_\kappa\left({\mathcal{H}_\lambda}^{\kappa\tau\pi}\right)\Buddd\sigma\alpha\theta\beta \ .
\end{split}
\ee
\subsubsection{Edge constraints and equations of motion}
The constraints and equations of motion on the edges are significantly modified by these extra couplings. The edge spin conservation equation now takes the form
\be \label{eq:fullstedge1}
\begin{split}
&\wt{\mathcal{D}}^{\mu\lambda[\sigma}\Ksddu\mu\lambda{\alpha]}+\psud\sigma\lambda\psud\alpha\rho \wt \nabla_\mu \wt{\mathcal{S}}^{\mu\lambda\rho}+\psud{[\sigma}\nu\psud{\alpha]}\lambda\wt{\Pi}^{\nu\lambda}-\wt n_\lambda\wt n_\mu \mathcal{D}^{\lambda\mu[\sigma}\wt{n}^{\alpha]}-\wt n_\lambda \psud\alpha\mu\psud\sigma\nu\mathcal{S}^{\lambda\nu\mu}\\
&-4\wt n_\kappa \wt n_\rho\mathcal{I}^{\rho\kappa\beta\mu}\Ksddu\beta\mu{[\sigma}\wt n^{\alpha]}-2\psud\alpha\tau\psud\sigma\beta\wt\nabla_\mu\left(\hsud\mu\lambda\wt n_\kappa\mathcal{H}^{\beta\kappa\tau\lambda}\right)-2\wt n_\lambda \puu\mu{\sigma}\pud\alpha\tau\overline\nabla_\kappa{\mathcal{H}_\mu}^{\kappa\tau\lambda}=0 \ .
\end{split}
\ee
The transverse components of the edge monopole stress tensor read
\be \label{eq:fullstedge2}
\begin{split}
\wt{\mathcal{B}}^{\alpha\sigma}=&\wt{\mathcal{D}}^{\mu\lambda(\sigma}\Ksddu\mu\lambda{\alpha)}+\psud{(\sigma}\nu\psud{\alpha)}\lambda\wt \Pi^{\nu\lambda}-\wt n_\lambda\wt n_\mu \mathcal{D}^{\lambda\mu(\sigma}\wt{n}^{\alpha)}+4\wt\nabla_\rho\left(\hsud\rho\nu\wt n_\kappa \mathcal{I}^{\mu\kappa\nu\lambda}\right)\psud{(\alpha}\mu \psud{\sigma)}\lambda  \ ,\\
&+4\wt\nabla_\gamma\left(\wt n_\kappa\wt n_\rho\hsud{(\gamma}\beta \hsud{\lambda)}\mu\mathcal{I}^{\beta\kappa\rho\mu}\wt n^{\nu}\right)\psud{(\alpha}\nu \psud{\sigma)}\lambda-4\wt n_\kappa \wt n_\beta \mathcal{I}^{\beta\kappa\nu\lambda}\Kddu\nu\lambda{(\sigma}\wt n^{\alpha)}\\
&+4\psud{(\alpha}\tau\psud{\sigma)}\beta\wt n_\kappa \mathcal{H}^{\mu\kappa\tau\lambda}\Kdud\lambda{\beta}\mu \ ,
\end{split}
\ee
while the mixed components take the form
\be \label{eq:fullstedge3}
\begin{split}
{\wt{\mathcal{P}}^{\mu}}_{~\nu}\psuu\nu\alpha \hsud\sigma\mu=&\wt{\mathcal{S}}^{\mu\alpha\lambda}\Ksdud\mu\sigma\lambda+\hsud\sigma\nu\psud\alpha\lambda\wt\Pi^{\nu\lambda}-2\wt n^\mu \wt n_\lambda \wt n^\alpha \hsud\sigma\nu\overline\nabla_\kappa {\mathcal{I}_\mu}^{\kappa\nu\lambda}-2\wt n_\kappa \mathcal{I}^{\beta\kappa\nu\lambda}\hsud\sigma\beta\Kddu\nu\lambda\alpha\\
&+4\wt\nabla_\rho\left(\hsud\rho\nu\wt n_\kappa \mathcal{I}^{\mu\kappa\nu\lambda}\right)\psud{(\alpha}\mu\hsud{\sigma)}\lambda+4\wt\nabla_\gamma\left(\wt n_\kappa\wt n_\rho\hsud{(\gamma}\beta\hsud{\lambda)}\mu\mathcal{I}^{\beta\kappa\rho\mu}\wt n^\nu\right)\psud\alpha\nu\hsud\sigma\lambda\\
&+4\psud{\alpha}\tau\hsud{\sigma}\beta\wt n_\kappa \mathcal{H}^{\mu\kappa\tau\lambda}\Kdud\lambda{\beta}\mu+2\psud\alpha\tau\hsud\sigma\beta\wt\nabla_\mu\left(\hsud\mu\lambda\wt n_\kappa \mathcal{H}^{\beta\kappa\tau\lambda}\right) \ .
\end{split}
\ee
Given this, the equations of motion at the edges take the form
\be \label{eq:fullshapeedge}
\begin{split}
&\wt\nabla_\lambda\left(\wt{\mathcal{T}}^{\lambda\sigma}+\hsud\lambda\mu\left(\psud\sigma\nu\wt{\Pi}^{\mu\nu}-\wt{\Pi}^{\sigma\mu}\right)-\hsud\lambda\nu\wt\nabla_\mu\wt{\mathcal{D}}^{\mu\nu\sigma}-2{\wt{\mathcal{S}^{\mu}}_\alpha}^{~\sigma}\Ksduu\mu\lambda\alpha+\wt n_\mu\hsud\lambda\nu\mathcal{D}^{\mu\nu\sigma}\right) \\
&-4\wt\nabla_\lambda\left(\hsud{(\lambda}\mu\hsud{\sigma)}\beta\wt\nabla_\alpha\left(\hsud\alpha\nu\wt n_\kappa \mathcal{I}^{\beta\kappa\nu\mu}\right)-\wt n_\kappa\hsud{(\sigma}\tau\hsud{\beta)}\nu\mathcal{I}^{\tau\kappa\mu\nu}\Ksdud\beta\lambda\mu\right)\\
&+4\wt\nabla_\lambda\left(\psud\sigma\nu\hsud\lambda\mu\wt\nabla_\gamma\left(\wt n_\kappa \wt n_\alpha \hsud{(\gamma}\beta\hsud{\mu)}\rho\mathcal{I}^{\beta\kappa\alpha\rho}\wt n^\nu\right)+\hsud{(\lambda}\beta\hud{\sigma)}\nu\wt n^\mu\rho\overline\nabla_\kappa{\mathcal{I}_\mu}^{\kappa\nu\beta}\right)\\
&-\wt\nabla_\lambda\left(4\wt n_\kappa \hsud{(\lambda}\alpha\hsud{\sigma)}\beta\mathcal{H}^{\mu\kappa\alpha\nu}\Kdud\nu\beta\mu+2\hsud\lambda\alpha\hsud\sigma\rho\wt\nabla_\mu\left(\hsud\mu\lambda\wt n_\kappa \mathcal{H}^{\rho\kappa\alpha\nu}\right)\right)\\
&=\left(\wt{\mathcal{S}}^{\mu\lambda\rho}-\wt{\mathcal{D}}^{\mu\lambda\rho}\right)\Buddd\sigma\mu\lambda\rho+2\wt{\mathcal{Q}}^{\mu\nu\lambda\rho}\nabla_\nu\Buddd\sigma\rho\mu\lambda+4\wt n_\kappa\wt n_\alpha\wt n^\nu\mathcal{I}^{\beta\kappa\alpha\mu}\hsud{(\gamma}\beta\hsud{\lambda)}\mu\Buddd\sigma\lambda\gamma\nu\\
&~~~~+\wt n_\lambda\left(\mathcal{T}^{\lambda\sigma}+\hdu\mu\lambda\left(\pdu\nu\sigma\Pi^{\mu\nu}-\Pi^{\sigma\mu}\right)-\hud\lambda\nu\overline\nabla_\mu\mathcal{D}^{\mu\nu\sigma}-2{{\mathcal{S}^{\mu}}_\alpha}^{\sigma}\Kduu\mu\lambda\alpha\right)\\
&~~~~-4\wt n_\lambda\overline\nabla_\rho\left(\Kddu\nu\lambda{(\sigma}\mathcal{I}^{\lambda)\rho\nu\lambda}+\huu\mu\rho\hud{(\sigma}\nu\hud{\lambda)}\alpha\overline\nabla_\kappa{\mathcal{I}_\mu}^{\kappa\nu\alpha}\right)-4\wt n_\lambda\Kudd{(\lambda}\kappa\rho\mathcal{I}^{\sigma)\kappa\tau\alpha}\Kddu\alpha\tau\rho\\
&~~~~+4\wt n_\kappa\mathcal{H}^{\beta\kappa\alpha\mu}\Buddd\sigma\beta\mu\alpha+4\wt n_\alpha \mathcal{H}^{\nu\kappa\tau\lambda}\Kdud\lambda{(\sigma}\nu\Kdud\kappa{\alpha)}\tau-2\wt n_\beta \overline\nabla_\theta\left(\hud\theta\pi\puu\sigma\beta\pud\sigma\tau\overline\nabla_\kappa{\mathcal{H}_\alpha}^{\kappa\tau\pi}\right) \ .
\end{split}
\ee

\subsubsection{Spacetime stress tensor}
The spacetime stress tensor acquires new components due to these couplings. It takes the form of \eqref{eq:stgen} but with the components
\be \label{eq:stact2full1}
\begin{split}
T^{\mu\nu}=\mathcal{T}^{\mu\nu}-4\Kudd{(\mu}\kappa\rho\mathcal{I}^{\nu)\kappa\sigma\lambda}\Kddu\lambda\sigma\rho+4\mathcal{H}^{\sigma\kappa\tau\lambda}\Kdud\lambda{(\mu}\sigma\Kdud\kappa{\nu)}\tau+2{\mathcal{P}^{\lambda}}{_\rho}\puu\rho{(\mu}\hdu\lambda{\nu)}+\mathcal{B}^{\mu\nu} \ ,
\end{split}
\ee
\be \label{eq:stact2full2}
\begin{split}
T^{\mu\nu\rho}=&2\mathcal{D}^{\rho(\mu\nu)}-\mathcal{D}^{\mu\nu\rho}+2\mathcal{S}^{(\mu\nu)\rho}+4\Kddu\lambda\sigma{(\mu}\mathcal{I}^{\nu)\rho\sigma\lambda}-4{\mathcal{I}_\alpha}^{\kappa\beta(\mu}\overline\nabla_\kappa\left(\hud{\nu)}\beta\huu\alpha\rho\right)\\
&+4\Kdud\lambda{(\mu}\alpha\mathcal{H}^{\nu)\rho\alpha\lambda}-4\mathcal{H}^{\sigma\kappa\tau(\mu}\overline\nabla_\kappa\left(\pud{\nu)}\tau\pud\rho\sigma\right) \ ,
\end{split}
\ee
\be \label{eq:stact2full3}
\begin{split}
T^{\mu\nu\rho\lambda}=-4\mathcal{Q}^{\lambda\rho(\mu\nu)}-4\mathcal{I}^{\lambda\rho(\mu\nu)}-4\mathcal{H}^{\lambda\rho(\mu\nu)} \ .
\end{split}
\ee
One observes that there are extra contributions to the tangential components of the monopole part of the stress tensor. One may explicitly check that choosing $\mathcal{I}^{\lambda\rho\mu\nu}=\alpha_1\huu\lambda{[\rho}\huu{\nu]}\mu$ yields the same stress tensor as choosing $\mathcal{D}^{\mu\nu\rho}=\alpha_1 (\huu\mu\nu K^\rho-\Kuuu\mu\nu\rho)$ and $\mathcal{Q}^{\lambda\rho\mu\nu}=\alpha_1\huu\lambda{[\rho}\huu{\nu]}\mu$. This means that the stress tensor due to the contribution proportional to $\mathcal{R}$ in the action \eqref{eq:seven} is equivalent to the stress tensor obtained once the Gauss-Codazzi equation \eqref{eq:GC} is used. Consequently, both the surface and edge dynamics are equivalent, as expected. The same holds true for the equations of motion that arise due to the term proportional to $\lambda_3$ in \eqref{eq:sodd} and that proportional to $\alpha_5$.

\subsection{Comparison between spacetime and gauge variational principles} \label{app:comp}
In this section we compare the variational principle of Sec.~\ref{sec:shapeequations} with the gauge variational principle employed previously in the literature (see e.g.~\cite{Capovilla:1994bs, Arreaga:2000mr, Armas:2013hsa}). We show that the two principles are equivalent, modulo certain constraints, including the diffeomorphism constraints \eqref{eq:bt} and \eqref{eq:pt}. For simplicity, we do not consider couplings to the intrinsic, outer and background curvatures, though these can be straightforwardly included. The variation of the action \eqref{eq:varL} can be recast into gauge formulation language by using the tangent and normal vectors. Consider for example the couplings to $\huu\mu\nu$ appearing in \eqref{eq:varL}. One may write a general variation as
\be \label{eq:vargu}
\begin{split}
\delta S[\Gdd\mu\nu, X^\mu]&=-\frac{1}{2}\int_{\mathcal{W}}\sqrt{|\gamma|}d^p\sigma\mathcal{T}_{\mu\nu}\delta\left(\gamma^{ab}\eud\mu a\eud\nu b\right) \\
&=-\frac{1}{2}\int_{\mathcal{W}}\sqrt{|\gamma|}d^p\sigma\left(\mathcal{T}_{ab}\delta\gamma^{ab}+2{\widehat{\mathcal{T}}^{a}}_{~~b}\edu\mu b\delta\eud\mu a\right) \ .
\end{split}
\ee
A few remarks are now in place. Gauge variational principles only consider the first term in the second line above. For Lagrangian variations for which $\delta_\xi X=0$ and $\delta_\xi \eud\mu a=0$ the two variational principles are equivalent but if the mapping functions are allowed to vary as well, they differ. In the second term in the second line above we have defined $\widehat{\mathcal{T}}^{ab}=\mathcal{T}^{ab}-\gamma^{ab}\widehat{\mathcal{L}}$ with $\widehat{\mathcal{L}}=\mathcal{L}/\sqrt{|\gamma|}$ being the Lagrangian density. The reason for this technical detail of reasonable importance is that the determinant of the induced metric is computed with $\gamma_{ab}$ and not with $\hdd\mu\nu$. For any action there will be a contribution to the surface stress tensor of the form $\gamma_{ab}\widehat{\mathcal{L}}\delta\gamma^{ab}$, which by using the tangential vectors, can be turned into the form $\gamma_{\mu\nu}\widehat{\mathcal{L}}\delta\gamma^{\mu\nu}-2\widehat{\mathcal{L}}\gamma_{\mu\nu}\euu\mu a\delta \eud\mu a$. If the variation is Lagrangian then we have the equivalence $\gamma_{ab}\widehat{\mathcal{L}}\delta\gamma^{ab}=\gamma_{\mu\nu}\widehat{\mathcal{L}}\delta\gamma^{\mu\nu}$ but otherwise this equivalence is not valid and hence we must in general subtract this component from the second term in \eqref{eq:vargu}. The same type of considerations apply to the other terms in \eqref{eq:varL}. 

Performing this tedious exercise for all terms, we find the variation
\be \label{eq:everything}
\begin{split}
\delta S[\Gdd\mu\nu, X^\mu]&=\int_{\mathcal{W}}\sqrt{|\gamma|}d^p\sigma\left(-\frac{1}{2}\mathcal{T}_{ab}\delta\gamma^{ab}+{\mathcal{D}^{ab}}_i\delta {K_{ab}}^{i}+{\mathcal{S}^{a}}_{ij}\delta {\omega_{a}}^{ij}\right) \\
&-\int_{\mathcal{W}}\sqrt{|\gamma|}d^p\sigma\left({\widehat{\mathcal{T}}^{a}}_{~~b}+2{\mathcal{D}^{ca}}_i {K_{cb}}^{i}+{\mathcal{S}^{a}}_{ij}{\omega_b}^{ij}\right)\edu\mu b\delta\eud\mu a\\
&+\int_{\mathcal{W}}\sqrt{|\gamma|}d^p\sigma\left({\mathcal{P}^{a}}_i-{\mathcal{S}^{b}}_{ji}{K_{b}}^{aj}\right){n_\mu}^{i}\delta {e^\mu}_{a}\\
&+\int_{\mathcal{W}}\sqrt{|\gamma|}d^p\sigma\left({\mathcal{B}^{j}}_i -{\mathcal{D}^{ab}}_i {K_{ab}}^{j}\right){n^{\mu}}_j\delta {n_\mu}^{i} \ .
\end{split}
\ee
Here we have allowed for arbitrary variations of the action \eqref{eq:varL} and not only Lagrangian variations, that is, we have also allowed for variations of the embedding map. On the other hand, the variational principle \eqref{eq:varL} does not allow for internal rotations of the normal vectors $\omega^{ij}=0$ because the action has been formulated in terms of spacetime indices. This implies, using \eqref{eq:varnupL}, that the last variation in \eqref{eq:everything} is proportional to the variation of the background metric, namely ${n^{\mu}}_j\delta {n_\mu}^{i}=-{n^{\lambda}}_j{n^{\rho i}}\delta g_{\lambda\rho}$ and hence symmetric in the indices $i,j$. The first line in \eqref{eq:everything} is the gauge formulation of the variational principle as encountered in \cite{Armas:2013hsa}, hence, for the two principles to be equivalent in full generality one must require
\be \label{eq:cnew}
{\mathcal{T}}^{ab}=\gamma^{ab}\widehat{\mathcal{L}}-2{\mathcal{D}^{c(a}}_i {K_{c}}^{b)i}-{\mathcal{S}^{(a}}_{ij}{\omega}^{b)ij} \ ,~~2{\mathcal{D}^{c[a}}_i {K_{c}}^{b]i}+{\mathcal{S}^{[a}}_{ij}{\omega}^{b]ij}=0 \ ,
\ee
\be \label{eq:cnew1}
{\mathcal{P}^{a}}_i={\mathcal{S}^{b}}_{ji}{K_{b}}^{aj} \ ,~~{\mathcal{B}^{ij}}={\mathcal{D}^{ab(i}} {K_{ab}}^{j)} \ .
\ee
These last two conditions are the constraints \eqref{eq:pt} and \eqref{eq:bt} when no curvature moments are present while the two conditions \eqref{eq:cnew} are new constraints that are only obtained when performing a variation of the mapping functions and requiring the result to be equivalent to that obtained when Lagrangian variations are performed instead. 

If variations of \eqref{eq:everything} would consider the fields ${e^\mu}_{a}$ and ${n_\mu}^{i}$ to be independent of $\gamma^{ab}$, ${K_{ab}}^{i}$ amd ${\omega_{a}}^{ij}$, such as in \cite{Guven:2004wd}, the constraints \eqref{eq:cnew}-\eqref{eq:cnew1} would follow as furthers requirements on the invariance of $S[\Gdd\mu\nu, X^\mu]$. In such case, the second constraint in \eqref{eq:cnew} would be the consequence of invariance under intrinsic local Lorentz transformations, that is, invariance under the infinitesimal transformation $X^{\mu}\to {\Lambda^{\mu}}_b \sigma^{b}$ where ${\Lambda^{\mu}}_b$ is an anti-symmetric tangential matrix of constant coefficients.

The two conditions \eqref{eq:cnew} and the first condition in \eqref{eq:cnew1} are obtained by allowing the mapping functions to vary while the last condition in \eqref{eq:cnew1} is obtained for pure Lagrangian variations. These constraints are left unknown if the gauge variational principle (first line in \eqref{eq:everything}) is used as a starting point. If only the first line in \eqref{eq:everything} is taken into account, then the equations of motion \eqref{eq:genshape} can be obtained as well as the spin conservation equation \eqref{eq:st} when $\mathcal{Q}^{\mu\nu\lambda\rho}=0$ (see e.g. \cite{Armas:2013hsa, Armas:2014rva}). This implies that tangential diffeomorphism invariance in a spacetime formulation, besides \eqref{eq:cnew}-\eqref{eq:cnew1}, implies that the surface, in a gauge formulation, must be reparametrisation invariant and invariant under infinitesimal variations of the normal vectors.

\subsection{Actions for defects with edges and instructive examples} \label{app:DCFTedge}
In this section we provide further details on the actions for CFTs coupled to defects with edges and their corresponding displacement operator, whose surface contribution was analysed in Sec.~\ref{eq:WardDCFT}. We then give concrete well-known examples of actions, including DCFT actions, with the purpose of illustrating the correctness of these Ward identities.

\subsubsection{Ward identities for defects with non-trivial edges}
We begin with the Ward identity for tangential diffeomorphisms \eqref{eq:difftang}. According to \eqref{eq:conditionsdefect}, this Ward identity required the identifications
\be \label{eq:conditions2}
\wt{\mathcal{E}}^{\mu}\hsud\nu\mu=\wt{n}_\mu\wt{B}^{\mu\nu} \ , \quad \wt{\mathcal{E}}^{\lambda\nu}\psud\mu\lambda=\wt{n}_\alpha\wt{B}^{\alpha\lambda\nu}\psud\mu\lambda \ .
\ee
From the first condition above we obtain the requirement
\be \label{eq:firstreq2}
\hsud\alpha\sigma\left(\wt\nabla_\lambda \wt{\mathbb{T}}^{\lambda\sigma}-\wt\Sigma^{\mu\lambda\rho}\Buddd\sigma\mu\rho\lambda-\wt n_\lambda \mathbb{T}^{\lambda\sigma}\right)=\wt{\mathcal{E}}^{\mu}\hsud\alpha\mu~~,
\ee
where $\tilde \Sigma^{\mu\lambda\rho}$ was defined in \eqref{eq:trueSigmas}, $\mathbb{T}^{\lambda\sigma}$ was introduced in \eqref{eq:truestress2}, while $\wt{\mathbb{T}}^{\lambda\sigma}$ is a modification of \eqref{eq:truestress}, namely,
\be
\wt{\mathbb{T}}^{\lambda\sigma}=\wt{\mathcal{T}}^{\lambda\sigma}+{{\wt{\mathcal{P}}}^{\mu}}_{~\nu}\psuu\nu\sigma\hsud\lambda\mu-\hsdu\mu\lambda\wt{\Pi}^{\sigma\mu}-\hsud\lambda\nu\wt\nabla_\mu\wt{\mathcal{D}}^{\mu\nu\sigma} - \wt{\mathcal{S}}^{\mu}{_\alpha}{^\sigma}  \Ksduu\mu\lambda\alpha+\wt n_\mu\hsud\lambda\nu\mathcal{D}^{\mu\nu\sigma}-{\wt{\mathcal{V}}^{\sigma}}_{~~\nu}\hsuu\nu\lambda~~.
\ee
In turn, the second condition in \eqref{eq:conditions2} implies a modification of the constraints \eqref{eq:st1}, in particular we now obtain
\be \label{eq:st1a}
\begin{split}
&\wt{\mathcal{D}}^{\mu\lambda[\sigma}\Ksddu\mu\lambda{\alpha]}+\psud\sigma\lambda\psud\alpha\rho\wt \nabla_\mu \wt{\mathcal{S}}^{\mu\lambda\rho}-\wt n_\lambda\wt n_\mu \mathcal{D}^{\lambda\mu[\sigma}\wt{n}^{\alpha]}-\wt n_\lambda \psud\alpha\mu\psud\sigma\nu\mathcal{S}^{\lambda\nu\mu}=\wt{\mathcal{E}}^{\mu\nu}\psud{[\sigma}\mu\psud{\alpha]}\nu \ ,\\
&\wt{\mathcal{B}}^{\alpha\sigma}-\wt{\mathcal{D}}^{\mu\lambda(\sigma}\Ksddu\mu\lambda{\alpha)}+\wt n_\lambda\wt n_\mu \mathcal{D}^{\lambda\mu(\sigma}\wt{n}^{\alpha)}-\psud{(\sigma}\nu\wt{\mathcal{V}}^{\alpha)\nu} =\wt{\mathcal{E}}^{\mu\nu}\psud{(\sigma}\mu\psud{\alpha)}\nu \ ,\\
&{\wt{\mathcal{P}}^{\sigma}}{_{\alpha}} -\wt{\mathcal{S}}^{\mu}{_\alpha}{^\lambda}\Ksdud\mu\sigma\lambda-{\wt{\mathcal{V}}^{\alpha}}_{~~\mu}\hsuu\mu\sigma=\wt{\mathcal{E}}^{\mu\nu}\psud\alpha\mu\hsud\alpha\nu  \ ,
\end{split}
\ee
 where we have set $\wt{\Pi}^{\mu\nu}=0$ in \eqref{eq:st1} since we did not consider, for simplicity, couplings to the background curvatures in Sec.~\ref{eq:WardDCFT}.
 
We now consider the Ward identity for reparametrisation invariance for which the surface contribution was obtained in \eqref{eq:surfaceinv}. The edge contribution is similar to its surface counterpart, in particular, it reads
 \be \label{eq:edgeinv}
\begin{split}
&\left\langle\left(\wt{\mathcal{D}}^{\nu}\hsud\sigma\nu+\hsuu\alpha\sigma\left({\wt{\mathbb{S}}^{\mu}}_{~~\nu}+{\wt{\mathbb{B}}^{\mu}}_{~~\nu}+{\wt{\mathbb{N}}^{\mu}}_{~~\nu}-{\wt{\mathbb{P}}^{\mu}}_{~~\nu}-{\wt{\mathcal{C}}^{\mu}}_{~~\nu}-{\wt{\mathcal{V}}^{\mu}}_{~~\nu}\right)\Gamma^{\nu}_{\mu\alpha}\right)\boldsymbol{\chi}\right\rangle\\
&+\langle\left(\wt\nabla_\lambda\left({\wt{\mathbb{N}}^{\lambda}}_{~~\nu}-{\wt{\mathbb{P}}^{\lambda}}_{~~\nu}-{\wt{\mathcal{C}}^{\lambda}}_{~~\nu}\right)\hsuu\sigma\nu\boldsymbol{\chi}\right\rangle \\
&=\left\langle\left(\hsud\sigma\nu\wt\nabla_\lambda\left(\wt{\mathbb{T}}^{\lambda\nu}-{\wt{\mathcal{E}}^{\mu}}_{~~\rho}\hsud\lambda\mu\psuu\rho\nu-\hsud\lambda\rho\wt{\mathcal{V}}^{\mu\rho}\psud\nu\mu\right)-\hsud\sigma\nu\wt{\Sigma}^{\mu\lambda\rho}\Buddd\nu\mu\rho\lambda-\tilde n_\lambda \mathbb{T}^{\lambda\nu}\hsud\sigma\nu\right)\boldsymbol{\chi}\right\rangle
\end{split}
\ee
where we have defined
\be \label{eq:defs1a}
\begin{split}
&{\wt{\mathbb{N}}^{\mu}}_{~~\nu}={{\wt{\mathcal{T}}}^{\mu}}_{~~\nu}-\frac{\mathcal{L}_e}{\sqrt{|h|}}\hsud\mu\nu+2{\wt{\mathcal{D}}^{\lambda\mu}}_{~~~\rho} \Ksddu\nu\lambda\rho+{\wt{\mathcal{S}}^{\mu}}_{~~\lambda\rho}\osduu\nu\lambda\rho \ ,~~{\wt{\mathbb{P}}^{\mu}}_{~~\nu}={\wt{\mathcal{P}}^{\mu}}_{~~\nu} +{\wt{\mathcal{S}}^{\alpha}}_{~~\lambda\nu}\Ksduu\alpha\mu\lambda  \ ,\\
&\wt{\mathbb{B}}^{\lambda\rho}=\left(\wt{\mathcal{B}}^{\lambda\rho}-\wt{\mathcal{D}}^{\mu\nu(\lambda}\Ksddu\mu\nu{\rho)}+\wt n_\nu\wt n_\mu \wt{\mathcal{D}}^{\nu\mu(\lambda}\wt n^{\rho)}\right) \ ,\\ 
&\wt{\mathbb{S}}^{\alpha\beta}=\psud\alpha\lambda\psud\beta\rho\left(\wt{\mathcal{D}}^{\mu\nu[\lambda}\Ksddu\mu\nu{\rho]}+\wt\nabla_\mu\wt{\mathcal{S}}^{\mu\lambda\rho}-\wt n_\nu\wt n_\mu \mathcal{D}^{\nu\mu[\lambda}\wt{n}^{\rho]}-\wt n_\nu \psud\rho\mu\psud\lambda\alpha\mathcal{S}^{\nu\alpha\mu}+\wt{\mathcal{V}}^{[\lambda\rho]}\right) \ .
\end{split}
\ee
Furthermore, the Ward identity for diffeomorphism invariance in the presence of defects with non-trivial edges reads
\be \label{eq:warddiffedge}
\begin{split}
&\delta_{\xi}\left\langle\boldsymbol{\chi}\right\rangle-\int_{\mathcal{M}}\sqrt{|g|}d^D x\left\langle\nabla_\mu T^{\mu\nu}_{\text{b}}\boldsymbol{\chi}\right\rangle\xi_\nu+\int_{\mathcal{W}}\sqrt{|\gamma|}d^p\sigma\left\langle\left(B^{\mu\nu}\pud\lambda\nu\nabla_\lambda\xi_\mu-\mathcal{D}_\lambda\xi^{\lambda}\right)\boldsymbol{\chi}\right\rangle\\
&+\int_{\mathcal{W}}\sqrt{|\gamma|}d^p\sigma\left\langle\left(\left({\mathbb{N}^{\mu}}_{\nu}-{\mathbb{P}^{\mu}}_{\nu}-{\mathcal{C}^{\mu}}_{\nu}\right)\partial_\mu\xi^\nu+\left({\mathcal{V}^{\mu}}_\nu-{\mathbb{B}^{\mu}}_{\nu}-{\mathbb{S}^{\mu}}_{\nu}\right)\Gamma^{\nu}_{\mu\alpha}\xi^\alpha\right)\boldsymbol{\chi}\right\rangle\\
&-\int_{\mathcal{W}}\sqrt{|\gamma|}d^p\sigma\left\langle\overline\nabla_\lambda\left({\mathcal{E}^{\mu}}_\nu\hud\lambda\mu\puu\nu\sigma+\hud\lambda\nu\mathcal{V}^{\mu\nu}\pud\sigma\mu\right)\xi_\sigma\boldsymbol{\chi}\right\rangle\\
&+\int_{\partial\mathcal{W}}\sqrt{|h|}d^{p-1}\wt\sigma\left\langle\left(\wt n_\alpha \wt B^{\alpha\nu\mu}\pud\lambda\nu\nabla_\lambda\xi_\mu-\wt{\mathcal{D}}_\lambda\xi^{\lambda}\right)\boldsymbol{\chi}\right\rangle\\
&+\int_{\partial \mathcal{W}}\sqrt{|h|}d^{p-1}\wt \sigma\left\langle\left(\left({\wt{\mathbb{N}}^{\mu}}_{~~\nu}-{\wt{\mathbb{P}}^{\mu}}_{~~\nu}-{\wt{\mathcal{C}}^{\mu}}_{~~\nu}\right)\partial_\mu\xi^\nu+\left({\wt{\mathcal{V}}^{\mu}}_{~~\nu}-{\wt{\mathbb{B}}^{\mu}}_{\nu}-{\wt{\mathbb{S}}^{\mu}}_{~~\nu}\right)\Gamma^{\nu}_{\mu\alpha}\xi^\alpha\right)\boldsymbol{\chi}\right\rangle\\
&-\int_{\partial \mathcal{W}}\sqrt{|h|}d^{p-1}\wt\sigma\left\langle\wt\nabla_\lambda\left({\wt{\mathcal{E}}^{\mu}}_{~~\nu}\hsud\lambda\mu\psuu\nu\sigma+\hsud\lambda\nu\wt{\mathcal{V}}^{\mu\nu}\psud\sigma\mu\right)\xi_\sigma\boldsymbol{\chi}\right\rangle=0~~.
\end{split}
\ee
Finally, the spacetime stress tensor associated with the edges takes the same form as in \eqref{eq:fullstress} but with all quantities replaced by their \emph{tilde} definitions. On the other hand the edge displacement operator takes the form
\be
\begin{split}
\textbf{D}^{\sigma}_{\text{e}}=&\Big[\wt\nabla_\lambda\wt{\mathbb{T}}^{\lambda\sigma}-\wt{\Sigma}^{\mu\lambda\rho}\Buddd\sigma\mu\rho\lambda-\wt n_\lambda\left(\mathbb{T}^{\lambda\sigma}-{\mathcal{E}^{\mu}}_\rho\hud\lambda\mu\puu\rho\sigma-\hud\lambda\rho\mathcal{V}^{\mu\rho}\pud\sigma\mu\right)\\
&-\left(\wt{\mathcal{D}}^{\sigma}+g^{\alpha\sigma}\left({\wt{\mathbb{S}}^{\mu}}_{~~\nu}+{\wt{\mathbb{B}}^{\mu}}_{~~\nu}+{\wt{\mathbb{N}}^{\mu}}_{~~\nu}-{\wt{\mathbb{P}}^{\mu}}_{~~\nu}-{\wt{\mathcal{C}}^{\mu}}_{~~\nu}-{\wt{\mathcal{V}}^{\mu}}_{~~\nu}\right)\Gamma^{\nu}_{\mu\alpha}\right)\\
&-\wt\nabla_\lambda\left({\wt{\mathbb{N}}^{\lambda\sigma}}-{\wt{\mathbb{P}}^{\lambda\sigma}}-{\wt{\mathcal{C}}^{\lambda\sigma}}\right)-\wt\nabla_\lambda\left({\wt{\mathcal{E}}^{\mu}}_{~~\nu}\hsud\lambda\mu\psuu\nu\sigma+\hsud\lambda\nu\wt{\mathcal{V}}^{\mu\nu}\psud\sigma\mu\right)\Big]\widehat{\delta}_e(x) \ .\end{split}
\ee
This completes the analysis of the Ward identities for defects with non-trivial edges.


\subsubsection{Instructive examples}
We now provide a few examples of well defined actions and show how to extract the different objects introduced here in order to verify the Ward identities. Most of these examples have been discussed elsewhere. We will focus on classes of actions that obey $\mathbb{S}^{\mu\nu}=\mathbb{B}^{\mu\nu}=\mathbb{P}^{\mu\nu}=0$ and similarly for the edge counterparts. This class of actions, in the context of DCFTs and in flat space, is the one analysed in \cite{Billo:2016cpy}.

\paragraph{Free scalar field.} Consider a free bulk scalar field  coupled to surface and edge defects in curved space such that 
\be
\begin{split}
S[\Gdd\mu\nu,X^\mu,\wt X^\mu, \phi]=&\frac{1}{2}\int_{\mathcal{M}}\sqrt{|g|}d^Dx\left(\nabla_\mu\phi\nabla^\mu\phi+\tau R\phi^2\right)+\tau_1\int_{\mathcal{W}}\sqrt{|\gamma|}d^p\sigma~\phi(X) \\
&+\tau_2\int_{\partial\mathcal{W}}\sqrt{|h|}d^{p-1}\wt\sigma~\phi(\wt X)~~,
\end{split}
\ee
where $\tau=(D-2)/4(D-1)$ and $\tau_1,\tau_2$ are arbitrary constants. The surface defect can be made conformal if $p=(D-2)/2$, in which case the edge defect cannot be made simultaneously conformal. However, it is nevertheless interesting to keep it in order to show the correctness of the Ward identities. We now evaluate the non-zero quantities of interest, in particular the bulk stress tensor reads
\be
T^{\mu\nu}_{\text{b}}=\nabla^\mu\phi\nabla^\nu\phi-\frac{1}{2}\Guu\mu\nu\nabla_\lambda\phi\nabla^\lambda\phi+\tau\left(R^{\mu\nu}-\frac{1}{2}R\Guu\mu\nu\right)\phi^2-\tau\left(\nabla^\mu\nabla^\nu-\Guu\mu\nu\square\right)\phi^2~~.
\ee
The equation of motion for the field $\phi$ takes the form
\be
\square\phi-\tau R\phi=-\tau_1\widehat\delta(x)-\tau_2\widehat\delta_e(x)~~.
\ee
Using the equation of motion for $\phi$ in the divergence of the bulk stress tensor one obtains
\be \label{eq:divscalar}
\nabla_\mu T^{\mu\nu}_{\text{b}}=-\tau_1\nabla^\nu\phi \widehat\delta(x)-\tau_2\nabla^\nu\phi \widehat\delta_e(x)~~.
\ee
From here one readily obtains that 
\be\label{eq:ees}
\mathcal{E}^{\mu}=-\tau_1\nabla^\mu\phi~~,~~\wt{\mathcal{E}}^{\mu}=-\tau_2\nabla^\mu\phi~~,~~\mathcal{E}^{\mu\nu}=\wt{\mathcal{E}}^{\mu\nu}=0~~. 
\ee
The remaining quantities of interest, such as the surface and edge stress tensors,  are easily obtained from the action and read
\be \label{eq:dds}
\mathcal{T}^{\mu\nu}=\tau_1\huu\mu\nu \phi~~,~~\wt{\mathcal{T}}^{\mu\nu}=\tau_2\hsuu\mu\nu \phi~~,~~\mathcal{D}^\mu=\tau_1\nabla^\mu\phi~~,~~\wt{\mathcal{D}}^\mu=\tau_2\nabla^\mu\phi~~.
\ee
In this specific case, the Ward identity for diffeomorphism invariance \eqref{eq:warddiffedge} implies that
\be \label{eq:wardscalar}
\nabla_\mu T^{\mu\nu}_{\text{b}}=-\mathcal{D}^{\mu}\widehat\delta(x)-\wt{\mathcal{D}}^{\mu}\widehat\delta_e(x)~~.
\ee
Introducing \eqref{eq:dds} into \eqref{eq:wardscalar} leads to \eqref{eq:divscalar} and hence the Ward identity \eqref{eq:wardscalar} is satisfied, as expected. Furthermore, the Ward identities for reparametrisation invariance \eqref{eq:surfaceinv} and \eqref{eq:edgeinv} imply that
\be \label{eq:eomscalar}
\hud\sigma\nu\left(\overline\nabla_\lambda\mathcal{T}^{\lambda\nu}-\mathcal{D}^{\nu}\right)=0~~,~~\hsud\sigma\nu\left(\wt\nabla_\lambda\wt{\mathcal{T}}^{\lambda\nu}-\wt n_\lambda \mathcal{T}^{\lambda\nu}-\widehat{\mathcal{D}}^{\nu}\right)=0~~,
\ee
which upon using \eqref{eq:dds} are seen to be satisfied since $\wt n_\lambda \mathcal{T}^{\lambda\nu}\hsud\sigma\nu=0$. Finally, consider the Ward identity for tangential diffeomorphisms which essentially reduces to \eqref{eq:firstreq1} and \eqref{eq:firstreq2}. Upon using \eqref{eq:ees}, these are seen to be equivalent to \eqref{eq:eomscalar} and hence satisfied.

\paragraph{Minimally coupled $p$-form.} Consider the action of a background $p$-form gauge field coupled to a defect of dimensionality $p$. The action takes the simple form
\be
\begin{split}
S[\Gdd\mu\nu, X^\mu, A_{\mu_1...\mu_p}]=&\frac{1}{2(p+1)!}\int_{\mathcal{M}}\sqrt{|g|}d^Dx\left(F_{\mu_1...\mu_{p+1}}F^{\mu_1...\mu_{p+1}}\right)~~\\
&-\tau_1\frac{\lambda}{p!}\int_{\mathcal{W}}A_{\mu_1...\mu_p}\eud{\mu_1}{a_1}...\eud{\mu_p}{a_p}d\sigma^{a_1}\wedge...\wedge d\sigma^{a_p}~~,
\end{split}
\ee
where $F_{\mu_1...\mu_{p+1}}=dA_{\mu_1...\mu_p}$ and $\tau_1$ is an arbitrary constant. For the defect to be conformal one requires that $p=(D-2)/2$. The coupling to the defect is simply the pull-back of the gauge field onto the surface. The equation of motion for the gauge field reads
\be
\nabla_\mu F^{\mu\mu_1...\mu_p}=-\frac{\tau_1}{p!}\epsilon^{\mu_1...\mu_p}_{||}\widehat\delta(x)~~,
\ee
which can be used to deduce that the r.h.s. of the bulk stress tensor conservation equations leads to a Lorentz force term localised on the defect, that is
\be\label{eq:congauge}
\nabla_\mu T^{\mu\nu}_{\text{b}}=-\frac{\tau_1}{p!}\epsilon^{\mu_1...\mu_p}_{||}{F_{\mu_1...\mu_p}}^{\nu}\widehat\delta(x)~~.
\ee
Consider now the diffeomorphism constraint for this specific case. From \eqref{eq:warddiffedge}, we obtain
\be
\nabla_\mu T^{\mu\nu}_{\text{b}}=\Guu\lambda\nu\overline\nabla_\mu {\mathcal{C}^{\mu}}_{\lambda}+{\mathcal{C}^{\mu}}_\alpha \Gamma^{\alpha}_{\mu\lambda}\Guu\lambda\nu-\mathcal{D}^{\nu}~~.
\ee
Extracting the relevant terms from the action leads to 
\be \label{eq:wardgauge}
{\mathcal{C}^{\mu}}_{\nu}=-\frac{\tau_1}{(p-1)!}\epsilon^{\mu\mu_2...\mu_p}_{||}A_{\nu\mu_2...\mu_p}~~,~~\mathcal{D}^\mu=-\frac{\tau_1}{p!}\epsilon^{\mu_1...\mu_p}_{||}\partial^{\mu}A_{\mu_1...\mu_p}~~.
\ee
Inserting these quantities into \eqref{eq:wardgauge} leads to \eqref{eq:congauge}, as expected.

\paragraph{Coupling between vector fields and extrinsic curvature.} We now consider a case for which a background vector field $\phi^{\mu}$ couples to the extrinsic curvature of the defect. We assume that there is some well-defined bulk action given in terms of this vector field and that gives rise to a non-trivial equation of motion. Here, for the purposes of exemplifying the different terms in the Ward identities, we do not require the exact form of the bulk action as we just want to test the Ward identity for surface reparametrisations. Therefore, consider the defect action
\be
S[\Gdd\mu\nu,X^\mu, \phi^\mu]=\int_{\mathcal{W}}\sqrt{|\gamma|}d^p\sigma\left(\alpha+\tau_1\phi^\mu \phi^\nu\Kddu\mu\nu\rho K_\rho\right)~~,
\ee
where for simplicity we have assumed that the background vector field when restricted to the surface is purely tangential, that is $\phi^{\mu}(X)=\hud\mu\nu\phi^\nu(X)$. From this action we compute the relevant quantities, namely
\be \label{eq:nn0}
\begin{split}
&\mathcal{T}^{\mu\nu}=\alpha \huu\mu\nu+\tau_1\huu\mu\nu\phi^\lambda\phi^\alpha \Kddu\lambda\alpha\rho K_\rho-2\tau_1\phi^\lambda\phi^\alpha\Kddu\lambda\alpha\rho\Kuud\mu\nu\rho~~,\\
&\mathcal{B}^{\mu\nu}=2\tau_1\phi^{\lambda}\phi^{\alpha}\Kddu\lambda\alpha{(\mu}K^{\nu)}~~,~~\mathcal{D}^\mu=2\tau_1\partial^\mu\phi^\lambda\phi^\alpha\Kddu\lambda\alpha\rho K_\rho~~, \\
&\mathcal{D}^{\mu\nu\rho}=\tau_1\phi^\mu\phi^\nu K^\rho+\tau_1\huu\mu\nu \phi^\lambda\phi^\alpha\Kddu\lambda\rho\rho~~,
\end{split}
\ee
from which we can derive that $\mathbb{B}^{\mu\nu}=\mathbb{S}^{\mu\nu}=0$ and
\be \label{eq:nn}
{\mathbb{N}^{\mu}}_\nu=2\tau_1\phi^\mu\phi^\lambda\Kddu\nu\lambda\rho K_\rho~~.
\ee
The Ward identity for reparametrization invariance \eqref{eq:surfaceinv} for this specific case takes the form
\be \label{eq:wardvector}
\begin{split}
\mathcal{D}^\nu\hud\sigma\nu+{\mathbb{N}^{\mu}}_{\nu}\Gamma^{\nu}_{\mu\lambda}\huu\lambda\sigma+\hud\sigma\nu\overline\nabla_\lambda\mathbb{N}^{\mu\nu} = \hud\sigma\nu\overline\nabla_\lambda\mathbb{T}^{\lambda\nu}-\hud\sigma\nu\Sigma^{\mu\lambda\rho}\Buddd\nu\mu\rho\lambda \ ,
\end{split}
\ee
for which the non-manifestly covariant part can be made manifestly covariant using \eqref{eq:nn0}-\eqref{eq:nn} and yields
\be
\mathcal{D}^\nu\hud\sigma\nu+{\mathbb{N}^{\mu}}_{\nu}\Gamma^{\nu}_{\mu\lambda}\huu\lambda\sigma=2\tau_1\phi^\alpha\Kddu\lambda\alpha\rho K_\rho\overline\nabla^\sigma\phi^\lambda~~.
\ee
Using this, together with \eqref{eq:nn0}-\eqref{eq:nn}, one can verify that \eqref{eq:wardvector} is satisfied.

%

\addcontentsline{toc}{section}{References}
\footnotesize
\providecommand{\href}[2]{#2}\begingroup\raggedright\endgroup

\end{document}